\documentclass{aa}  

\usepackage{graphicx,blindtext}
\usepackage{txfonts}
\usepackage{txfonts}
\usepackage{hyperref}
\usepackage{xcolor}
\hypersetup{colorlinks,citecolor=blue,linkcolor=blue,urlcolor=blue}
\usepackage{ulem}
\def\msun{M$_\odot$}
\def\lsun{L$_\odot$}

{}
{}
{}

\newcommand{\iso}[1]{$^{#1}$}

\begin{document}

   \title{Nucleosynthetic yields of intermediate-mass primordial to extremely metal-poor stars}

   \author{P. Gil-Pons
          \inst{1,2}
          \and
          C.L. Doherty \inst{3,4}  \and S.W. Campbell\inst{4,5} \and J. Guti\'errez \inst{1,2}
          }

   \institute{EETAC, Universitat Polit\`ecnica de Catalunya, Campus Baix Llobregat, C3, 08840 Castelldefels, Spain.\\
        \email{pilar.gil@upc.edu}
         \and
             Institut d'Estudis Espacials de Catalunya, Ed- Nexus Campus Nord, Barcelona, Spain.
        \and
            Konkoly Observatory, Hungarian Academy of Sciences, 1121 Budapest.
        \and 
         School of Physics and Astronomy, Monash University, Victoria 3800, Australia
       \and 
       ARC Centre of Excellence for All Sky Astrophysics in Three Dimensions (ASTRO-3D), Australia
             }

   \date{Received ; accepted }

 
  \abstract
   {Stellar models and nucleosynthetic yields of primordial to extremely metal-poor (EMP) stars are crucial to interpret the surface abundances of the most metal-poor stars observed and, ultimately, to better understand the earliest stellar populations. In addition, they are key ingredients of Galactic chemical evolution models.}  
    {We aim to better characterise the evolution and fates, and determine updated nucleosynthetic yields of intermediate-mass stars between primordial and EMP metallicity ($Z=10^{-10}, 10^{-8}, 10^{-7}, 10^{-6}$, and $10^{-5}$). We also probed uncertainties in the nucleosynthesis of the oldest intermediate-mass stars, namely those related to the treatment of convection and convective boundaries and those related to wind prescriptions during the asymptotic giant branch (AGB) phase.}
  {We analyse the evolution of models from their main sequence, through the thermally pulsing AGB (TP-AGB), to the latest stages of their evolution, using the Monash-Mount Stromlo stellar evolution code \texttt{MONSTAR}. The results were post-processed with the code \texttt{MONSOON}, which allowed for the determination of the nucleosynthetic yields of 77 species up to \iso{62}Ni. By comparing them to similar calculations existing in the literature, we inspected the effects of input physics on the nucleosynthesis of EMP models. }
   {From the evolutionary point of view, as reported in former works, we identified proton ingestion episodes (PIEs) in our lowest-mass lowest-metallicity models. Models of $Z=10^{-10}$ and $Z=10^{-8}$ in a narrow initial mass range around 5 \msun{} experience the cessation of thermal pulses, and their final fates as type-I1/2 supernovae cannot be discarded. However, the initial mass range of models eventually leading to the formation of type-I1/2 and electron-capture supernovae is considerably reduced compared to former works. All the models of initial mass $\gtrsim$ 6-7 \msun{} experience a corrosive second dredge-up and, analogously to those experiencing PIEs, undergo significant metal enrichment in their envelopes. The associated increase in their opacities allows them to develop a solar-like TP-AGB or TP-super-AGB, ultimately becoming white dwarfs. Except for those undergoing the cessation of thermal pulses, all of our models show the nucleosynthetic signatures of both efficient third dredge-up and hot-bottom burning, with the activation of the NeNa cycle and the MgAlSi chains. This leads to the creation of vast amounts of CNO, with typical [N/Fe]~>~4), and the characteristic abundance signature [N/Fe]~>~[C/Fe]~>~[O/Fe]. Our nucleosynthetic yields present dramatic differences with respect to recent results existing in the literature for intermediate-mass models of similar metallicities. The reason for these discrepancies lay in the poorly known input physics related to stellar winds and, above all, the treatment of convection and convective boundaries.}
  {}


\maketitle
%

\section{Introduction}

The first generation of stars (Pop III) provided the first elements heavier than lithium in the Universe. Although Pop III star formation theory is still uncertain, it is thought that the Pop III initial mass function (IMF) was biased towards massive stars compared to the current-day IMF (e.g. \citealt{abel02,sharda2022}). As the metallicity of the Universe increased due to successive generations of stars, cooling via metal lines in the star formation process allowed for the formation of lower-mass stars, and the proportion of low- and intermediate-mass stars increased (\citealt{chon2021,sharda2022}). In particular, second stellar generations are expected to have formed from pristine gas mixed with the ejecta from one or a few primordial stars (\citealt{umeda2003,nomoto2013,marassi2014,frebel2015,hartwig2019}) -- and thus any surviving stars can be excellent tracers of the nucleosynthesis of the first generations, providing a fossil record of the first stars.

The amount of observational data of extremely metal-poor (EMP) stars has increased significantly since the early 1990s due to observing programmes such as the HK survey \citep{beers1992}, the Hamburg-ESO survey \citep{christlieb2002}, SkyMapper \citep{keller2007}, the Sloan Extension for Galactic Understanding and Exploration (SEGUE; \citealt{yanny2009}), the Large Sky Area Multi-Object Fibre Spectroscopic Telescope (LAMOST; \citealt{cui2012}), and the PRISTINE survey \citep{starkenburg2014}. In the very near future, the James Webb Space Telescope \citep{zackrisson2011}, which is now in orbit, will further expand our knowledge of the first and early generations of stars.

To understand precisely when low- and intermediate-mass stars started to contribute significantly to the evolution of the baryonic universe, it is essential to have a quantitative understanding of how they evolved and ended their lives, as well as the nature and amounts of the matter they ejected as a function of their initial mass and metallicity, that is, their contribution to the chemical evolution of the Galaxy and other environments.\\
Compared to higher metallicities, the low-metallicity regime of low- and intermediate-mass stellar modelling has been less well-studied. Work overlapping with the scope of the current study includes the PopIII ($Z \sim 0.0$) study by
\citealt{chieffi2001} ($M_{ini}=4-8$~\msun{}),
the metal-poor intermediate-mass star study of \citet{iwamoto2009} 
(at a single metallicity $Z = 10^{-5}$), and the range of low-mass ($1-3 \rm{M}_{\odot}$), low-metallicity models by \citet{campbell2008}. \citet{gilpons2018} reviewed the literature and presented new computations to explore the uncertainties in EMP asymptotic giant branch (AGB) and super-AGB modelling, the key phase for chemical enrichment in low and intermediate-mass stars. We hereafter use the acronym (S)AGB to refer both to AGB and super-AGB stars. 

The current study aims to expand the mass and metallicity range of EMP and lower metallicity modelling, in particular, to provide a self-consistent grid of models. Moreover, we aim to provide an analysis of the evolution, final fates, and nucleosynthetic yields of our models. Our grid ranges from `primordial' ([Fe/H]~$=-10$) to EMP ([Fe/H]~$=-3$) models in the mass range 3.0~to~8.5~M$_{\odot}$.
Comparison between our theoretical gas yields and surface abundance observations of the most metal-poor stars may help to constrain the properties of the first stellar populations (e.g. \citealt{frebel2015}) and even the evolution of the primitive IMF (\citealt{suda2013}, \citealt{ishigaki2018}). They may provide insight on the Milky Way formation history (e.g. \citealt{gibson2003}, \citealt{white2000}, \citealt{tumlinson2010}, \citealt{sestito2021}), and on (ultra-faint) dwarf galaxies, whose metallicities lay between the very metal-poor and the extremely metal-poor regime \citep{roederer2017,simon2019,skuladottir2020}. Our models are also highly relevant in the context of Galactic Chemical Evolution (e.g. \citealt{chiappini1997},
\citealt{tsujimoto2012}, \citealt{brusadin2013},
\citealt{spitoni2017}, \citealt{prantzos2018}, \citealt{millan2019}, \citealt{kobayashi2020}),
and could give insight on the pollution history of the intracluster medium. Recently, a new very low metallicity system was discovered -- a stellar stream with [Fe/H]~$\simeq -3.8$ \citep{martin2022a}. Thought to be the remnant of a disrupted globular cluster, this system is also in the metallicity range of our modelling.\\
This work is organised as follows: Section \ref{sec:codes} describes the structural evolution and nucleosynthetic postprocessing codes used. Section \ref{sec:evolution} details the main aspects of the evolution and final fates of our models, highlighting the effects of initial metallicity. Section \ref{sec:yields} describes the main nucleosynthetic evolution trends and chemical yields,  Section \ref{sec:summary} summarises our key results, and finally, Section \ref{sec:conclusions} presents the main conclusions derived from our work.

\section{Description of codes and input physics}\label{sec:codes}

\subsection{Structure evolution code}\label{sec:code_str}

We performed our calculations with the Monash-Mount Stromlo code \texttt{MONSTAR} (\citealt{frost1996}, \citealt{campbell2008}), described in GP21. \texttt{MONSTAR} only includes the isotopes relevant for the structural evolution, that is to say those involved in important energy-producing reactions.  
Nuclear reaction rates are from \citet{harris1983}, \citet{caughlan1988} and \texttt{NACRE} \citet{angulo1999}, except for \iso{14}N + p which is from \texttt{REACLIB} \citep{champagne2005}. Hot-CNO cycle reactions and neutron captures, which are important for calculating proton-ingestion episodes (PIEs), are not included in our structure evolution code. 
We give in Appendix \ref{sec:reactions} additional details on the reactions considered in our structure evolution codes and, in Section \ref{sec:pies}, a brief discussion on the use of these reactions in the calculation of PIEs.

The treatment of the structure and composition (mixing and burning) equations follows a partially simultaneous approach \citep{lattanzio2015}, in which solutions are calculated at each iteration until the model is converged. By carefully adapting the time-step based on a range of physical and composition variables, this method is approximately equivalent to a simultaneous solution approach \citep{stancliffe2006}. Following PIEs is quite time-consuming since the time steps get very small because physical changes occur on very short timescales. In such cases, time steps in our calculations can reach values as small as $10^{-8}-10^{-7}$ yr. 

\subsubsection{Treatment of mixing, convection, and convective boundaries}
\label{subsec:convection}

Convection is treated following the Mixing-Length Theory (MLT; \citealt{bom58}) in the specific implementation presented in \cite{coxgiuli1968}. The MLT is a rather crude approximation for the intrinsically multi-dimensional phenomenon of convection. It relies on a proper calibration of $\alpha_{MLT}$, the ratio of the mixing length to pressure scale height. After calibration to the solar parameters, a value for $\alpha_{MLT}=1.75$ was determined. At present, it is impossible to rely on calibrations with observations at the extremely low metallicities we are considering since there are so few (or zero) well-characterised stars in this regime. Given this unavoidable uncertainty, a full parametric study covering different values of $\alpha_{MLT}$, as proposed by \cite{chieffi2001}, would be of interest. However, such a study is beyond the scope of the present work.

The Schwarzschild criterion combined with a search for `convective neutrality' was used to locate convective boundaries \citep{lat86}.  
In this method, convective boundaries are located by extrapolating the value of the quotient of the radiative and the adiabatic gradients ($\nabla_{rad}/\nabla_{ad}$) from the two convective mesh points closest to the Schwarzschild boundary to the first adjacent radiative mesh point. 
\citet{frost1996} showed how this implementation enhanced the efficiency of the third dredge-up (TDU) episode. 

Mixing is instantaneous and homogeneous, except during proton-ingestion episodes (PIEs). In such cases, evolutionary timesteps become shorter than the convective turnover times, so we use a time-dependent diffusive approach \citep{campbell2008}.

\subsubsection{Opacities}
For the present work, we updated the conductive opacities \citep{cassisi2007,potekhin2015}, and the variable-composition low-temperature molecular opacities, extending the metallicity range considered in \cite{constantino2014} and GP21, to cover values between $Z=10^{-10}$ and ${\rm Z=10^{-5}}$. Low-temperature opacity data were from \texttt{AESOPUS} (\citealt{lederer2009,marigo2009}). Interior stellar opacities are from \citet{igl96}. 

\subsubsection{Mass-loss rates}
Mass-loss rates in our low-metallicity models are insignificant before the end of core-He burning. During the early AGB and the thermally pulsing AGB (TP-AGB) phase, we used the prescription by \citet{bloecker1995} with $\eta=0.02$, as proposed by \citet{ventura2001} based on observations of Li-rich giants in the Large Magellanic Cloud. This wind prescription is commonly used by authors computing very low-Z models (e.g. \citealt{herwig2004} and \citealt{ritter2018}). GP21 showed that this prescription yielded a better agreement between theory and observations of EMP stars.
We note that the wind prescription by \citet{bloecker1995} accounts for the effects of radiation pressure on dust grains. This might represent a problem because dust formation in our extremely metal-poor stars is uncertain and perhaps inefficient (V21). However, no existing wind prescription for primordial to EMP stars is a reliable option, since there are no direct observational constraints. Generally, the current prescriptions used for mass-loss for extremely metal-poor stars are not based on first principles -- they are empirical fits to observations of stars of considerably higher metallicities. However, if the situation arises whereby metal-poor stars undergo strong self-pollution, for example, through third dredge-up or proton-ingestion episodes (see Sec.~\ref{sec:yields}), and reach globular cluster or even near-solar metallicities\footnote{In terms of metallicity Z -- the stars will still be Fe-poor.}, then the mass-loss rates calibrated to more metal-rich populations may be valid.

\subsection{Nucleosynthesis code}

The stellar evolution results were postprocessed to obtain the gas yields of 77 species. We used the code \texttt{MONSOON} \citep{can93,lug04}, whose most recent updates were described in \citet{doherty2014a} and GP21. 
Abundance variations due to nuclear reactions and time-dependent convection are calculated using a 'donor-cell' scheme, and space and temporal resolution aim to resolve regions of strong abundance variations. Nuclear reaction rates are from \texttt{JINA} \citep{cyb10}, and additional p- and $\alpha-$captures are from \cite{iliadis2001}, \cite{hale2002}, \cite{hale2004}, and \cite{karakas2006}. A `g' particle is used as a proxy for s-process elements \citep{lug04}, and neutron-captures on isotopes not considered in our network are accounted for using the neutron-sink approach \citep{jorissen1989, lugaro2003, herwig2003}. 

i-process nucleosynthesis could occur during proton-ingestion episodes (see, e.g. \citealt{campbell2010,dardelet2014,Hampel2016,choplin2021}, and references therein). Given the relatively limited number of isotopes and reaction rates we are considering, and the fact that our neutron sink is adapted to the s-process, we do not expect to obtain reliable results for i-process nucleosynthesis. The extension to heavy elements nucleosynthesis will be analysed in a future work. 
Initial abundances were solar-scaled from \cite{grevesse1996}.

\section{Evolution of the models}\label{sec:evolution}

\begin{figure}
    \centering
    \includegraphics[width=0.88\linewidth]{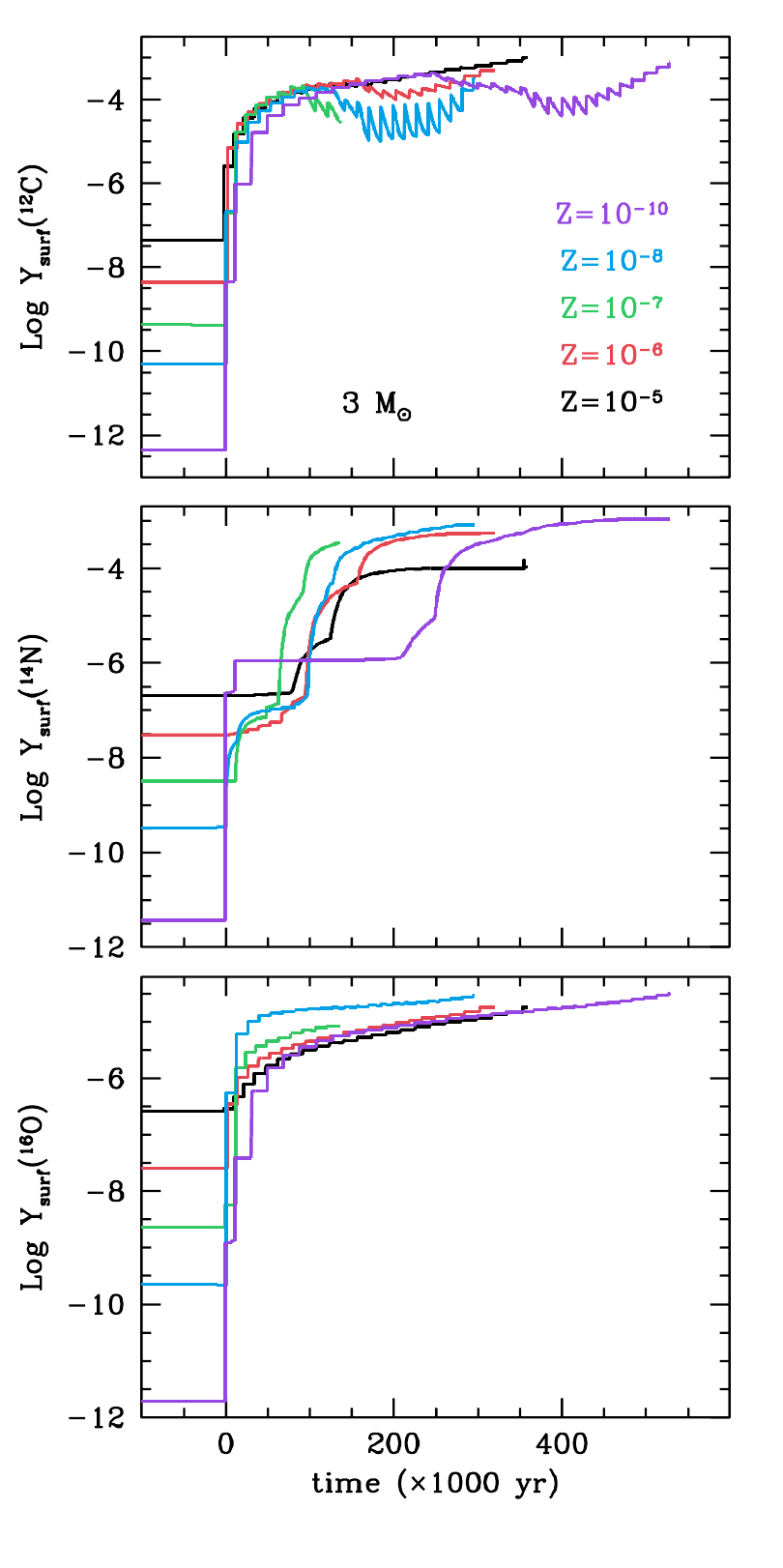}
    \caption{Surface C, N, and O molar fractions of our 3 \msun models at different metallicities. Zero time marks the beginning of the TP-AGB phases.}
    \label{fig:cno}
\end{figure}

\begin{table*}
\begin{center}
\caption{Relevant structure and composition parameters for our models. }
\end{center}
\centering
\vspace{-1cm}
\begin{tabular}{llllllllllll}\\
\hline
\hline
&  & & CHB & ${\rm CHeB_{begin}}$ 
& ${\rm CHeB_{end}}$ 
& ${\rm CCB_{begin}}$ & pre-SDU & post-SDU &   \\
${\rm M_{ZAMS}}$ & 
${\rm \tau_{CHB}}$ & ${\rm \tau_{CHeB}}$ &
${\rm M_{cc}}$ & 
${\rm M_{HexC}}$ & 
${\rm M_{HexC}}$ & 
${\rm M_{C-ign}}$ & $\rm M_{HexC}$ &
${\rm M_{HexC}}$ &
$X_{\rm SDU}(C)$ & $X_{\rm SDU}(N)$ & $X_{\rm SDU}(O)$\\
\msun & 
Myr & Myr &
\msun & \msun & \msun & \msun & \msun & \msun & & & \\[1pt]
\hline \\[-4pt]
\multicolumn{12}{c}{{\bf ${\bf Z=10^{-10}}$}}\\
\hline
\hline \\[-4pt]
3.0 & 197.0 & 13.6 & 0.23 & 0.44 & 0.60 & -- & 0.67 & 0.77 &  5.4$\times 10^{-12}$ &  5.2$\times 10^{-11}$ &  3.1$\times 10^{-11}$ \\
4.0 & 95.6 &  8.0 & 0.38 & 0.54 & 0.77 & -- & 0.86 & 0.85 & 3.8$\times 10^{-12}$ & 6.3$\times 10^{-11}$ & 2.4$\times 10^{-11}$\\
5.0 &  57.5 &  6.0 & 0.55 & 0.64 & 0.94 & -- & 0.98 & 0.90 &  5.7$\times 10^{-9}$ &  5.5$\times 10^{-10}$ &  2.3$\times 10^{-11}$\\
6.0 &  38.7 &  4.9 & 0.77 & 0.74 & 1.12 & -- & 1.16 & 0.95 &  3.4$\times 10^{-7}$ &  8.4$\times 10^{-10}$ &  1.5$\times 10^{-10}$ \\
7.0 &  28.5 &  3.7 & 0.92 & 0.83 & 1.30 & 0.53 & 1.33 & 1.00 &  5.7$\times 10^{-6}$ &  1.3$\times 10^{-9}$ &  1.1$\times 10^{-8}$ \\
7.5 &  25.2 &  3.2 & 1.00 & 0.87 & 1.40 & 0.41 & 1.04 & 1.04 &  1.8$\times 10^{-5}$ &  3.0$\times 10^{-8}$ &  7.3$\times 10^{-8}$\\     8.0 &  22.6 &  2.8 & 1.25 & 0.93 & 1.52 & 0.31 & 1.15 & 1.11 &  2.2$\times 10^{-5}$ &  7.9$\times 10^{-8}$ &  9.5$\times 10^{-8}$  \\ 
8.5 &  20.5 &  2.5 & 1.34 & 0.96 & 1.64 & 0.21 & 1.50 & 1.16 &  1.4$\times 10^{-4}$ &  1.5$\times 10^{-5}$ &  3.5$\times 10^{-6}$\\
\\ [-4pt]
\multicolumn{12}{c}{{\bf ${\bf Z=10^{-8}}$}}\\
\hline
\hline\\[-4pt]
3.0 & 190.2 & 21.4 & 0.23 & 0.37 & 0.69 & -- & 0.81 & 0.81 &  5.9$\times 10^{-10}$ &  4.7$\times 10^{-9}$ &  3.5$\times 10^{-9}$ \\
4.0 & 89.8 & 13.6 & 0.37 & 0.42 & 0.86 & -- & 0.89 & 0.87 &  4.5$\times 10^{-10}$ &  5.4$\times 10^{-9}$ &  2.9$\times 10^{-9}$ \\
5.0 & 53.5 &  9.9 & 0.61 & 0.47 & 1.02 & -- & 1.04 & 0.91 &  7.0$\times 10^{-9}$ &  5.9$\times 10^{-9}$ &  2.4$\times 10^{-9}$\\
6.0 &  36.9 &  8.1 & 0.96 & 0.51 & 1.18 & -- & 1.20 & 0.96 &  5.1$\times 10^{-7}$ &  6.3$\times 10^{-9}$ &  2.4$\times 10^{-9}$\\
7.0 & 28.3  & 6.0  & 1.31 & 1.15 & 1.68 & -- & 1.68 & 1.02 & 9.8$\times 10^{-6}$ & 6.5$\times 10^{-9}$ &  2.9$\times 10^{-8}$\\
7.5 &   25.5 &  5.1 & 1.57 & 0.61 & 1.39 & 0.42 & 1.30 & 1.06 &  3.7$\times 10^{-5}$ &  4.5$\times 10^{-8}$ &  2.4$\times 10^{-7}$\\    8.0 &  23.2 &  4.3 & 1.84 & 0.66 & 1.50 & 0.31 & 1.39 & 1.12 &  4.9$\times 10^{-5}$ &  6.1$\times 10^{-7}$ &  3.6$\times 10^{-7}$ \\ 
8.5 &  21.3 &  3.7 & 2.11 & 0.75 & 1.56 & 0.21 & 1.17 & 1.17 &  2.0$\times 10^{-4}$ &  2.6$\times 10^{-5}$ &  7.1$\times 10^{-6}$ \\
\\ [-4pt]
\multicolumn{12}{c}{{\bf ${\bf Z=10^{-7}}$}}\\
\hline
\hline\\[-4pt]
3.0 & 178.0 & 17.8 & 0.23 & 0.57 & 0.77 & -- & 0.79 & 0.83 & 6.4$\times 10^{-9}$ & 4.2$\times 10^{-8}$ & 4.0$\times 10^{-8}$ \\
4.0 &  86.5 & 14.2 & 0.44 & 0.62 & 0.95 & -- & 0.97 & 0.85 & 4.8$\times 10^{-9}$ & 5.0$\times 10^{-8}$ & 3.3$\times 10^{-8}$ \\
5.0 &  55.5 & 13.8 & 0.84 & 0.50 & 1.12 & -- & 1.13 & 0.91 &  7.4$\times 10^{-9}$ &  5.5$\times 10^{-8}$ &  2.9$\times 10^{-8}$ \\
6.0 &  39.7 &  9.3 & 1.35 & 0.58 & 1.29 & -- & 1.30 & 0.96 & 2.1$\times 10^{-7}$ & 5.8$\times 10^{-8}$ & 2.6$\times 10^{-8}$ \\
7.0 &  30.8 &  6.3 & 1.80 & 0.66 & 1.49 & 0.43 & 1.50 & 1.04 &  1.6$\times 10^{-5}$ &  7.6$\times 10^{-8}$ &  8.3$\times 10^{-8}$ \\
7.5 &  27.7 &  5.4 & 2.05 & 0.72 & 1.62 & 0.30 & 1.36 & 1.11 &  5.0$\times 10^{-5}$ &  6.4$\times 10^{-7}$ &  4.0$\times 10^{-7}$ \\
8.0 &  25.2 &  4.7 & 2.30 & 0.76 & 1.74 & 0.22 & 1.75 & 1.18 &  5.8$\times 10^{-4}$ &  8.5$\times 10^{-5}$ &  4.3$\times 10^{-5}$ \\
8.5 &  23.0 &  4.0 & 2.55 & 0.79 & 1.85 & 0.12 & 1.30 & 1.24 &  1.4$\times 10^{-3}$ &  2.7$\times 10^{-4}$ &  5.7$\times 10^{-4}$\\
\\
\multicolumn{12}{c}{{\bf ${\bf Z=10^{-6}}$}}\\
\hline
\hline\\[-4pt]
3.0 & 170.8 & 19.3 & 0.28 & 0.62 & 0.81 & -- & 0.81 & 0.83 & 7.4$\times 10^{-8}$ & 3.7$\times 10^{-7}$ & 4.5$\times 10^{-7}$\\
4.0 &  91.3 & 16.8 & 0.72 & 0.66 & 0.99 & -- & 0.87 & 0.87 & 5.8$\times 10^{-8}$ & 4.5$\times 10^{-7}$ & 3.8$\times 10^{-7}$  \\
5.0 &  59.2 & 14.7 & 1.18 & 0.55 & 1.19 & -- & 0.92 & 0.91 &  5.3$\times 10^{-8}$ &  4.9$\times 10^{-7}$ &  3.4$\times 10^{-7}$ \\
6.0 &  42.6 &  9.7 & 2.23 & 0.65 & 1.42 & -- & 1.03 & 0.99 &  1.9$\times 10^{-6}$ &  5.2$\times 10^{-7}$ &  3.2$\times 10^{-7}$ \\
7.0 &  33.0 &  6.6 & 2.03 & 0.75 & 1.66 & 0.31 & 1.15 & 1.11 &  4.1$\times 10^{-5}$ &  7.1$\times 10^{-7}$ &  5.7$\times 10^{-7}$\\
7.5 &  29.3 &  5.7 & 2.21 & 0.80 & 1.79 & 0.22 & 1.81 & 1.17 &  4.8$\times 10^{-4}$ &  6.7$\times 10^{-5}$ &  2.7$\times 10^{-5}$\\
8.0 & 26.4  & 4.9  & 2.42 & 1.59 & 2.13 & 0.18 & 2.14 & 1.25 &  1.9$\times 10^{-3}$ &  1.2$\times 10^{-4}$ &  7.5$\times 10^{-4}$\\
\\
\end{tabular}
\tablefoot{${\rm M_{ZAMS}}$ represents the initial mass of our models. 
${\rm \tau_{CHB}}$ and ${\rm \tau_{CHeB}}$ represent, respectively, the duration of the core H-burning and the core He-burning phases. 
${\rm M_{\rm cc}}$ represents the maximum size of the convective core during core H-burning (CHB).  
${\rm {M_{\rm HexC}}}$ in columns 5, 6, 8 and 9 refer to the size of the H-exhausted core at the beginning, at the end of core He-burning (CHeB), and just before and after the second dredge-up. M$_{\rm C-ign}$ gives the internal mass point of C ignition.
$X_\mathrm{SDU}(C)$, $X_\mathrm{SDU}(N)$ and $X_\mathrm{SDU}(O)$ are surface abundances at the end of the second dredge-up. All masses are given in solar units.
The end of CHB was taken when central H-abundance $X_{\rm c}(H) < 10^{-8}$.
The beginning of CHeB was taken when $\rm L_{\rm He}$= 100 L$_{\odot}$ and
the end of CHeB was taken when central the He-abundance $X_{\rm c}(\text{He}) < 10^{-8}$.}
\label{tab:ref1}
\end{table*}

\begin{table*}[hbt!]
 \caption{Main characteristics of the TP-(S)AGB of our models.}
    \centering
    \begin{tabular}{lcccccccccccc}
    \hline
    \hline
    $M_\mathrm{ini}$ & $N_{TP}$ & $\tau_{TP-(S)AGB}$ 
    & $\Delta t_{IP}$ & $M_{c,ini}$ & $M_{c,f}$ & $M_{env,f}$ & $\langle T_{HeBS}\rangle$ & 
    $\langle T_{HBS}\rangle$ & $\langle 
    T_{BCE}\rangle$ & $M_{dredge}^{tot}$ & 
    $\langle\lambda\rangle$ & $\langle\dot 
    M_{wind}\rangle$\\[1pt]
    \msun & & Myr & yr & \msun & \msun & \msun & $MK$ & $MK$ & $MK$ &  \msun& &\msun/yr\\[1pt]
       \hline \\
            \multicolumn{13}{c}{{\bf ${\bf Z=10^{-10}}$}}\\[1pt]
    \hline
    \hline\\[-4pt]
        $3^{\bigstar}$  & 35(2) & 0.78 & 17990 & 0.77 & 0.79 & < 0.1 & 313  & 82  & 36 &  0.14 & 0.87 & $2.8\times 10^{-6}$\\
        $4^{\bigstar}$  & 41(1) & 0.33 & 7712 & 0.85 & 0.87 & < 0.1 &  312 & 99  & 58 &  0.12 & 0.76 & $9.5\times 10^{-6}$\\
        $5^{\clubsuit}$  & 482 & 0.59 & 1471 & 0.89 & 1.14 & 3.08 & 223  & 157  & 26 &  -- & -- & $1.3\times 10^{-6}$\\
        $6$  & 77 & 0.99 & 2927 & 0.95 & 0.97 & < 0.1 & 331  & 111  & 86 &  0.19 & 0.80 & $2.9\times 10^{-5}$\\
        $7$  & 69 & 0.13 & 1658 & 1.00 & 1.02 & < 0.1 & 329  & 116  & 103 &  0.05 & 0.71 & $4.7\times 10^{-5}$\\
        $7.5$  & 82 & 0.09 & 1053 & 1.04 & 1.06 & 0.18 & 327  & 123  & 118 &  0.04 & 0.70 & $6.6\times 10^{-5}$\\
        $8.0$  & 103 & 0.05 & 488 & 1.11 & 1.12 & 0.10 & 323  & 135  & 130 &  0.02 & 0.62 & $1.3\times 10^{-4}$\\
        $8.5$  & 102 & 0.03 & 285 & 1.16 & 1.17 & 0.21 & 328  & 136  & 132 &  0.01 & 0.57 & $2.3\times 10^{-4}$\\
        \\
                \multicolumn{13}{c}{{\bf ${\bf Z=10^{-8}}$}}\\[1pt]
    \hline
    \hline\\[-4pt]
        $3^{\bigstar}$  & 27(2) & 0.34 & 15206 & 0.81 & 0.83 & < 0.1 & 316  & 85  & 47 &  0.11 & 0.87 & $6.1\times 10^{-6}$\\
        $4^{\bigstar}$  & 36(1) & 0.30 & 6719 & 0.87 & 0.88 & < 0.1 & 318  & 104  & 61 &  0.11 & 0.77 & $1.0\times 10^{-5}$\\
        $5^{\clubsuit}$  & 402 & 0.54 & 1418 & 0.91 & 1.14 & 3.11 & 230  & 154  & 24 &  -- & -- & $1.8\times 10^{-6}$\\
        $6$  & 58 & 0.17 & 2693 & 0.96 & 0.97 & 0.12 & 322  & 114  & 81 &  0.29 & 0.70 & $2.9\times 10^{-5}$\\
        $7$  & 69 & 0.11 & 1442 & 1.02 & 1.03 & < 0.1 & 323  & 120  & 107 &  0.07 & 0.57 & $5.3\times 10^{-5}$\\
        $7.5$  & 82 & 0.10 & 983 & 1.06 & 1.07 & 0.17 & 329  & 123  & 118 &  0.03 & 0.70 & $6.4\times 10^{-5}$\\
        $8.0$  & 95 & 0.05 & 492 & 1.12 & 1.13 & 0.18 &  330 & 138  & 133 &  0.02 & 0.63 & $1.3\times 10^{-4}$\\
        $8.5$  & 97 & 0.04 & 296 & 1.17 & 1.18 & 0.23 &  331 & 134  & 130 &  0.01 & 0.60 & $2.0\times 10^{-4}$\\
        \\
              \multicolumn{13}{c}{{\bf ${\bf Z=10^{-7}}$}}\\[1pt]
    \hline
    \hline\\[-4pt]
        $3^{\bigstar}$  & 24 & 0.38 & 12700  & 0.82 & 0.84 & < 0.1 & 310 & 94 & 46 & 0.10 & 0.75 & $5.6\times 10^{-6}$ \\
        $4$  & 38 & 0.28 & 78590 & 0.87 & 0.88 & < 0.1 & 317 & 105 & 58 & 0.19 & 0.69 & $1.1\times 10^{-5}$\\   
        $5$ &  60 & 0.07 & 3929 & 0.91 & 0.94 & < 0.1 & 311 & 109 & 76 & 0.06 & 0.57 & $1.6\times 10^{-5}$\\    
        $6$  & 58 & 0.05 & 2688 & 0.97 & 0.98 & < 0.1 & 327 & 114 & 86 & 0.06 & 0.73 & $1.1\times 10^{-5}$\\
        $7$  & 72 & 0.01 & 1082 & 1.05 & 1.07 & < 0.1 & 326 & 122 & 116 & 0.04 & 0.70 & $5.6\times 10^{-5}$\\
        $7.5$ & 85 & 0.06 & 534 & 1.11 & 1.12 & 0.09 & 324 & 131 & 126 & 0.05 & 0.64 & $1.0\times 10^{-4}$\\
        $8.0$ & 72 & 0.02 & 285 & 1.18 & 1.19 & 0.25 & 336 & 130 & 125 & 0.007 & 0.61 & $3.2\times 10^{-4}$\\
        $8.5$ & 87 & 0.002 & 153 & 1.23 & 1.24 & 0.24 & 347 & 131 & 126 & 0.009 & 0.50 & $4.5\times 10^{-4}$ \\
        \\
            \multicolumn{13}{c}{{\bf ${\bf Z=10^{-6}}$}}\\[1pt]
    \hline
    \hline\\[-4pt]
        $3$  & 26 & 0.42 & 14140 & 0.82 & 0.84 & < 0.1 & 317  & 89  & 42 &  0.08 & 0.80 & $2.3\times 10^{-6}$\\
        $4$  & 36 & 0.30 & 8424 & 0.87 & 0.88 & 0.22 & 325  & 96   & 65  & 0.15  & 0.86  & $9.7\times 10^{-6}$\\
        $5$  & 45 & 0.27 & 5070 & 0.89 & 0.92 & < 0.1 & 326  & 104  & 75 & 0.06  & 0.82 & $1.5\times 10^{-5}$\\
        $6$  & 54 & 0.13 & 2128 & 0.99 & 1.01 & 0.16 & 327  & 114  & 93 & 0.05  & 0.77 & $3.7\times 10^{-5}$\\
        $7$  & 65  & 0.06 & 668 & 1.11 & 1.12 & 0.27 & 350  & 127  & 121 & 0.02  & 0.81 & $9.4\times 10^{-5}$\\
        $7.5$  & 75  & 0.02 & 260 & 1.17 & 1.18 & 0.27 & 353  & 132  & 129 &  0.006 & 0.47 & $2.7\times 10^{-4}$\\
        $8.0$  & 92  & 0.01 & 139 & 1.24 & 1.25 & 0.34 & 349  & 131  & 125 &  0.005 & 0.47 & $4.0\times 10^{-4}$\\
       \\
        \end{tabular}
    \tablefoot{ Models marked with a star (${\bigstar}$) experience proton-ingestion episodes. Thermal pulses gradually decrease in intensity and vanish in the models marked with a club symbol (${\clubsuit}$). In these cases models fail to converge while their convective envelopes have very high masses ($\gtrsim$ 3 \msun{}), and thus their fates are unknown (see text for details). 
$M_\mathrm{ini}$ corresponds to the initial mass. $N_{TP}$
is the number of thermal pulses and, in parenthesis, the number of proton-ingestion episodes.
$\tau_{TP-(S)AGB}$ , and $\Delta t_{IP}$ are, respectively, the duration of the TP-(S)AGB (given from the first thermal pulse until the end of our computations), and the interpulse period. $M_{c,ini}$, $M_{c,f}$ and $M_{env,f}$ are the masses of the H-exhausted cores prior to the TP-(S)AGB, the masses of the H-exhausted cores, and the remnant envelopes at the end of the TP-(S)AGB. $\langle T_{HeBS}\rangle$, $\langle T_{HBS}\rangle$,  $\langle T_{BCE}\rangle$ are the temperatures at the times of
maximum luminosity for each pulse, given at the centre of the He-burning
shell, at the centre of the H-burning shell, and at the base of the convective envelope, and averaged over the number of thermal pulses in each
sequence. $M_{dredge}^{tot}$ is the total mass dredged-up
$\langle\lambda \rangle$ is averaged over the number of thermal
pulses in each case, and $\langle\dot M_{wind}\rangle$ is the average
mass-loss rates due to winds, that is, the envelope mass lost over the
duration of the (S)AGB phase. }
    \label{tab:evol1}
\end{table*}

\begin{figure*}
    \centering
    \includegraphics[width=0.90\linewidth]{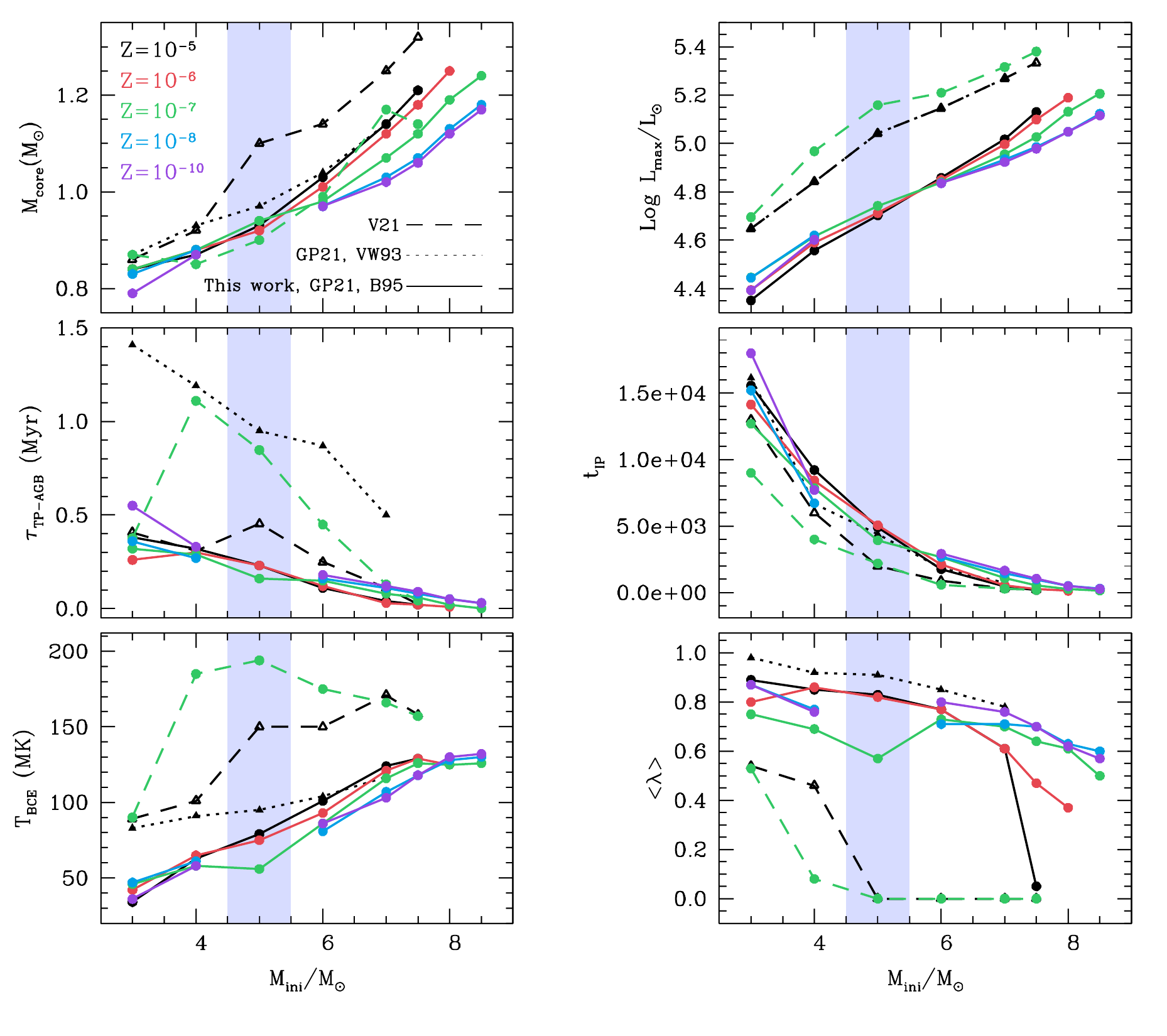}
    \caption{Key TP-(S)AGB parameters for our intermediate-mass models of different metallicities.    ${\rm M_{core}}$ is the core mass at the end of the TP-(S)AGB calculations, ${\rm Log \: L_{MAX}}$ the maximum luminosity, 
    $\tau_{\rm{TPAGB}}$ the duration of the TP-(S)AGB phase, 
    ${\rm t_{IP}}$ the average interpulse period, 
    ${\rm T_{BCE}}$ the maximum temperature at the base of the convective envelope, and $<\lambda>$ is the average of the TDU parameter.
    ${\rm N_{TP}}$ is the number of thermal pulses. We note that 5 \msun{} models of $Z=\times 10^{-8}$ and $Z=10^{-10}$ are not shown because they do not follow a standard TP-AGB evolution (see main text for details). For the sake of comparison, ${\rm Z=10^{-5}}$ models from GP21 with different wind prescriptions (\citealt{bloecker1995}, B95, in solid and \citealt{vassiliadis1993}, VW93 in dotted lines), and the models from \citet{ventura2021}  are also included (V21, dashed lines). 
    We note that, for the latter, maximum (instead of average) values of $\lambda$ are shown.
    Shaded areas in light indigo represent the approximate mass limits for which the final fates of $Z=10^{-10}$ and $Z=10^{-8}$ models are unknown (see main text for details).}
\label{fig:tpagb}
\end{figure*}

\subsection{Evolution before the TP-AGB}

The evolution during core H- and He-burning of intermediate-mass metal-poor stars was thoroughly described in \citet{gilpons2018}. The trends detailed for the lowest metallicity cases ($Z=10^{-10}$ and $10^{-8}$) are similar to those found in \cite{chieffi2001} and \cite{siess2002}. All the above works reported the onset of H-burning through the pp-chains in primordial models and the gradual increase in the relevance of the CNO-cycle as small amounts of C (yielding abundances $X_C \sim 10^{-8}$) form in the core by 3-$\alpha$ reactions. 

Our model stars do not climb the first giant branch after the end of core H-burning. Instead, they experience their first giant ascent and associated envelope pollution event after the end of core He-burning. This first pollution event is still referred to as a second dredge-up (SDU) in analogy to the equivalent process in higher-metallicity ($Z \gtrsim 10^{-3}$) stars, which do undergo the first dredge-up event at the beginning of the RGB. During the SDU, the base of the stellar convective envelope reaches down to regions previously processed by H-burning. Thus, the envelope becomes enhanced in helium and other H-burning products (see Sec.~\ref{sec:sdu} for details).

Models of masses $\gtrsim$ 5-6 \msun{} experience a significant enrichment in C, unlike the typical SDU in which C is depleted through the CN cycle. This occurs through `corrosive' SDU, in which the base of the convective envelope reaches further down than in a standard SDU and not only allows for He and CN-product surface enhancement, but also the mixing of the envelope with matter processed by He-burning \citep{gilpons2013,doherty2014b}. 
This process is favoured at very low metallicity because the higher structural temperatures allow for the occurrence of He-burning very close to the base of the H-burning shell.
Due to corrosive SDU, the more massive models also reach very high He surface abundances. 

Regarding the trends of core mass with metallicity, we find that H-exhausted core masses after the SDU tend to increase with increasing metallicity (just as core masses do at the end of core-H and core-He burning; e.g. \cite{gilpons2021}. The behaviour of the base of the convective envelope (BCE) during the SDU determines the core masses at the beginning of the TP-AGB and, ultimately, the temperatures in nuclearly active regions, the efficiency of the TDU, the wind efficiency, and the duration of the TP-AGB itself (see Figure \ref{fig:cno}). It also determines whether or not a normal TP-AGB ensues. These differences can drastically affect  nucleosynthetic evolution and yields.

\subsection{Main characteristics of the TP-AGB phase}

\begin{figure}
    \centering
    \includegraphics[width=1\linewidth]{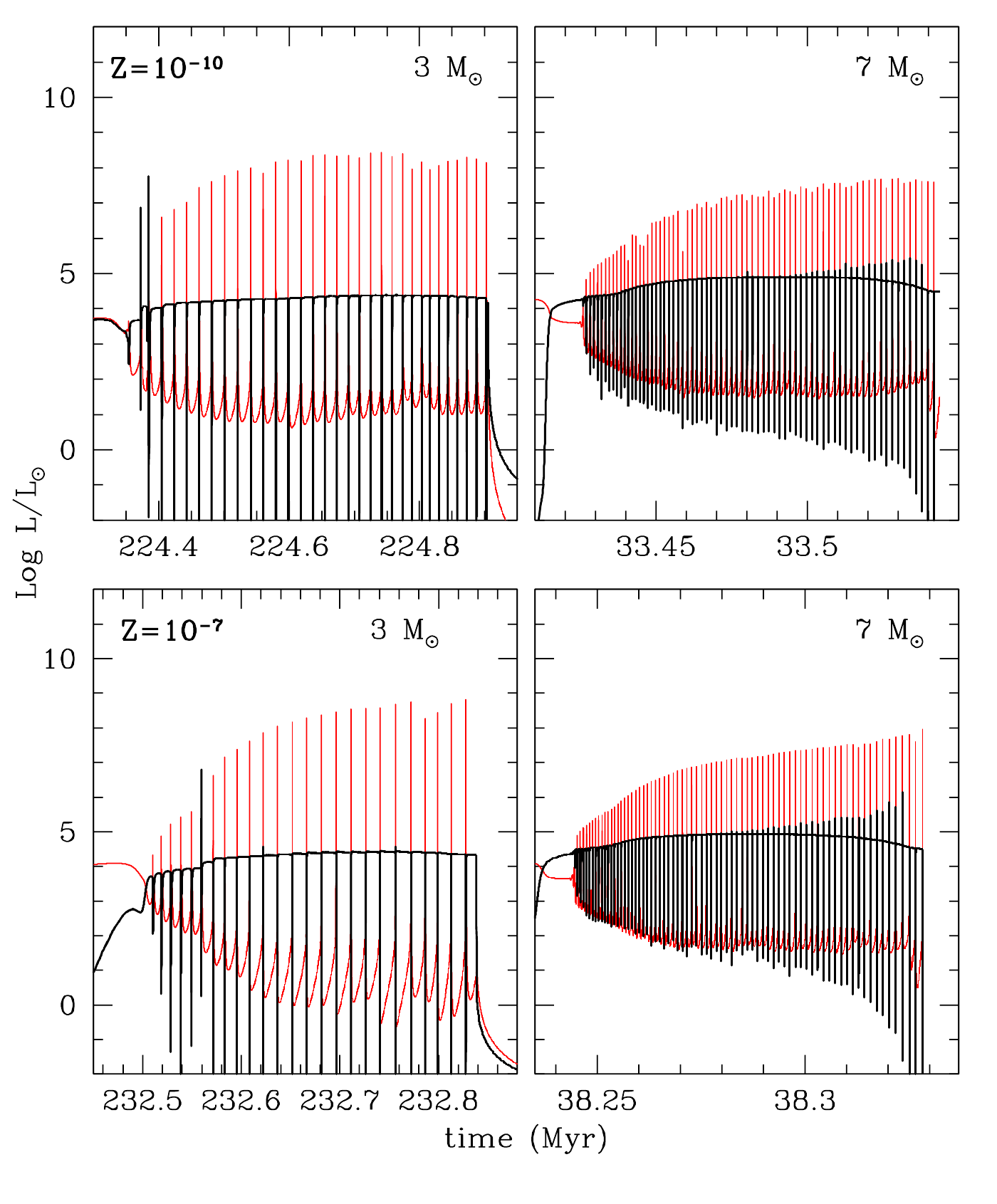}
    \caption{Luminosities  H-burning, $L_{H}$, in black and to He-burning, $L_{He}$, in red, for some selected models of metallicity $Z=10^{-10}$ (upper panel), $Z=10^{-7}$ (lower panel), during the TP-(S)AGB phase.}
    \label{fig:dsfs}
\end{figure}

Most of our models experience a normal thermally-pulsing AGB or super-AGB phase (TP-(S)AGB hereafter), efficient third dredge-up (TDU) or hot-TDU, as well as hot-bottom burning. The main characteristics of the TP-(S)AGB phase are shown in Table \ref{tab:evol1} and Figure \ref{fig:tpagb}. The trends for the most massive cases are quite clear in the metallicity range we are considering. As metallicity decreases, H-exhausted core masses become lower, and TDU efficiency, represented by the $\lambda$ parameter\footnote{$\lambda=\frac{\Delta M_\mathrm{dredge}}{\Delta M_\mathrm{core}}$, where $\Delta M_\mathrm{dredge}$ is the H-exhausted core mass dredged up by the convective envelope after a thermal pulse, and $\Delta M_\mathrm{core}$ is the amount by which the core has grown during the previous interpulse period.},
is higher. Highly efficient TDU (as well as SDU) is attributed to the lower entropy barrier characteristic of low-metallicity stellar structures (see, e.g. \citealt{fujimoto2000}). 
Average values of $\lambda$ range between 0.47 for our 8~\msun{} model of $Z=10^{-6}$, and 0.87, for our 3~\msun{} models of the lowest metallicity values (Table~\ref{tab:evol1}).

Efficient dredge-up does not only affect the surface composition but also hampers core growth and lowers the temperature of active regions. 
In addition, the interpulse periods and the duration of the TP-(S)AGB are longer for the most metal-poor massive AGB and super-AGB stars. Values of the average ${\rm T_{BCE}}$ range between 36~MK for the 3~\msun{} model of $Z=10^{-10}$ and up to 138~MK for our most massive models of the same metallicity.  

On the other hand, GP21 reported increasingly higher core masses, ${\rm T_{BCE}}$, and shorter interpulse periods for massive AGB and super-AGB stars with decreasing Z, for metallicities between ${\rm Z=10^{-3}}$ and ${\rm Z=10^{-5}}$. 
Thus, there is a change in trends in the characteristics of TP-(S)AGB models at ${\rm Z=10^{-5}}$. \cite{siess2007} also reported a similar tendency variation for core masses of super-AGB models at $Z=10^{-4}$. The overall compactness and higher temperature of the lowest metallicity models are probably the main reasons for this trend change.

For models of initial mass $\lesssim$ 5 \msun{} the tendency is altered. Core masses, both at the beginning and end of the TP-AGB, are very mildly affected by initial metallicity (except for the primordial 3 \msun{} case). In these cases, the occurrence or absence of dual shell flashes (e.g. \citealt{campbell2008}), and varying TDU efficiencies drastically affect the stellar structure. Furthermore, as we see in the next section, they determine the occurrence of a `normal' TP-AGB and even the final fate of some models. We note that values for the 5 \msun{} models of Z~$=10^{-8}$ and $10^{-10}$ are not shown in Figure \ref{fig:tpagb} because their thermal pulses vanish. Models fail to converge shortly afterwards while keeping very massive envelopes at the end of our calculations (see Section \ref{sec:cessation}). 

Once the total stellar mass has reduced to about 1.6-1.7 \msun{} the high T$_{BCE}$ $\gtrsim$ 30 MK values cannot be sustained, and the HBB ceases. 
Our calculations end with increasingly lower envelope masses as the metallicity decreases. 
Values are as low as $10^{-5}$~\msun{}.
This supports the hypothesis that the instability that halts the evolution might be related to an Fe-opacity peak near the base of the convective envelope in advanced stages of the TP-(S)AGB \citep{lau2012}.

\subsection{Comparison to previous models}\label{sec:comparison}

Our results differ considerably from those from \citealt{ventura2021} (V21), who presented yields for low- and intermediate-mass stars of $Z = 3\times 10^{-5}$ and $Z = 3\times10^{-7}$. This is not surprising, given the important differences between the codes and the input physics used. V21 used \texttt{ATON} (see \citealt{ventura2013} and references therein), whereas we used \texttt{MONSTAR}, described in Section~\ref{sec:codes}.

Detailed comparisons between the results of the \texttt{ATON} code and similar versions of our code \texttt{MONSTAR}, which we refer to here collectively as \texttt{MONASH}, have been made previously over a range of masses and metallicities (see, e.g. \citealt{doherty2014b,ventura2016a}).  
Specifically, \cite{ventura2015a} reported that the treatment of convection and its boundaries in the envelopes of AGB stars during the third dredge-up is the main cause of differences between the surface chemistry results of the two codes.  
\cite{ventura2018} also used \texttt{ATON} and \texttt{MONASH} to study dust production in solar-metallicity AGB stars, and found a good agreement for surface composition evolution in the case of low-mass AGB stars.
However, differences arose for models of initial mass between 3 and 5 \msun{}, as \texttt{MONASH} produced higher carbon yields than \texttt{ATON}. Discrepancies became increasingly important for models of initial masses $\gtrsim$ 4-5 \msun{}, as the variations between the HBB efficiencies of the two codes grew.  These authors confirmed the conclusion by \cite{ventura2015a} that the former model discrepancies were mostly due to differences in convection and convective boundary modelling, which affects the luminosity and the duration of AGB models experiencing HBB.

In light of this, the differences between the results of the current study and V21 are also most likely due to the treatment of convection and its boundaries. A further substantial factor is likely the difference in stellar wind mass-loss rates.
With respect to convection and its boundaries, V21 used the Full Spectrum of Turbulence model \citep{can91}, together with exponentially decaying overshooting on convective boundaries. Our code implements the mixing-length theory and a search for convective neutrality (see \ref{subsec:convection}), respectively.
In terms of wind prescriptions, V21 used the one proposed by \citet{vassiliadis1993} (VW93), which for the considered metallicities, yields lower mass-loss rates than the prescription used in this work (B95 with $\eta=0.02$).
These different treatments hardly affect the H-exhausted core masses during the main central burning stages. However, they lead to critical divergences in the TP-(S)AGB structural evolution and nucleosynthesis, which become apparent in Figure \ref{fig:tpagb}.

\begin{figure}[t]
    \centering
    \includegraphics[width=1.05\linewidth]{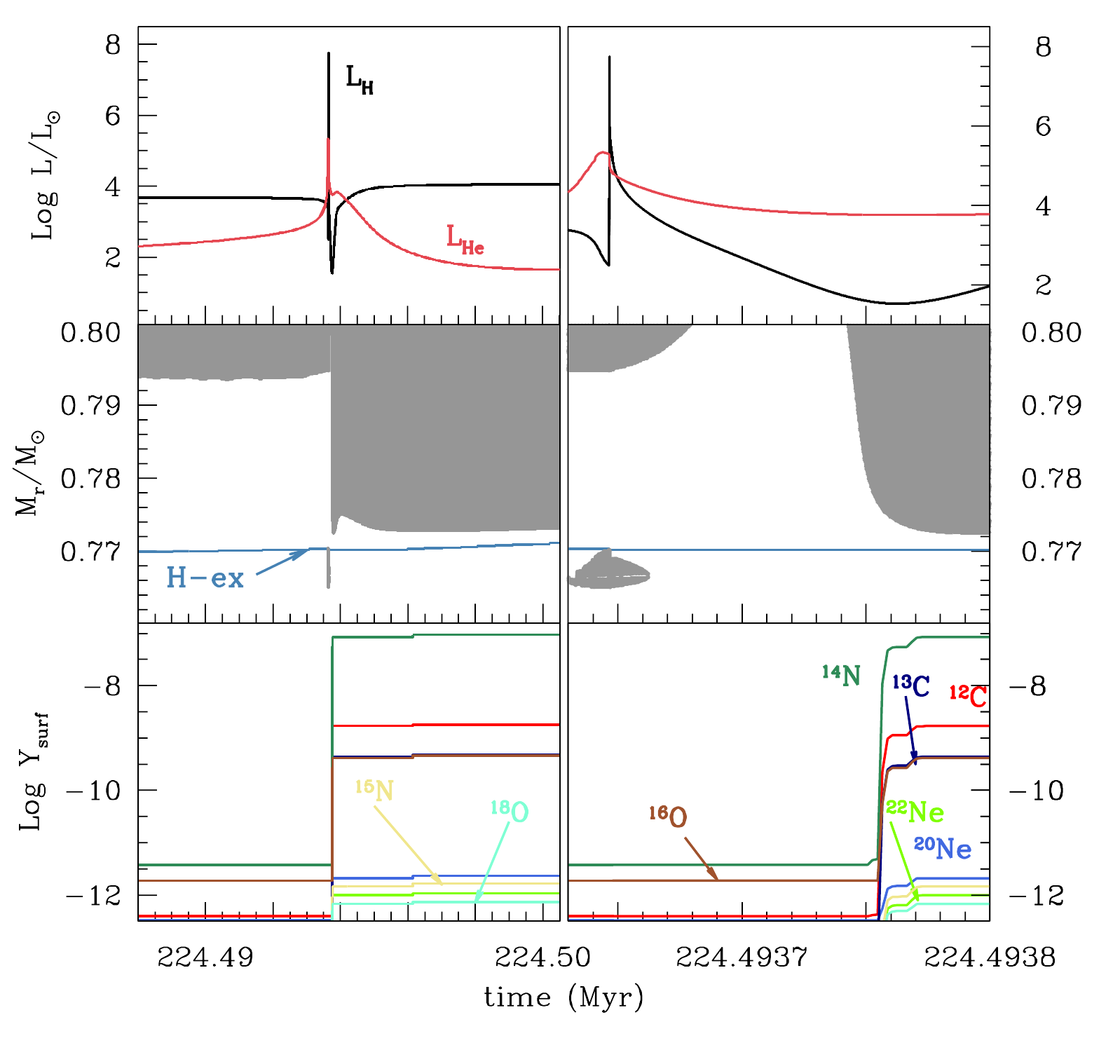}
    \caption{{\bf Upper panels:} Luminosities of H-burning $L_{H}$ (black) and He-burning $L_{He}$ (red), for the 3~\msun{} $Z=10^{-10}$ model, during the occurrence of the first dual-shell flash. Right hand panels are a zoomed version of the left panels. {\bf Middle panels:} Convective zones (grey) and H-rich region (blue). {\bf Lower panels:} Logarithm of surface abundances of the most abundant metals. }
    \label{fig:dsf1}
\end{figure}

\begin{figure}[t]
    \centering
    \includegraphics[width=1.05\linewidth]{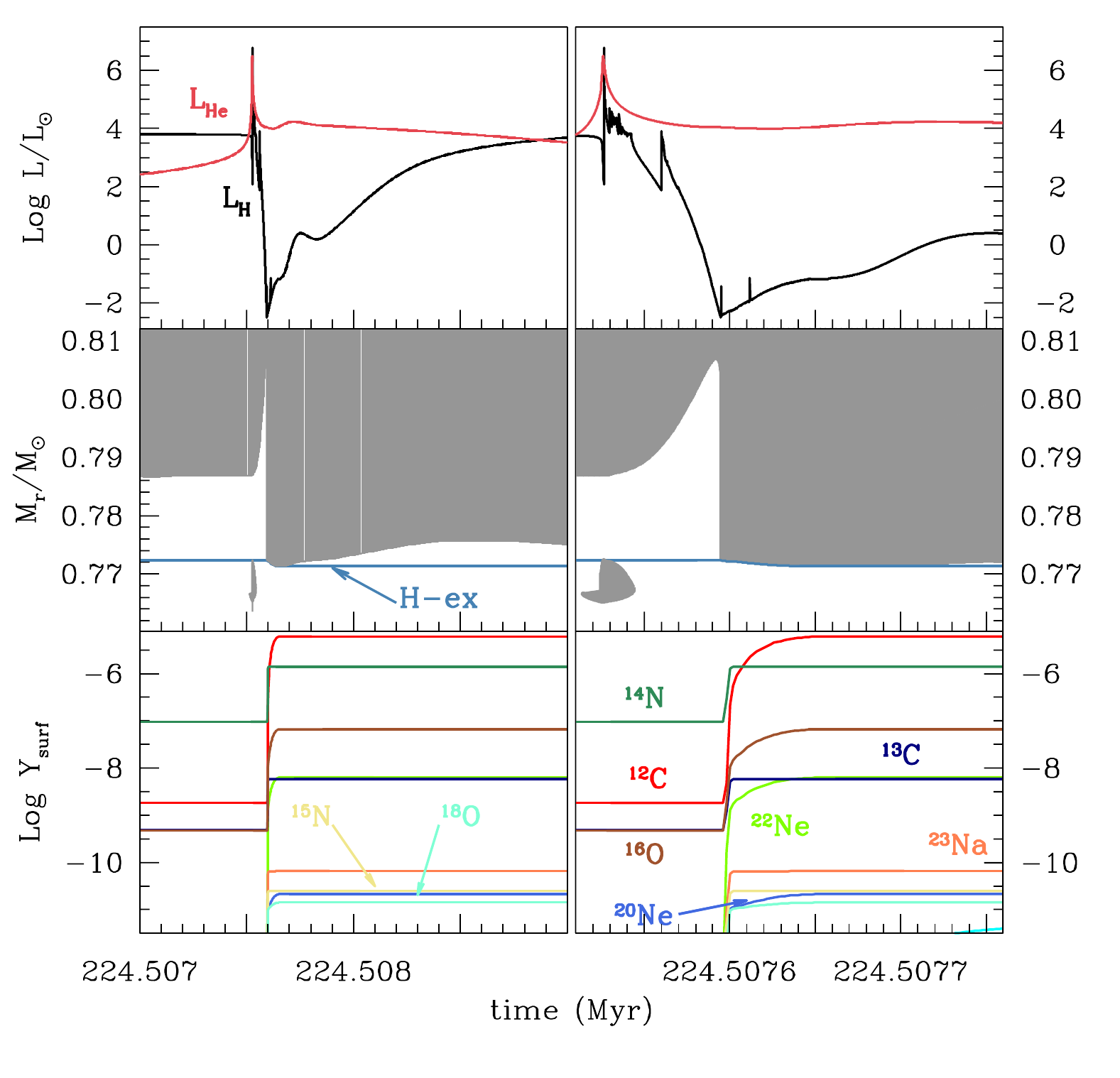}
    \caption{Same as Figure \ref{fig:dsf1}, but during the second dual-shell flash.  }
    \label{fig:dsf2}
\end{figure}

The main discrepancies between the results from V21 and the present work can be summarised as follows. Firstly, 
the different treatments of convection and, specially, of convective boundaries produce very different TDU efficiencies. Our models systematically yield a more efficient TDU than those of V21. TDU is negligible for all their models of initial mass $\geq 5$~\msun{}. On the other hand, all the average values of $\lambda$ in our models are above $\simeq 0.5$, regardless of their initial metallicity, and all our models of initial mass $\lesssim 7$~\msun{} have $\lambda_{\rm max} \geq 0.9$ (panel 6 of Fig.~\ref{fig:tpagb}). We note that the issue of large uncertainties associated with TDU has been frequently addressed in the literature, for instance, in \cite{karakas2014}, and references therein. As discussed above, differences between our code and \texttt{ATON} have been explored previously.

Secondly, TP-(S)AGB cores in the present work grow more slowly and reach lower final masses than those reported in V21. The FST formalism used by V21 favours higher temperatures and consequently allows for more efficient nuclear burning than the mixing-length recipes. Together with the different treatment of convective boundaries, it is also responsible, as mentioned above, for less efficient TDU and, thus, for a lower degree of core erosion. In addition, the lower wind rates of VW93 also favour longer TP-(S)AGB phases and higher final core masses. Quantitatively, our H-exhausted cores grow $\lesssim 0.03$~\msun{} when calculated with the wind prescription by B95, or at most 0.11~\msun{} when VW93 was used for the ${\rm Z=10^{-5}}$ models in GP21. The cores in V21 grow up to 0.3~\msun{} for the 4 and 5~\msun{} cases of $Z=3\times 10^{-7}$. Besides altering core growth and surface composition, TDU also allows for the cooling of nuclearly active regions. This causes a longer average interpulse period, lower maximum temperatures at the BCE, and surface luminosities in our models compared to those of V21.
    
Another important difference between the results from V21 and the present work is that the trend of increasing temperature at the BCE with decreasing metallicity reported in V21 and \citet{dellagli2019} is not reproduced in our most massive models ($\geq$ 6\msun{}). The reason is again the higher TDU efficiency, and thus the cooling of nuclearly active regions, with decreasing metallicity. TIn addition, the use of the MLT for the treatment of convection and the lower core masses in our models also lead to temperatures at the BCE and maximum luminosities that are significantly lower than those in the models by V21. HBB is much more efficient and can  potentially lead to more advanced nucleosynthesis.  
    
Even when comparing V21 models with those from GP21 which were computed with the same mass-loss prescription, quite large differences remain. The duration of the TP-AGB is longer in the models by GP21, however the final core masses and maximum luminosities in V21 are higher, as well as the ${\rm T_{BCE}}$ which reach up to 190~MK for their 5~\msun{} model of $Z=10^{-7}$. As we subsequently see in Sec.~\ref{sec:abund}, these differences translate into critical divergences in nucleosynthetic yields. The differences identified, even when using the same wind prescription, highlight the relevance of the treatment of convection and convective boundaries and their crucial effects on the overall evolution of model stars.

\begin{figure}[t]
    \centering
    \includegraphics[width=1.0\linewidth]{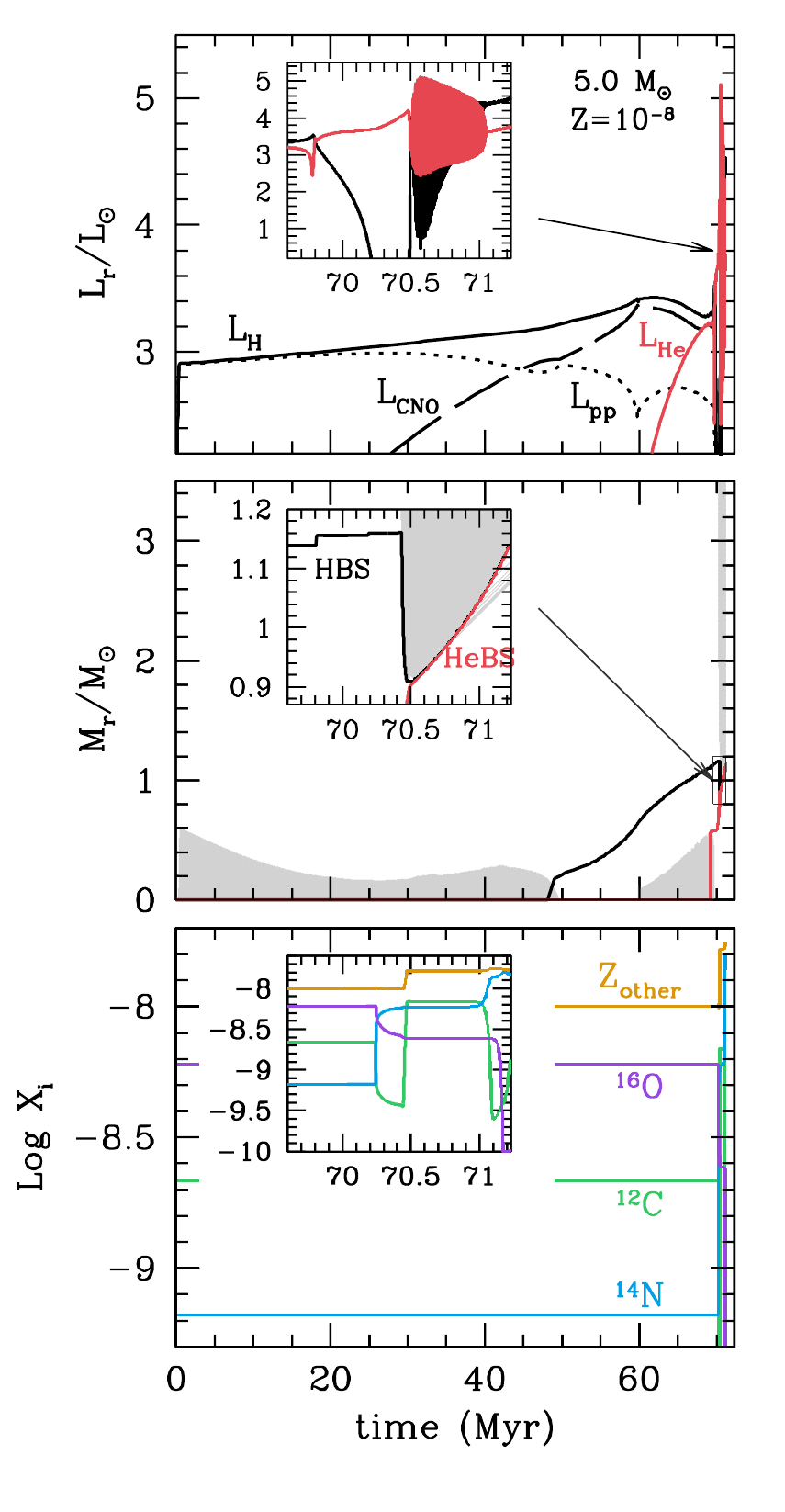}
    \caption{Details of the evolution of our 5~\msun model of $Z=\times 10^{-8}$. Upper panel shows the temporal evolution of luminosities associated with H-burning ($L_{H}$, black), and He-burning luminosity ($L_{He}$, red). In the case of H-burning, the contributions due to the pp-chains ($L_{pp}$) and the CNO-cycle ($L_{CNO}$) are shown, respectively, in black dotted and dashed lines. Middle panel shows the evolution of convective regions (grey areas), and the location of the H-burning shell (HBS) and the He-burning shell (HeBS) in black and red lines respectively. Lower panel shows the evolution of surface abundances of \iso{12}C (green), \iso{14}N (blue), \iso{16}O (green), and $Z_{other}$ (gold), which represents all the species of mass higher than \iso{16}O. 
    .} 
    \label{fig:sn15}
\end{figure}
\subsection{Occurrence of AGB proton-ingestion events}\label{sec:pies}

A subset of our models start the TP-AGB with proton ingestion episode(s) during the first one or two thermal pulses. These are known as `dual shell flashes' (DSF, see, e.g. \citealt{cassisi1996,fujimoto2000,chieffi2001,herwig2003,campbell2008,choplin2021}). During a thermal pulse, the He-burning convective zone extends outwards beyond the H-He discontinuity and causes the ingestion of protons into the He-burning region. This region is characterised by temperatures $\sim 10^8$~K, so the protons burn very rapidly and a H-flash ensues\footnote{This occurs while the He-flash is still ongoing, hence the term `dual flash'.}, with luminosities which can reach $10^8$ \lsun.

Our 3 and 4~\msun{} models of $Z=10^{-8}$ and $10^{-10}$, as well as our 3~\msun{} model of $Z=10^{-7}$ all undergo strong DSFs (Table~\ref{tab:evol1}). These stars are at the lowest-mass and lowest-metallicity end of our grid.
The first DSF in our $Z=10^{-10}$, 3~\msun{} model, which occurs during the first thermal pulse, reaches a H-burning peak luminosity of $5\times10^7$~\lsun{}. 
After the DSF subsides, the base of the convective envelope moves in, mixing up H-shell burning products. In this case the envelope is not able to reach the H-exhausted core, however there is an increase in \iso{14}N of 4 orders of magnitude, and an increase in \iso{12}C and \iso{16}O surface abundances of 3 orders of magnitude (see Figure \ref{fig:dsf1}). Despite these large relative changes, the total envelope metallicity is still low, with $Z_{env}\simeq 10^{-6}$.
The peak H-burning luminosity of a second DSF, which occurs during the second thermal pulse, is $6 \times 10^6$~\lsun{}. In this case, the base of the convective envelope is able to reach the H-exhausted core, and mixes up He-burning products, finally driving envelope metallicities to $Z_{env}\simeq 10^{-4}$. The DSF events drive carbon surface abundances values well above those of nitrogen and oxygen (Fig.~\ref{fig:dsf2}). With a high C/O ratio this temporarily favours higher surface opacities and radii and, ultimately, stronger stellar winds and a shorter TP-AGB. Later this model reaches temperatures at the base of the convective envelope high enough for the occurrence of hot-bottom burning, which then primarily converts the C to N (Fig.~\ref{fig:cno}). The DSF events have also been suggested as possible sites for s-process and i-process neutron-capture nucleosynthesis (e.g. \citealt{fujimoto2000,campbell2010,herwig2011,dardelet2014,Hampel2016,choplin2021}).

We simulated DSF proton-ingestion episodes with our 1D hydrostatic code, which solves in a partially simultaneous way the structure and composition equations (Sec.~\ref{subsec:convection}), and includes a limited set of isotopes and reaction rates. Therefore it is important to highlight a few caveats. Firstly, PIEs are, by nature, multi-dimensional convective-reactive phenomena, where the burning timescales are similar to the mixing timescales. Thus their proper analysis must be done by carefully simulating small enough timescales, in order to mimic the coupling between structure and mixing/burning. Other works more focused on PIEs (e.g. \citealt{cristallo2009}, \citealt{cristallo2011}, \citealt{cristallo2016}, \citealt{cirillo2022}) use codes which simultaneously solve the structure and composition equations.
Secondly, the nuclear reaction rates in our structural-evolution code should involve the treatment of the hot CNO-cycle and neutron-captures, which can make non-negligible energy contributions. Specifically, \cite{cristallo2009} showed that the \iso{13}C($\alpha$,n)\iso{16}O is the most important reaction in terms of the energetics of PIEs. 
Thirdly, our current nucleosynthesis code uses a neutron sink which is adapted for the s-process. Thus it is likely to not be accurate given the high neutron density conditions that occur during PIEs. Finally, rather than 1D hydrostatic codes, 3D hydrodynamical models should be used to follow convective-reactive phenomena such as PIEs (e.g. \citealt{mocak2010,stancliffe2011,woodward2015,jones2016a}).

Despite these caveats, our PIE calculations serve two main purposes: (a) They allow us to identify the metallicity and mass-ranges for which PIEs occur, and (b) they help us to determine whether metal pollution due to these episodes is enough to allow for further evolution as `normal' AGB stars, that is, to experience intense thermal pulses and end their lives as white dwarfs.
We also note that, in terms of light element production which we focus on in our study, the details of the PIEs are often not important to the final yields of 3-4~\msun models (the only models that experience PIEs in our grid), since subsequent TDU and HBB tend to erase the elemental signature of the PIEs (see eg. Figure~3 of \citealt{campbell2008}). In terms of heavy elements, the PIEs will likely be important, and this is the subject of our future work.

Our models of initial masses $\gtrsim$ 4-5~\msun{} experience mild proton-ingestion episodes at advanced stages of the TP-(S)AGB phase (see $L_{H}$ spikes in Fig.~\ref{fig:dsfs}). As the thermal pulses become increasingly intense, the associated expansion and cooling of the regions near the HeBS becomes more significant. The BCE is able to advance further inwards after the pulse, and mix protons into high temperature regions, where He-burning is still active.
Peak H-luminosities in this case are $\lesssim 10^5-10^6$~\lsun{}, and are not accompanied by variations in surface abundances of light elements different from those of a normal TP flash and its subsequent TDU episode. 
We note that the H-luminosity peaks we report are different from the ones described in \citet{jones2016a}. In our case, the advance inwards of the BCE occurs when the pulse-driven convective zone has already disappeared, whereas the H-luminosity peaks in \citet{jones2016a} were  a proton-ingestion episode in the He-burning convective zone. The H-luminosity spikes in our model stars stop once the T$_{BCE}$ becomes $\lesssim$ 30 MK.

\subsection{Final fates}\label{sec:cessation}

\begin{figure}
    \centering
    \includegraphics[width=1.0\linewidth]{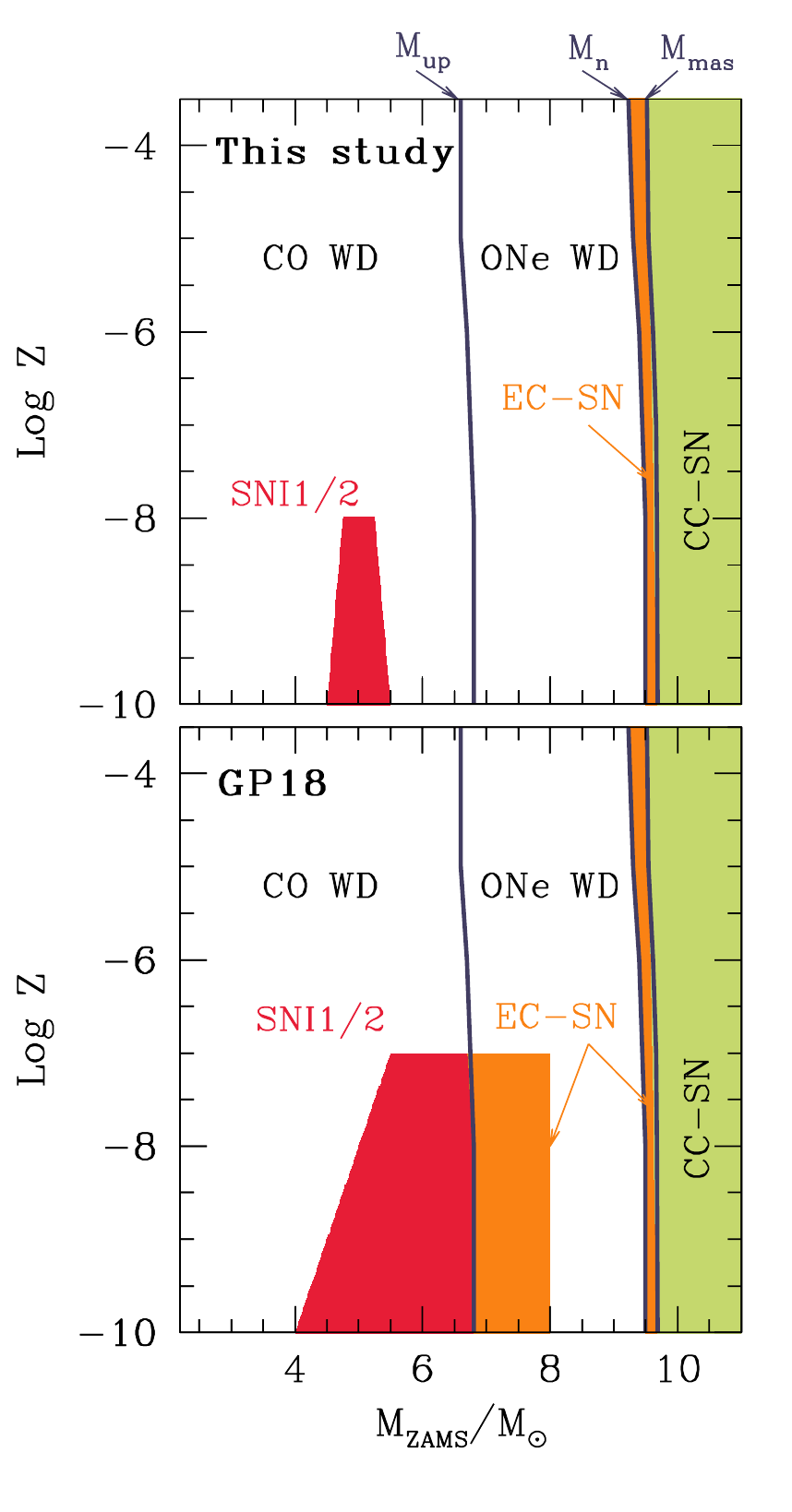}
    \caption{{\bf Upper panel:} Expected fates of our model stars at different metallicities, as resulting from the present work (variable composition low-temperature opacities and mass-loss rates from \citealt{bloecker1995}). {\bf Lower panel:} Expected fates according to the calculations by \citet{gilpons2018} (constant composition low-temperature opacities and mass-loss rates from \citealt{vassiliadis1993}).}
    \label{fig:fates}
\end{figure}

In Figure~\ref{fig:fates} we show the expected fates of our stellar model calculations in the initial-mass and metallicity plane. For comparison, the region which might lead to supernovae I1/2 (SNe~I1/2), according to \citet{gilpons2018}, is also shown in the lower panel of the figure. As can be seen, our current calculations predict smaller mass/metallicity ranges for SNe~I1/2 and electron-capture supernovae (EC-SNe).

The models that likely lead to SN~I1/2 have metallicities $Z=10^{-10}$ and $Z=10^{-8}$, and initial masses around 5~\msun{}. These models experience a deep SDU which allows for a vast increase in helium surface abundance, reaching a value ${\rm X_s(He)}\simeq 0.37$. However, CNO abundances after the SDU remain low ($Z_{CNO} \sim 10^{-8}$). In addition, these models are too massive to experience a proton-ingestion enrichment episode (Sec.~\ref{sec:pies}). Thus no significant surface CNO increase is achieved during the early thermal pulses. The combination of high helium envelope abundances and low CNO abundances at the H-burning shell determines a peculiar behaviour during the TP-AGB.

Due to the scarcity of CNO elements, H-burning through the CNO-cycle is inefficient. It thus proceeds at considerably higher temperatures, and the conditions for the occurrence of He-flashes are achieved more quickly, giving shorter interpulse periods. High temperatures are also associated with relatively low degeneracy and with weak thermal pulses which, in this case, have luminosities ${\rm L_{He}\lesssim 10^5}$ \lsun{} (Fig.~\ref{fig:sn15}). This is much lower than the typical ${\rm L_{He}\sim 10^8}$ \lsun{} of other TP-AGB stars (Fig.~\ref{fig:tpagb}).
Such weak thermal pulses are unable to drive TDU episodes and feed the base of the convective envelope with carbon, which hampers the onset of the HBB. \cite{tashibu2017} also reported the absence of TDU in their 5~\msun{} models of $Z=0$ and ${\rm Z=10^{-5}}$. \cite{gilpons2018} noted the same for models of metallicities up to ${\rm Z=10^{-7}}$.
In addition, the dearth of metals in the envelope also drives very weak
stellar winds. As the TP-AGB proceeds, thermal pulses become increasingly weaker and finally vanish. This behaviour was described by \cite{lau2007} and \cite{gilpons2018} for somewhat higher initial mass models.  \cite{lau2007} concluded that the exhaustion of thermal pulses should eventually lead to the growth of the degenerate core prior to the complete ejection of the stellar envelope. Thus these models would become SNe~I1/2. However, \cite{gilpons2018} reported the re-onset of thermal instabilities and significant metal pollution of the envelopes, which shed doubt on the inevitability of explosive outcomes in these models.

The narrower mass/metallicity range for SNe~I1/2 we find here, as compared to \cite{gilpons2018} (Fig.~\ref{fig:fates}), is due to our updated input physics. Unlike the current study, \cite{gilpons2018} used constant-composition low-temperature opacities (\citealt{ferguson2005}) and mas-loss rates by \citet{vassiliadis1993}, which naturally led to weaker winds and thus to a higher probability of pulse cessation and, eventually, to the occurrence of SNe~I1/2. Our present calculations also lead to pulse exhaustion, but in a considerably reduced initial mass range. 
 
The mass limit $M_{up}$ (minimum value required for the onset of C-burning), $M_{n}$ (minimum initial mass required for the formation of a neutron star through an electron-capture supernova), and $M_{mass}$ (minimum initial mass for the formation of a core-collapse supernova) are not noticeably altered in the present work. $M_{up}$ is determined by the size of the CO core at the end of CHeB, and the role of low-T opacities or specific wind rates only becomes important later in the evolution. Neither of these alter significantly the depth of the second dredge-up and thus, $M_{n}$ and $M_{mass}$ remain practically the same as in \citet{gilpons2018}.

We emphasise that the initial mass boundaries leading to different final fates are strongly dependent on poorly known input physics (mostly related to convection and the determination of convective boundaries, and stellar wind mass-loss rates). Less efficient TDU and, as mentioned above, mass-loss rates would lead to the occurrence of SNe~I1/2 in a wider range than the one obtained here. On the other hand, if reality actually favoured strong winds and fast core growth, or for instance, if very efficient rotation played a crucial role, SNeI~1/2 might not occur at all.

\section{Nucleosynthetic evolution and yields} \label{sec:yields} 

\subsection{Main mixing episodes}

\subsubsection{Second dredge-up episode}
\label{sec:sdu}

All our model stars experience some degree of SDU, although its effects on the surface composition are rather mild for the lowest-Z, lowest-mass models. 
As a consequence of the SDU, stars increase surface He abundance.
For instance, the helium surface abundance increases from $X_{He} = 0.25$ to 0.29 in our 3~\msun{} model of $Z=10^{-6}$. Considering that the final He abundance (after the TP-(S)AGB) is 0.32, we can see that the a significant He enrichment occurs during the SDU. By comparison, our 3~\msun{} model of $Z=10^{-10}$ does not experience an efficient SDU (see Table \ref{tab:ref1} and left panels of Figure \ref{fig:surfab3ms_z1em10}). However, its final He surface abundance reaches 0.36, mostly due to the occurrence of a proton-ingestion episode in the early phases of the TP-AGB (Sec.~\ref{sec:pies}). SDU also causes a surface enhancement in CN-cycle products. \iso{14}N, \iso{13}C, \iso{18}O and \iso{23}Na increase, whereas \iso{12}C, \iso{16}O and \iso{22}Ne are depleted.
For models up to 5~\msun{}, the N abundance after the SDU is only $\lesssim  10^{-8}$ for models of $Z_{ini}$ between primordial and $Z=10^{-7}$, and it remains  $\sim  10^{-7}$ for our 5 \msun{} star of $Z=10^{-6}$. A standard SDU only mixes CN-processed matter with the envelope and therefore, it does not allow for a total metallicity variation. As a consequence, unless a PIE or efficient TDU occurs early during the TP-AGB, the lowest metallicty stars ($Z\leq 10^{-8}$) are not able to undergo a `normal' evolution and may end their lives as SNe~I1/2 (Sec.~\ref{sec:cessation}). 

Models of masses $\gtrsim$ 6 \msun{} experience the corrosive SDU (CSDU).
During this process the base of the convective envelope reaches further down than in a standard SDU, and does not only allow for He and CN-product surface enhancement, but also causes the mixing of the envelope with matter processed by He-burning \citep{gilpons2013,doherty2014b}. 
This process is favoured at very low metallicity because the higher structural temperatures allow for the occurrence of He-burning very close to the base of the H-burning shell.
Due to corrosive SDU, the more massive models also reach very high He surface abundances. 

In terms of surface He-enrichment, for instance, the 7.5~\msun{} model of $Z=10^{-6}$ reaches $X_{\text{He}} = 0.35$, and the 8.5~\msun{} of $Z=10^{-10}$ reaches a $X_{\text{He}} = 0.38$. Helium surface abundances variations during the TP-SAGB, on the other hand, are smaller than $\delta X_{\text{He}} \simeq 0.05$, because the characteristic duration of the TP-(S)AGB of the relatively massive stars which experience the CSDU are very short ($\tau_{TP-(S)AGB} \lesssim 0.1-0.2$~Myr; Table~\ref{fig:tpagb}).
The left panels of Figure \ref{fig:surfab75ms_z1em10} show the occurrence of the CSDU on the surface abundances of a 7.5 \msun{} star of $Z=10^{-10}$. 
We first see the effects of the BCE reaching CN-processed matter, that is, the same as a standard SDU described above. However, as the BCE reaches matter processed by He-burning, surface abundances of \iso{12}C and \iso{16}O increase.
It is important to realise the crucial effect of the CSDU on the yields of elements up to oxygen in the models considered. The bulk of envelope enhancement of C, N, and O occurs during this episode, rather than during the TP-(S)AGB phase.

\subsubsection{Effects of dual-shell flashes}
As detailed in Sec.~\ref{sec:pies}, our 3 and 4~\msun{} models of $Z=10^{-8}$ and $10^{-10}$, as well as our 3 \msun{} model of $Z=10^{-7}$, experience DSF proton-ingestion episodes early during the TP-AGB phase. As an example of the nucleosynthetic effects of these events, Figures~\ref{fig:dsf1} and~\ref{fig:dsf2} show the evolution of convection and surface abundances during the occurrence of the two DSFs of the primordial model of 3~\msun. The first H-flash causes an enhancement of all the most abundant isotopes (except H) by several orders of magnitude. This DSF leads to a post-DSF dredge-up of the He+H-burning region, causing an increase in surface abundances of isotopes formed in the pulse-driven zone (e.g. \iso{12}C, \iso{16}O, \iso{20,21,22}Ne), but also of those  H-burning (e.g. \iso{13}C, \iso{14}N, \iso{20}Ne, \iso{23}Na).
The (light element) metallicities of these models reach very high values. For example, in the $3$~\msun{} model of $Z=10^{-10}$ the metallicity reaches $Z\sim 10^{-4}$.  Such high metallcities allow for the occurrence of a normal TP-AGB phase. 

\subsubsection{Third dredge-up}\label{sec:tdu}
The efficiency of the TDU is crucial in determining the characteristics of the yields of our model stars.
Except for the $Z=10^{-8}$ and $Z=10^{-10}$ cases of initial masses $\simeq5$~\msun{}, whose final fates are uncertain, all our models experience both efficient TDU and HBB, and end their lives as white dwarfs (see Fig.~\ref{fig:fates}). TDU mixes He-burning products from the pulse-driven convective zone into the envelope, and thus causes surface enrichment in \iso{4}He, \iso{12}C and, to a lesser extent, in \iso{16}O. 
Enrichment in \iso{20,21,22}Ne is produced via the n- and $\alpha$-capture reactions as described in \cite{doherty2014b} and GP21. 
 The $\alpha$-captures on \iso{22}Ne allow for the formation of heavy magnesium isotopes during He-burning. 

In the low-Z models we are considering, H-burning also is partially active during the thermal pulses. Thus \iso{4}He, \iso{13}C, \iso{14}N and \iso{15}N are synthesised above the He-flash driven convective zone. These isotopes are eventually transported to the surface during subsequent TDU episodes (GP21).

\subsubsection{Hot bottom burning, NeNa cycle and MgAlSi chains}\label{sec:hbb}

\begin{figure}
    \centering
    \includegraphics[width=1.01\linewidth]{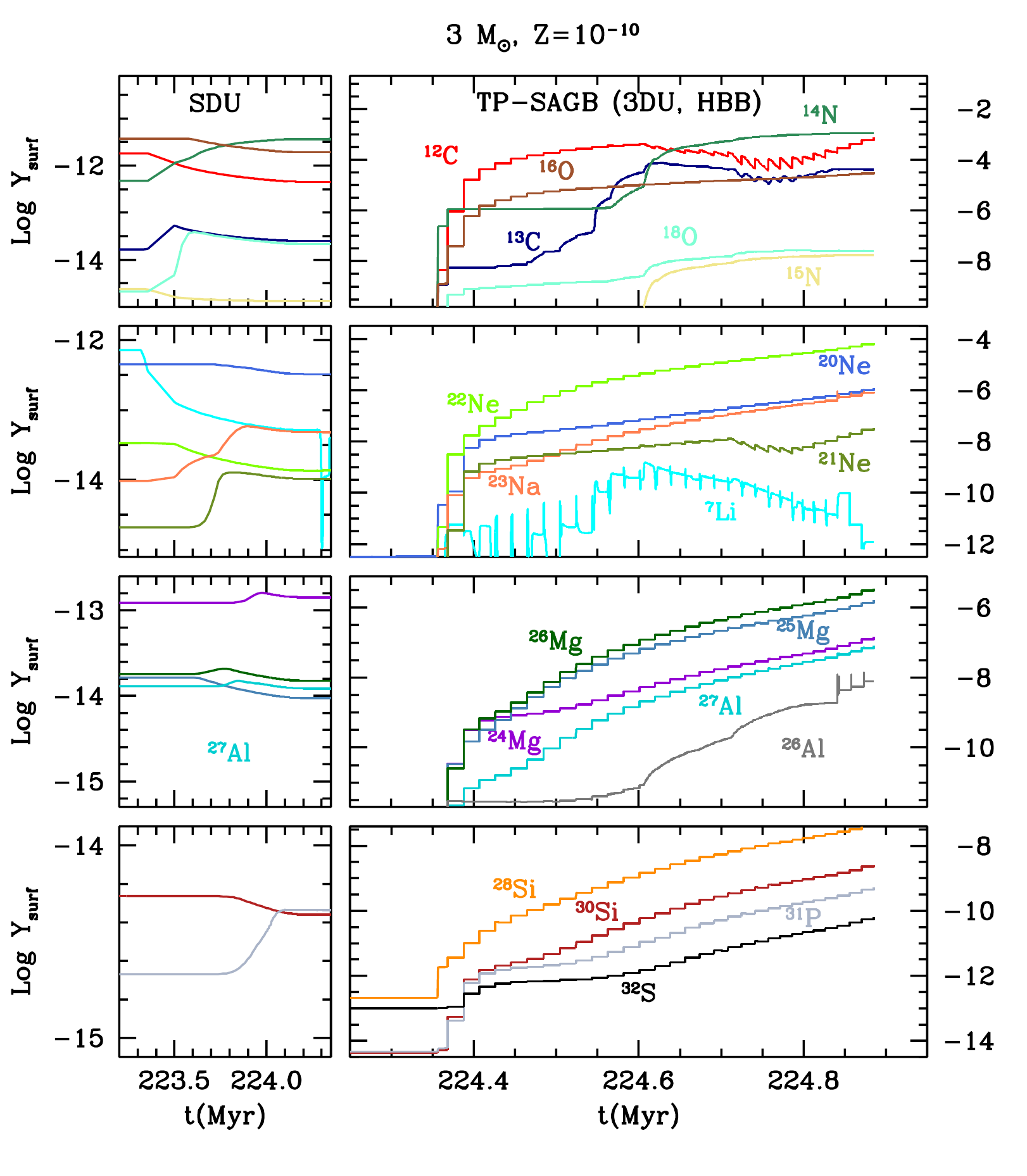}
    \caption{Evolution of the surface abundances of some selected isotopes for the  3 \msun, Z=10$^{-10}$ model. }
    \label{fig:surfab3ms_z1em10}
\end{figure}

\begin{figure}
    \centering
    \includegraphics[width=1.01\linewidth]{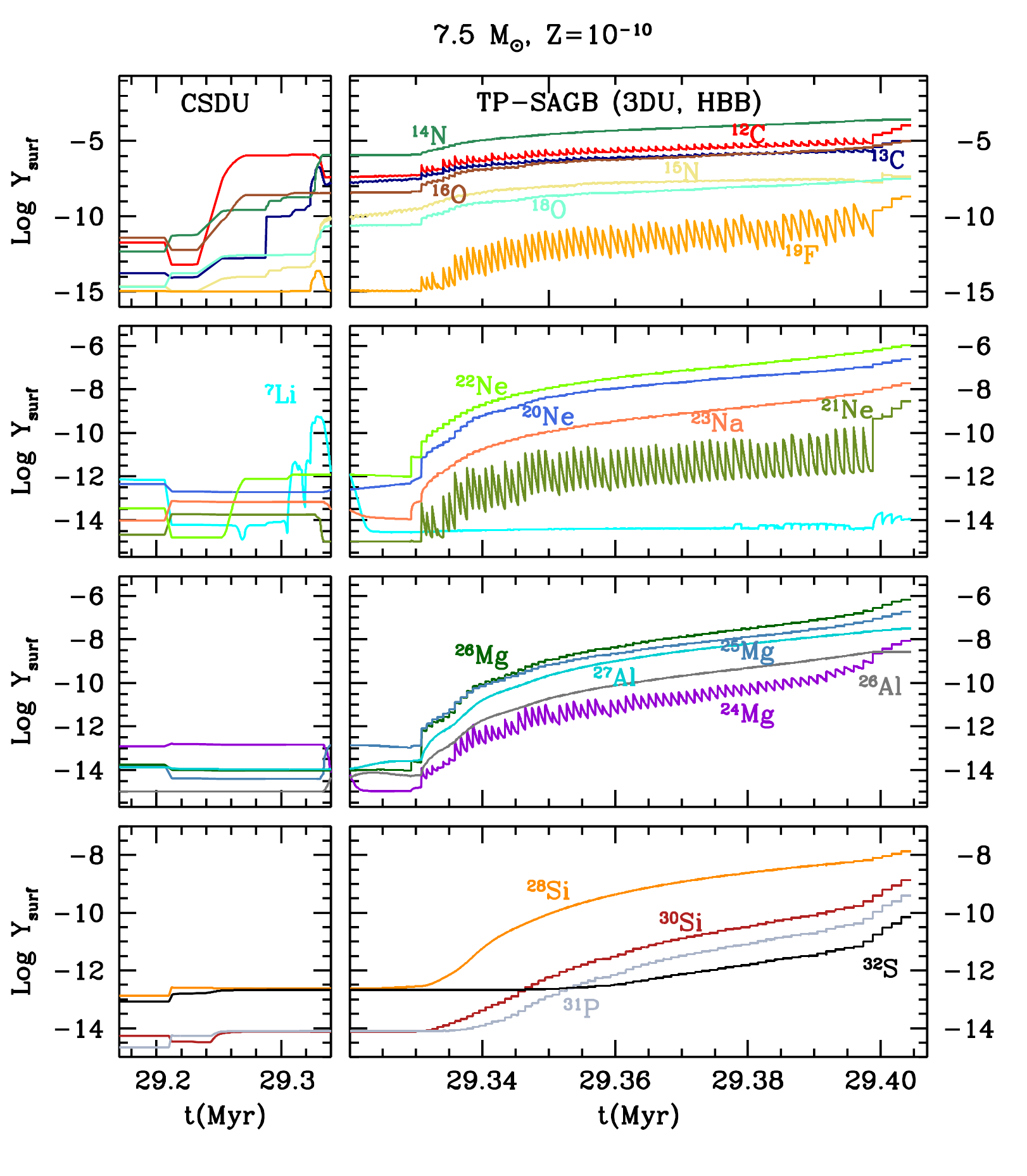}
    \caption{Evolution of the surface abundances of some selected isotopes for the  7.5 \msun, Z=10$^{-10}$ model. }
    \label{fig:surfab75ms_z1em10}
\end{figure}

In all our 3 \msun{} models, temperatures at the base of the convective envelope reach values of approximately 30 MK halfway through the TP-AGB, and then HBB ensues. The same happens in our 4 \msun{} stars of $Z=10^{-10}$ and $Z=10^{-8}$. In the higher metallicity models of the same mass HBB develops from the beginning of their TP-(S)AGB phase. The \iso{12}C transported to the envelope by the TDU is partially processed by the CN burning at the base of the convective envelope. \iso{4}He forms, together with \iso{14}N, \iso{13}C and \iso{15}N, and the surface \iso{12}C/\iso{13}C ratio remains nearly constant around 4, its equilibrium value, during most of the TP-(S)AGB evolution.  
In Figure \ref{fig:surfab3ms_z1em10} we show the evolution of surface abundances of the 3~\msun{} primordial model throughout the SDU and the TP-AGB phase.
We note that this star becomes a carbon star after the second DSF (C/O~$> 1.0$; $t \sim 224.35$~Myr). As mentioned above, efficient TDU during the entire TP-AGB phase replenishes the stellar envelope with Ne, Na, Mg, Al and Si isotopes. This figure also shows the onset of HBB halfway through the TP-AGB, reflected in a sudden increase in \iso{14}N together with some \iso{13}C. On the other hand, the 7.5~\msun{} model of Z~$=10^{-10}$ shown in Figure \ref{fig:surfab75ms_z1em10} undergoes both TDU and HBB from the beginning of the TP-SAGB phase. During its last 4 thermal pulses, when the envelope is relatively thin ($\lesssim$ 2 \msun) and the temperature at the BCE decreases below 30 MK, the HBB stops operating.

The NeNa cycle is activated in all our models,  except for part of the TP-AGB of the 3 \msun{} cases, and in those near 5 \msun{} of metallicities $Z=10^{-10}$ and $Z=10^{-8}$.
The \iso{22}Ne transported by the TDU feeds the NeNa cycle, which transforms this isotope into \iso{20}Ne.    
The MgAlSi chain is active at temperatures $\gtrsim$ 50 $MK$, and thus it is present in all the models of initial mass $\gtrsim$ 4 \msun{} which follow a normal TP-(S)AGB evolution. 
Proton captures and $\beta$-decays on Mg lead to the formation of Al and Si isotopes.

\subsection{Production factors \& comparison with V21}

\begin{figure*}[ht]
\begin{center}
\includegraphics[width=0.8\linewidth]{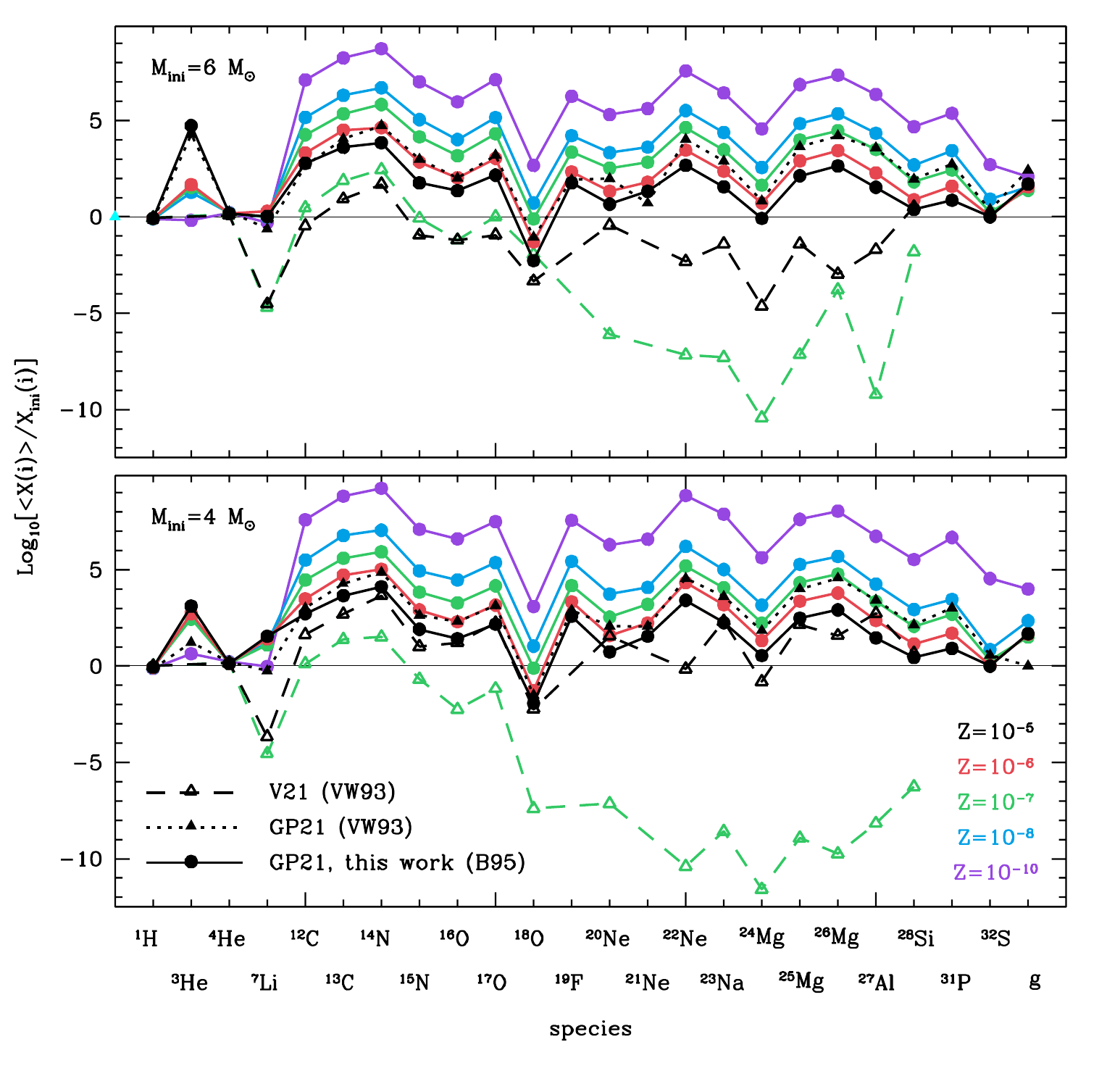}
\caption{Production factor of selected isotopes for the low metallicity models
        of 4, and 6 \msun. The ${\rm Z=10^{-5}}$ results (black) from GP21 are compared with
        the $Z=10^{-6}$ (red), the $Z=10^{-7}$ (green), the $Z=10^{-8}$ (blue), and the $Z=10^{-10}$ calculations (purple) of the current work. Dashed lines, corresponding to the results from V21, follow the same color code as above.  
        }
\label{fig:compareprod}
\end{center}
\end{figure*}

Figure \ref{fig:compareprod} shows the production factors of selected isotopes for some of our models. As a comparison we also show the production factors from the V21 models. 
The production factor for each species, $i$, is defined as 
\begin{equation}
    P_i=\log{\frac{\langle X_i\rangle}{X_{i,\mathrm{ini}}}}
,\end{equation}
where $\langle X_i\rangle = \frac{1}{t_{{\rm star}}}\sum_{j=1}^N X_{ij}(t)
\Delta t_j$ with $t_\mathrm{star}= \sum  \Delta t_j$ is the age of the star,
$\Delta t_j$ is the duration of time step $j,$ and $X_{ij}(t)$ is the mass fraction $i$ at that time.
A large disparity between our results and theirs can be seen, particularly with regards to their lowest Z models ($Z = 10^{-7}$) and for elements heavier than oxygen. Recall that V21 used a different prescription for convection and for the determination of convective boundaries, which prevented the occurrence of TDU in models of initial mass $\gtrsim 4$~\msun{} (Sec.~\ref{sec:comparison}). The presence or absence of TDU is crucial and has a direct effect on the production factors, not only because He-burning products such as \iso{12}C, \iso{16}O, \iso{20,22}Ne,  \iso{24,26}Mg and \iso{28}Si are not enhanced, but also because \iso{22}Ne and Mg isotopes fuel the activation of the NeNa cycle and the MgAlSi chains. The lack of these species therefore hampers the formation of Mg, Al and Si isotopes in the V21 models. 

\begin{figure*}[t]
    \centering
    \includegraphics[width=12cm]{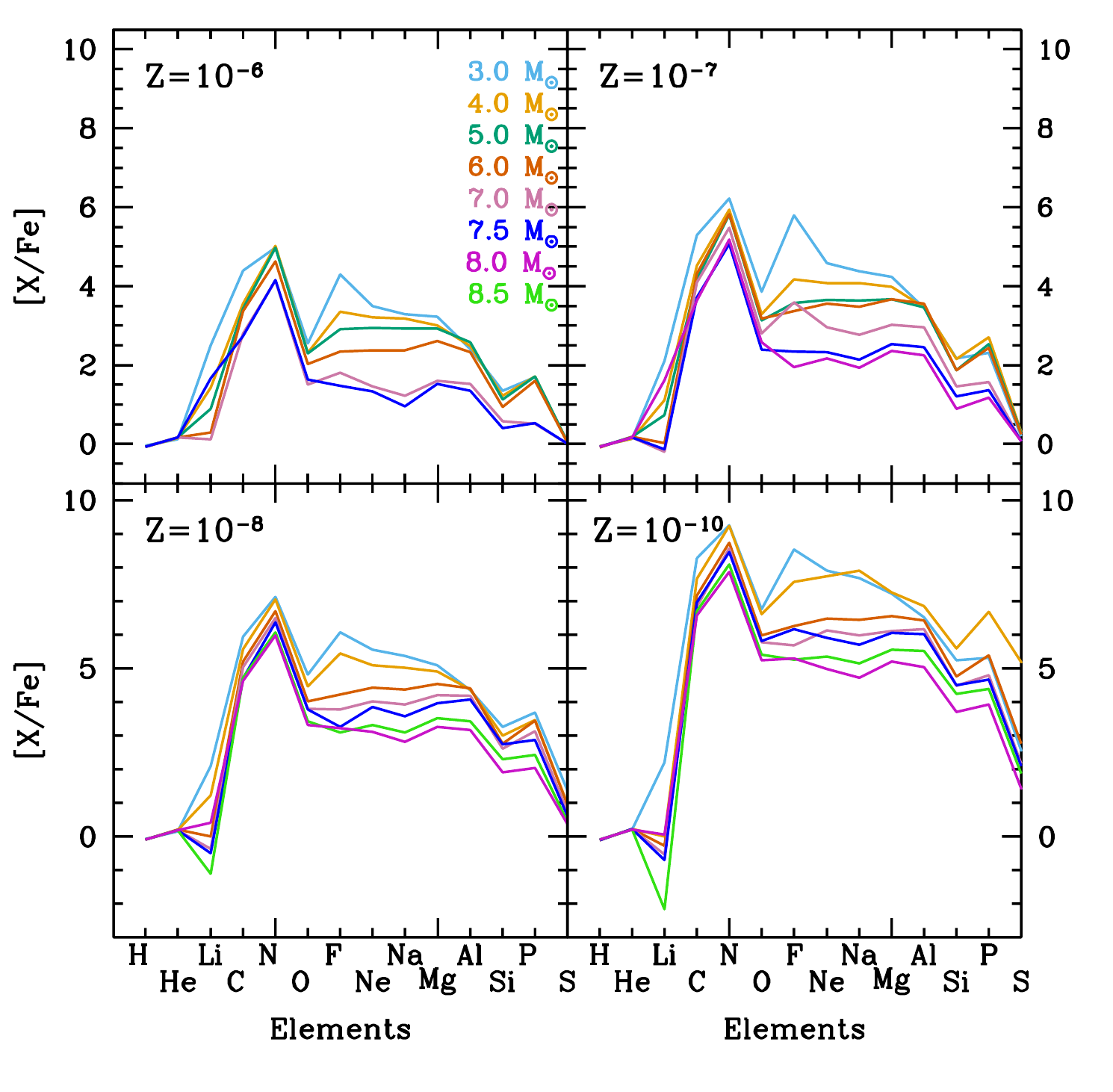}
    \caption{Abundance pattern in terms of the ejecta of selected species for intermediate-mass stars of metallicities $Z=10^{-6}$, $Z=10^{-7}$, $Z=10^{-8}$ and $Z=10^{-10}$. 
    For clarity, the vertical scales in the upper and lower panels are different.}
    \label{fig:feh}
\end{figure*}

From the structural point of view, the absence of C-enhancement in the V21 models of masses $\gtrsim$ 4 \msun{} leads to lower surface opacities, smaller stellar radii and ultimately, less efficient wind rates.
Inefficient or even absent TDU also favours more effective core growth. Together with their convection treatment, this leads to higher temperatures at the nuclearly active regions. In turn, this results in the net destruction of all isotopes up to \iso{28}Si, except \iso{4}He, \iso{12,13}C,  \iso{14,15}N, and \iso{16,17}O. It can also be seen in Figure \ref{fig:compareprod} that the higher temperatures at the BCE of the V21 models force very efficient destruction of \iso{7}Li, which is produced or only mildly depleted in our models.
On the other hand, the search for neutrality prescription for the determination of convective boundaries allows for a highly efficient TDU in our models. Moreover, V21 used the wind prescription by \cite{vassiliadis1993}, which yields considerably slower mass-loss rates than those of \cite{bloecker1995} with $\eta=0.02$, used in the present work. 

The vast discrepancies between the production factors presented here and those from V21 reach 10-15 orders of magnitude for isotopes more massive than \iso{17}O. V21 models strongly deplete all these isotopes whereas ours show considerable enhancements. Differences are naturally more modest (although still significant) when considering our lower mass cases of ${\rm Z=10^{-5}}$ (GP21) calculated with the mass-loss prescription of \cite{vassiliadis1993}. Indeed, equivalent mass-loss rate prescriptions and less efficient HBB (due to lower initial masses) favour similarities between the two studies. This can be seen when comparing the production factors from the 4~\msun{} cases shown in Figure~\ref{fig:compareprod} (GP21: VW93 case vs V21).

\subsection{Nucleosynthetic yields}\label{sec:abund}

The ejected mass of isotope $i$ is expressed in \msun{} and
calculated as follows: 
\begin{equation}
    M_i=\int_0^{t_\mathrm{end}}{ Y_i(t)A_i\dot M(t)dt}
    \label{eq:yields}
,\end{equation}
\noindent where $t_\mathrm{end}$ is the time at the end of our calculations,
 $Y_i(t)$ the number abundance at an arbitrary time $t$ of isotope $i$, $A_i$ is its mass number, and $\dot M(t)$ is the mass-loss rate at $t$.

The associated abundance in terms of the ejecta of species $X_i$, $[X_i/\text{Fe}]$ is calculated as:
\begin{equation}
    [X_i/\text{Fe}]=log_{10} \left (\frac{M_i/A_i}{M_{\text{Fe}}/A_{\text{Fe}}}\right )_* - log_{10}\left (\frac{N_i}{N_{\text{Fe}}}\right )_\odot .
\end{equation}

Figure \ref{fig:feh} shows the abundance patterns in terms of the ejecta for our model stars.

\subsubsection{CNO and F}
The main characteristics of the abundance pattern in terms of the ejecta from our models clearly reflects the above-mentioned efficiency of TDU+HBB: very high abundances [C/Fe], [N/Fe] and [O/Fe] -- with [N/Fe] dominating (Fig.~\ref{fig:feh}). This pattern is seen in a subset of metal-poor stars, known as N-rich EMP stars (NEMP; eg. \citealt{izzard2009,pols2012}, with [N/Fe]~>~1 and [N/C]~>~0.5), which includes one of the most metal-poor stars, \object{HE1327-2326} ([Fe/H] $ \lesssim -5.4$; \citealt{frebel2008}). 

NEMP stars appear to be more abundant at low metallicities ([Fe/H] $\lesssim$ -3), which was attributed to a primitive IMF biased towards high masses \citep{abia2001,lucatello2005, komiya2007}. In order to test this hypothesis,
\cite{pols2012} used a binary population synthesis code, and showed that such top-heavy IMF would imply a frequency of NEMP stars much higher than what is actually observed. Their calculations did not consider evolutionary models below [Fe/H]~$=-2.8$. However, the general patterns in relation to the abundances of C, N, and O of the lowest metallicity intermediate-mass stars are not peculiar enough to expect a significantly different conclusion from \cite{pols2012}. Indeed, the distinctive signature of combined TDU and HBB, which are reproduced regardless of the initial metallicity, are responsible for the [N/Fe] > [C/Fe] > [O/Fe] pattern. This points to different possibilities -- either intermediate-mass stars did not play a significant role as polluters of the early universe until [Fe/H] $\simeq -3$, or the number of observed stars below [Fe/H]~$\lesssim -4$ is still too small to allow a proper assessment of the contribution of these stars.   

In our model stars, envelope enhancements in He-burning products such as \iso{12}C and \iso{16}O are a consequence of   
PIEs (in stars of $M_{ini}\lesssim$ 4 \msun{} and metallicities $\lesssim 10^{-8}$), CSDU (in stars of $M_{ini}\gtrsim$ 6 \msun{}), and TDU.
\iso{16}O is formed via \iso{12}C($\alpha,\gamma$)\iso{16}O, \iso{13}C($\alpha$,n)\iso{16}O and \iso{13}N($\alpha$,p)\iso{16}O. 
In  addition, the occurrence of H-burning during the onset of thermal pulses, characteristic of very low metallicty models, gives rise to the efficient formation of CN-cycle products which are transported to the surface in subsequent thermal pulses. As a consequence, the TDU is not only responsible for surface enhancements of He-burning, but also of H-burning products similar to those produced during HBB such as \iso{4}He, \iso{13}C, \iso{14,15}N. 

As happens for higher metallicity models \citep{karakas2010}, fluorine abundance is particularly high in our 3 and 4~\msun{} models. It is formed during He-burning via \iso{14}N($\alpha$,$\gamma$)\iso{18}F($\beta^+$)\iso{18}O(p,$\alpha$)\iso{15}N($\alpha$,$\gamma$)\iso{19}F (e.g.  \citealt{forestini1992, abia2015, vescovi2022}). Protons are provided by \iso{14}N(n,p)\iso{14}C and neutrons via \iso{13}C($\alpha$,n)\iso{16}O. \iso{19}F is easily destroyed by \iso{19}F(p,$\alpha$)\iso{16}O and \iso{19}F($\alpha$,p)\iso{22}Ne. 
Fluorine was analysed in two CEMP stars \citep{muraguzman2020}. 
Observations of \object{HE1429-0551} ([Fe/H]~$=-2.53$) yielded 
[F/Fe]~$= +1.90$, 
and an upper limit of 
[Fe/F]~$< +1.00$ was determined for \object{HE1305+0007} ([Fe/H]~$=-2.28$).
These authors showed that \object{HE1429-0551} can be reasonably well explained by considering accretion from a primary TP-AGB companion. 
$Z=10^{-5}$ models between 4 and 6 \msun{} from GP21 also provide similar F abundances.

\subsubsection{NeNa elements}

The abundances relative to iron of the elements shown in Figure \ref{fig:feh} tend to increase with decreasing initial mass, regardless the metallicity. As seen in Table \ref{tab:evol1}, lower mass models have longer TP-AGB phases, tend to experience more efficient TDU and, overall, dredge-up more matter than higher massive models. Also, even our lowest mass cases experience some degree of HBB.

The isotopes \iso{20,21,22}Ne are created during thermal pulses by neutron and $\alpha$-captures, and efficiently transported by the TDU. 
They form via \iso{16}O(n,$\gamma$)\iso{17}O($\alpha$,n)\iso{20}Ne, \iso{16}O(n,$\gamma$)\iso{17}O($\alpha,\gamma$)\iso{21}Ne, \iso{18}F($\alpha$,p)\iso{21}Ne, \iso{18}O($\alpha$,n)\iso{21}Ne, \iso{20}Ne(n,$\gamma$)\iso{21}Ne and \iso{14}N($\alpha$,$\gamma$)\iso{18}F($\beta^+$,$\nu$)\iso{18}O($\alpha$,$\gamma$)\iso{22}Ne.
The latter fuels the NeNa cycle, which partially converts this isotope into \iso{20}Ne via \iso{22}Ne(p,$\gamma$)\iso{23}Na(p,$\alpha$)\iso{20}Ne. \iso{21}Ne is destroyed by p-captures to form \iso{22}Na, which $\beta$-decays on \iso{22}Ne, and is later converted into \iso{20}Ne, by the process described above.
However, the high efficiency of the TDU in our models allows for an increase in surface \iso{21,22}Ne along the TP-(S)AGB.
In addition, the temperatures in our models reach values high enough for the occurrence of \iso{20}Ne(p,$\gamma$)\iso{21}Na($\beta+$)\iso{21}Ne(p,$\gamma$)\iso{22}Na($\beta^+$)\iso{22}Ne. Therefore some \iso{22}Ne is recycled during the process. 

Surface \iso{23}Na is also enhanced during the TP-(S)AGB of our model stars. As \iso{21,22}Ne, it is mostly created during thermal pulses, in this case via \iso{22}Ne(n,$\gamma$)\iso{23}Ne($\beta^-$)\iso{23}Na, and then transported to the envelope by TDU. The rates of destruction (\iso{23}Na(p,$\alpha$)\iso{20}Ne) and formation (\iso{22}Ne(p,$\gamma$)\iso{23}Na) at the BCE during the interpulses are very similar, and thus, unlike what happens at higher metallicity models, in which p-captures on \iso{23}Na clearly dominate (see, e.g. \citealt{doherty2014b}), it is  only mildly depleted during each interpulse. 
In our most massive models ($M_{ini} \gtrsim$ 7 \msun) all the isotopes involved in the NeNa cycle increase, and reach equilibrium abundances practically from the beginning of the TP-SAGB.

It is also important to recall that the reactions involved in the NeNa cycle (and also in the MgAlSi chains) 
are affected by significant uncertainties \citep{arnould1999, izzard2006}, especially those leading to \iso{22}Ne, \iso{23}Na, and \iso{26}Al. This may compromise the validity of nucleosynthetic calculations and affect comparisons with observations.

\subsubsection{MgAlSi and higher mass elements}

At the characteristic BCE temperatures of our model stars, the reaction rate \iso{23}Na(p,$\gamma$)\iso{24}Mg is much lower than that of \iso{23}Na(p,$\alpha$)\iso{20}Ne. Thus there is practically no leakage from the NeNa cycle to feed the MgAlSi chain. However,
significant amounts of \iso{24}Mg form in the convective shells associated with TPs via alpha-captures on \iso{20}Ne in our more massive models of $Z\neq 10^{-10}$, and mostly via  \iso{22}Ne(n,$\gamma$)\iso{23}Ne($\beta^-$)\iso{23}Na(n,$\gamma$)\iso{24}Na($\beta^-$)\iso{24}Mg in the rest of our models. TPs also lead to the formation of \iso{25,26}Mg, via \iso{22}Ne($\alpha$,n)\iso{25}Mg, \iso{24}Mg(n,$\gamma$)\iso{25}Mg,  \iso{22}Ne($\alpha$,$\gamma$)\iso{26}Mg and \iso{25}Mg(n,$\gamma$)\iso{26}Mg. \iso{27}Al forms through \iso{26}Mg(n,$\gamma$)\iso{27}Mg($\beta^-$)\iso{27}Al. Si isotopes are synthesised via \iso{27}Al(n,$\gamma$)\iso{28}Al($\beta^-$)\iso{28}Si and \iso{28}Si(n,$\gamma$)\iso{29}Si(n,$\gamma$)\iso{30}Si. After another n-capture and a $\beta$ decay some \iso{30}Si is transformed into \iso{31}P which, after the same processes, is partially converted into \iso{32}S.

All the above isotopes are very efficiently transported to the convective envelope during the TDU.
At the temperatures characteristic of the BCE of our models Mg and Al isotopes are partially depleted during the interpulse via MgAlSi chains. \iso{24}Mg is destroyed very quickly (especially in our most massive models), as the fastest reaction in the MgAlSi chain is \iso{24}Mg(p,$\gamma$)\iso{25}Al, which is quickly followed by the $\beta$-decay of \iso{25}Al to form \iso{25}Mg. In our 3 \msun{} models, the MgAlSi chains are rather inefficient. The second and third fastest reactions are p-captures on \iso{25}Mg and \iso{26}Mg, which lead, respectively, to \iso{26}Al and \iso{27}Si. \iso{26}Al is quickly processed via \iso{26}Al(p,$\gamma$)\iso{27}Si, and this isotope $\beta$-decays to form \iso{27}Al.
In models of $M_{ini}\gtrsim$ 6 \msun{} of $Z\leq 10^{-8}$, and models of $M_{ini}\gtrsim$ 4 \msun{} of our higher metallicity cases the formation of \iso{28}Si is favoured by the fast occurrence of p-captures on \iso{26}Al followed by \iso{27}Si($\beta^+$)\iso{27}Al(p,$\gamma$)\iso{28}Si. 
Yet, the main source of this isotope remains the TDU of He-burning processed matter during thermal pulses.

Because the rate of the reaction \iso{27}Al(p,$\alpha$)\iso{24}Mg is very low at the BCE temperatures of our models, the recycling of \iso{24}Mg remains negligible.   
As mentioned above, \iso{30}Si, \iso{31}P and \iso{32}S are He-burning products which are dredged-up (TDU), but remain unaffected by the MgAlSi chain.
Finally, regardless of the efficiency of the MgAlSi chain, most of the envelope enhancement in the isotopes mentioned above is caused by very efficient TDU in our model stars. 

The abundance data of out models for the most abundant elements are shown in the Tables \ref{tab:abuns6} to \ref{tab:abuns10} \footnote{Detailed yields in electronic form can be found at  \url{http://dfa.upc.es/personals/pilar/research.php} and at the Strasbourg Astronomical Data Centre (CDS) \url{https://cdsweb.u-strasbg.fr/}.}.

\begin{figure}[t]
    \centering
        \vspace{0.04cm} \hspace{-0.02cm}
    \includegraphics[width=1.01\linewidth]{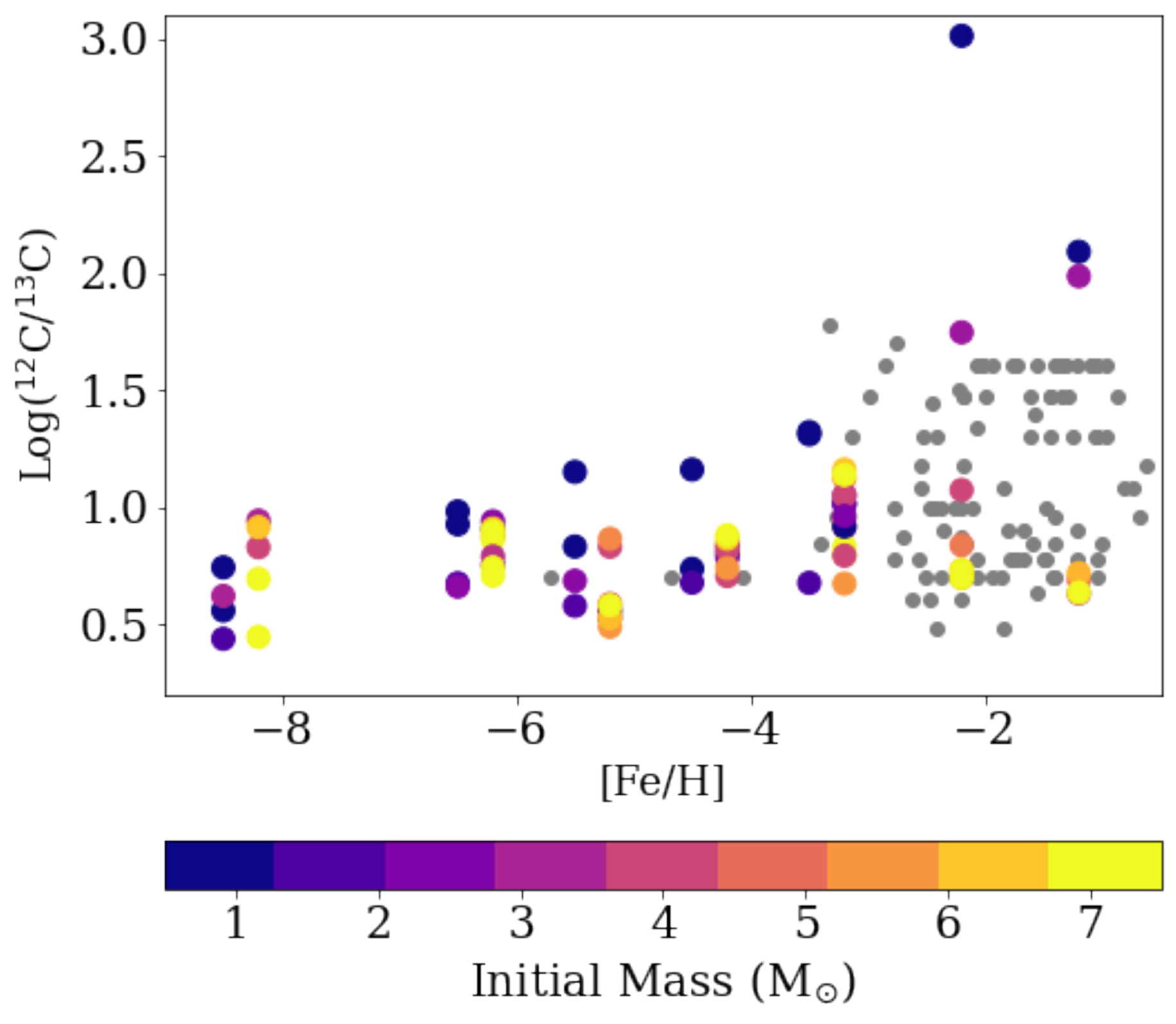}
    \caption{Carbon isotopic ratios in model yields (large circles and colour bar) and observations (small grey points) versus metallicity. The [Fe/H]~$\simeq -1$ models are from \cite{fishlock2014} for initial masses $\le 6$~\msun{}, and from \cite{doherty2014b} for initial masses $\ge 7$~\msun. [Fe/H]~$\simeq -2$ models are from \cite{karakas2010} for initial $\le 6$~\msun{}, and from \cite{doherty2014b} for initial masses $\ge 7$~\msun. [Fe/H]~$\simeq -3, -4$, $-5$, $-6$ and $-8$ models are from \cite{campbell2008} for initial masses $\le 3$~\msun{}, and from GP21 for initial masses $\ge 3$~\msun. The
    observed ratios are from the SAGA database \citep{suda2008}.}
    \label{fig:cisot}
\end{figure}

\subsection{Isotopic ratios}
Observed isotopic ratios help us probe the formation sites of different species. \iso{12}C/\iso{13}C gives information on the relative importance of massive and intermediate-mass stars, as \iso{12}C is formed due to He burning, both in massive and non-massive objects, whereas \iso{13}C is mostly formed in AGB stars due to CNO processing. Figure \ref{fig:cisot} shows \iso{12}C/\iso{13}C for different metallicities. Our models give a good agreement with the scarce observations available at the metallicity ranges we are considering. \cite{spite2021} recently reported surface \iso{12}C/\iso{13}C together with [C/Fe], [N/Fe] and [O/Fe] for HD140283, an EMP  turnoff star ([Fe/H]~$=-2.57$). Their \iso{12}C/\iso{13}C is 31, somewhat higher than our values for ${\rm Z=10^{-5}}$ stars, which are $\lesssim 20$, but well within the limits set by the ${\rm Z=10^{-5}}$ and the $Z=10^{-4}$ AGB and super-AGB models by \citet{karakas2010} and \citet{doherty2014b}. However, given the efficiency of our second dredge-up episode and, especially, of hot bottom burning, we cannot reproduce the observed [O/Fe] value in \cite{spite2021}, which is higher than their observed [N/H].    

The isotopic ratios \iso{25,26}Mg/\iso{24}Mg are also very useful to explore the origin of elements in stars (e.g. \citealt{kob11, romano2017}), to examine star formation rates \citep{vangioni2019}, and to probe Galactic chemical evolution \citep{fenner2003,melendez2007}. Most \iso{24}Mg is formed during C- and Ne-burning in massive stars, whereas AGB stars mostly contribute \iso{25,26}Mg.   
The former isotopes can be created (and destroyed) at the H-burning shell or at the base of the convective envelope, during HBB, through the MgAl chain (provided that the temperatures are $\gtrsim 90$~MK). In our models, these isotopes are mostly formed during the thermal pulses in the He-burning shell. Successive $\alpha-$captures on \iso{14}N lead to the formation of \iso{22}Ne. Additional $\alpha-$captures on the latter lead to \iso{25,26}Mg.  \citet{vangioni2019} showed that observations of Mg isotopic ratios between solar and down to [Fe/H]~$= -2.5$ could be better explained by a combination of massive and intermediate-star models.
To our knowledge, there are no determinations of observed \iso{25,26}Mg/\iso{24}Mg for metallicities [Fe/H]~$\lesssim -3$ at the moment.
However, we note that the trend to a decrease in \iso{26}Mg/\iso{24}Mg values with decreasing metallicities is clearly not reproduced by our intermediate-mass models. Perhaps, as happens for higher metallicities, the inclusion of intermediate-mass stellar yields might help to improve the agreement between future observations and theoretical results of Mg isotopic ratios. 
The low-mass models (M$_{ini}=0.85$, 1~\msun{}) from \cite{campbell2008}, which experience very efficient proton ingestion episodes, help to reproduce the carbon isotopic ratios shown in Figure \ref{fig:cisot}. They also help to reproduce Mg-isotopic ratios. In particular, the absence of  HBB and of the activation of the Mg-Al chain are responsible for the low \iso{26}Mg/\iso{24}Mg values compared to those of intermediate-mass stars of the same metallicities.

\section{Summary and discussion} \label{sec:summary}

We have computed the structural and nucleosynthetic evolution, as well as the gas yields, of intermediate-mass stars between primordial and extremely metal-poor metallicities. Our results significantly differ from analogous calculations existing in the literature (Sec.~\ref{sec:comparison}; Figs.~\ref{fig:tpagb} and~\ref{fig:compareprod}). This highlights the crucial effects of poorly known input physics on the evolution and nucleosynthetic yields of the most metal-poor intermediate-mass stars.

The most relevant input physics responsible for discrepancies are convection and the determination of convective boundaries,  mixing and stellar winds. Our calculations were computed with the Mixing-Length theory ($\alpha_{MLT}=1.75$), and convection limits were determined by the search for convective neutrality approach. Stellar winds followed the prescription by \cite{bloecker1995}, B95, (with $\eta=0.02$). 

In their review article \cite{gilpons2018} computed the (structural) evolution of stars in a similar mass and metallicity range to the ones considered in the present work. They used a previous version of the \texttt{MONSTAR} code, which included constant composition low-temperature opacities \citep{ferguson2005}. Their calculations used the mass-loss prescription from VW93 (\citealt{vassiliadis1993}), that, in the considered metallicity range, gives mass-loss rates significantly lower than those from B95 (even with $\eta=0.02$). As a consequence, these authors inferred that SNe~I1/2 could form up to a metallicity $Z=10^{-7}$, whereas the maximum Z for the occurrence of SNe~I1/2 in the current study is $10^{-8}$. 
The initial mass range to form SNe~I1/2 in \cite{gilpons2018} was also wider ($\lesssim 2$~\msun) than we found here ($\leq 1$~\msun), around a central initial mass of 5~\msun. In addition, in \cite{gilpons2018} EC-SNe were expected to form well below the initial mass required for a core-collapse supernova (CC-SN). Specifically, they reported an initial mass range between 7 and 8~\msun, and metallicities between primordial and $Z=10^{-7}$ for EC-SN formation. In the current study we do not expect EC-SNe to form in this lower-mass region. Instead we only find EC-SNe forming in a very narrow mass range ($\simeq 0.2$~\msun) just below the initial mass required to form CC-SNe (see Fig.\ref{fig:fates}).

\cite{schneider12} reported a critical metallicity for the formation of low-mass stars of $Z_{crit}=10^{-8}$, provided some pollution with the ejecta from a massive star. Therefore, although considerably reduced when calculations are performed with higher wind rates, the possibility of forming SNe~I1/2 cannot be discarded.  Such objects could eject considerable amounts of Fe to the interstellar medium in time scales $\leq 100$~Myr, that is, well below the typical 1 Gyr required for the formation of SNe~Ia. Local Fe injection in such short timescales could eventually lead to the birth of stars whose metallicity would be relatively high, considering their primitive origin. Perhaps they might be detected as relatively metal-rich objects belonging to a kinematic group of stars of much lower metallicities. 
Recent studies of stellar streams are providing a wealth of information related to the formation history of the Milky Way \citep{martin2022b}, and even  changing our current knowledge of globular clusters (GCs). The recent finding of C-9 \citep{martin2022a}, a stellar stream of surprisingly homogeneous low metallicity ([Fe/H]~$=-3.38$]) and characteristic chemical composition, points to its origins as the remnant of a disrupted GC.

We compared our results with V21 (\citealt{ventura2021}), a recent study reporting on the evolution and nucleosynthesis in low-and intermediate-mass stars of $Z\sim 10^{-5}$ and $Z\sim 10^{-7}$. This allowed us to further explore the critical effects of uncertain input physics. V21 used the Full Spectrum of Turbulence (FST) model, together with exponentially decaying overshooting, to determine convective boundaries. They applied the VW93 prescription for stellar winds, and as in the current work, used composition-dependent low-temperature opacities.  

A key difference in V21 and in this work is the efficiency of TDU (mostly  the method of determination of convective boundaries). The models of initial mass $\geq$ 4~\msun{} in V21 do not experience TDU, which naturally hampers surface C-enhancement. This has consequences on the envelope opacities, which remain relatively low, and thus favour relatively small radii and weaker stellar winds in the models from V21. In addition, the absence of TDU, together with the destruction of carbon by very efficient HBB, prevents the possibility of forming carbonaceous grains, which could also induce dust-driven stellar winds. Thus it is not only the choice of mass-loss prescription, but also the treatment of convection and its boundaries and, of course, the effect of variable-composition low-temperature opacities, which ultimately affect wind efficiency in these model stars. 

The final core masses in V21 reach higher values than those of our models and the temperatures at the base of their convective envelopes are considerably higher as well. The main reasons are the treatment of convection using the FST, the different determination of convective boundaries and,  the former, the absence of TDU in V21 models.  Not only are their cores not reduced after each thermal pulse, but also the cooling effect of the TDU is avoided. By comparing the results for the ${\rm Z=10^{-5}}$ from V21 and those from GP21 computed with the same mass-loss rates, we have been able to determine that the treatment of convection and its boundaries lead to wider discrepancies than the treatment of mass-loss rates (for the given prescriptions). These discrepancies affect the overall evolution and the nucleosynthesis of the considered models (see Sec.~\ref{sec:comparison}; Figs.~\ref{fig:tpagb} and~\ref{fig:compareprod}). 

The most metal-poor stars known at present seem to be the offspring of massive primordial stars (see, e.g. \citealt{iwamoto2005, umeda2003}). However, the set of observed stars of metallicity below [Fe/H] = -5 includes less than ten objects (\citealt{nordlander2019,aguado2018b,aguado2018a,bonifacio2015,keller2014,frebel2005,christlieb2004}). Thus it is as yet premature to discard the contribution of primitive low- and intermediate-mass stars to the metal enrichment of the early universe. As reported by \cite{osorio2022}, recent improvements in the calculations of H+Ca$^+$ collisional data \citep{belyaev2018} allow better determinations of non-LTE effects on the Ca II IR triplet lines. These lines can be used as a proxy for metallicity and, unlike the Ca J and K lines they are not affected by interstellar absorption. Because missions such as GAIA (e.g. \citealt{bokyung2022}), WEAVE \citep{dalton2012} and PFS \citep{tamura2016} will cover the CaII IR triplet, millions of spectra will be available. Thus, the new detections of extremely metal-poor stars are expected to proliferate in the near future. This will help to better constrain the relevance of primitive non-massive stars.    

\section{Conclusions} \label{sec:conclusions}

Computational results of the evolution and nucleosynthesis of the most metal-poor intermediate-mass stars show vast variations when calculated with different codes and different input physics. Discrepancies appear dramatic when considering the chemical production factors of model stars. They reach several orders of magnitude, and are particularly high for isotopes beyond \iso{19}F. On the other hand, the characteristic pattern of very high C and N, and relatively low O (compared to current observations of EMP stars), is reproduced in \cite{iwamoto2009}, PG21, V21, and in the present work. 

The fact that both the treatment of convection and its boundaries, and mass-loss rates are still so poorly constrained, added to the entanglement of both phenomena, and to the difficulties in comparing with observations, considerably limits the robustness of results in this area of stellar physics. However, as recently shown for the case of nitrogen \citep{johnson2022}, Galactic chemical evolution models combined with observations might help to impose constraints on the nucleosynthesis of specific elements. In particular, new detections (e.g. \citealt{osorio2022}), and detailed observations of extremely metal-poor stars using the James Webb telescope \citep{zackrisson2011} will considerably improve our knowledge of these elusive objects in the near future.

\begin{acknowledgements}
We thank the anonymous referee for their useful comments and clarifications. 
      This work was supported by the Spanish project \it{PID 2019-109363GB-100}, and by the German
      \it{Deut\-sche For\-schungs\-ge\-mein\-schaft, DFG\/} project
      number Ts~17/2--1. S.W.C. acknowledges federal funding from the Australian Research Council through a Future Fellowship (FT160100046) and Discovery Project (DP190102431). This research was supported by use of the Nectar Research Cloud, a collaborative Australian research platform supported by the National Collaborative Research Infrastructure Strategy (NCRIS). 
\end{acknowledgements}

%
\bibliographystyle{aa} 
\bibliography{primemp.bib} 

\begin{thebibliography}{139}
\expandafter\ifx\csname natexlab\endcsname\relax\def\natexlab#1{#1}\fi

\bibitem[{{Abel} {et~al.}(2002){Abel}, {Bryan}, \& {Norman}}]{abel02}
{Abel}, T., {Bryan}, G.~L., \& {Norman}, M.~L. 2002, Science, 295, 93

\bibitem[{{Abia} {et~al.}(2015){Abia}, {Cunha}, {Cristallo}, \& {de
  Laverny}}]{abia2015}
{Abia}, C., {Cunha}, K., {Cristallo}, S., \& {de Laverny}, P. 2015, \aap, 581,
  A88

\bibitem[{{Abia} {et~al.}(2001){Abia}, {Dom{\'{\i}}nguez}, {Straniero},
  {Limongi}, {Chieffi}, \& {Isern}}]{abia2001}
{Abia}, C., {Dom{\'{\i}}nguez}, I., {Straniero}, O., {et~al.} 2001, \apj, 557,
  126

\bibitem[{{Aguado} {et~al.}(2018{\natexlab{a}}){Aguado}, {Allende Prieto},
  {Gonz{\'a}lez Hern{\'a}ndez}, \& {Rebolo}}]{aguado2018b}
{Aguado}, D.~S., {Allende Prieto}, C., {Gonz{\'a}lez Hern{\'a}ndez}, J.~I., \&
  {Rebolo}, R. 2018{\natexlab{a}}, \apjl, 854, L34

\bibitem[{{Aguado} {et~al.}(2018{\natexlab{b}}){Aguado}, {Gonz{\'a}lez
  Hern{\'a}ndez}, {Allende Prieto}, \& {Rebolo}}]{aguado2018a}
{Aguado}, D.~S., {Gonz{\'a}lez Hern{\'a}ndez}, J.~I., {Allende Prieto}, C., \&
  {Rebolo}, R. 2018{\natexlab{b}}, \apjl, 852, L20

\bibitem[{{Angulo} {et~al.}(1999){Angulo}, {Arnould}, {Rayet}, {Descouvemont},
  {Baye}, {Leclercq-Willain}, {Coc}, {Barhoumi}, {Aguer}, {Rolfs}, {Kunz},
  {Hammer}, {Mayer}, {Paradellis}, {Kossionides}, {Chronidou}, {Spyrou},
  {degl'Innocenti}, {Fiorentini}, {Ricci}, {Zavatarelli}, {Providencia},
  {Wolters}, {Soares}, {Grama}, {Rahighi}, {Shotter}, \& {Lamehi
  Rachti}}]{angulo1999}
{Angulo}, C., {Arnould}, M., {Rayet}, M., {et~al.} 1999, Nuclear Physics A,
  656, 3

\bibitem[{{Arnould} {et~al.}(1999){Arnould}, {Goriely}, \&
  {Jorissen}}]{arnould1999}
{Arnould}, M., {Goriely}, S., \& {Jorissen}, A. 1999, \aap, 347, 572

\bibitem[{{Beers} {et~al.}(1992){Beers}, {Preston}, \& {Shectman}}]{beers1992}
{Beers}, T.~C., {Preston}, G.~W., \& {Shectman}, S.~A. 1992, \aj, 103, 1987

\bibitem[{{Belyaev} {et~al.}(2018){Belyaev}, {Voronov}, \&
  {Gad{\'e}a}}]{belyaev2018}
{Belyaev}, A.~K., {Voronov}, Y.~V., \& {Gad{\'e}a}, F.~X. 2018, \apj, 867, 87

\bibitem[{{Bloecker}(1995)}]{bloecker1995}
{Bloecker}, T. 1995, \aap, 297, 727

\bibitem[{{B{\"o}hm-Vitense}(1958)}]{bom58}
{B{\"o}hm-Vitense}, E. 1958, \zap, 46, 108

\bibitem[{{Bonifacio} {et~al.}(2015){Bonifacio}, {Caffau}, {Spite}, {Limongi},
  {Chieffi}, {Klessen}, {Fran{\c{c}}ois}, {Molaro}, {Ludwig}, {Zaggia},
  {Spite}, {Plez}, {Cayrel}, {Christlieb}, {Clark}, {Glover}, {Hammer}, {Koch},
  {Monaco}, {Sbordone}, \& {Steffen}}]{bonifacio2015}
{Bonifacio}, P., {Caffau}, E., {Spite}, M., {et~al.} 2015, \aap, 579, A28

\bibitem[{{Brusadin} {et~al.}(2013){Brusadin}, {Matteucci}, \&
  {Romano}}]{brusadin2013}
{Brusadin}, G., {Matteucci}, F., \& {Romano}, D. 2013, \aap, 554, A135

\bibitem[{{Campbell} \& {Lattanzio}(2008)}]{campbell2008}
{Campbell}, S.~W. \& {Lattanzio}, J.~C. 2008, \aap, 490, 769

\bibitem[{{Campbell} {et~al.}(2010){Campbell}, {Lugaro}, \&
  {Karakas}}]{campbell2010}
{Campbell}, S.~W., {Lugaro}, M., \& {Karakas}, A.~I. 2010, \aap, 522, L6

\bibitem[{{Cannon}(1993)}]{can93}
{Cannon}, R.~C. 1993, \mnras, 263, 817

\bibitem[{{Canuto} \& {Mazzitelli}(1991)}]{can91}
{Canuto}, V.~M. \& {Mazzitelli}, I. 1991, \apj, 370, 295

\bibitem[{{Cassisi} {et~al.}(1996){Cassisi}, {Castellani}, \&
  {Tornambe}}]{cassisi1996}
{Cassisi}, S., {Castellani}, V., \& {Tornambe}, A. 1996, \apj, 459, 298

\bibitem[{{Cassisi} {et~al.}(2007){Cassisi}, {Potekhin}, {Pietrinferni},
  {Catelan}, \& {Salaris}}]{cassisi2007}
{Cassisi}, S., {Potekhin}, A.~Y., {Pietrinferni}, A., {Catelan}, M., \&
  {Salaris}, M. 2007, \apj, 661, 1094

\bibitem[{{Caughlan} \& {Fowler}(1988)}]{caughlan1988}
{Caughlan}, G.~R. \& {Fowler}, W.~A. 1988, Atomic Data and Nuclear Data Tables,
  40, 283

\bibitem[{{Champagne}(2005)}]{champagne2005}
{Champagne}, A. 2005, in APS April Meeting Abstracts, APS Meeting Abstracts,
  K2.002

\bibitem[{{Chiappini} {et~al.}(1997){Chiappini}, {Matteucci}, \&
  {Gratton}}]{chiappini1997}
{Chiappini}, C., {Matteucci}, F., \& {Gratton}, R. 1997, \apj, 477, 765

\bibitem[{{Chieffi} {et~al.}(2001){Chieffi}, {Dom{\'{\i}}nguez}, {Limongi}, \&
  {Straniero}}]{chieffi2001}
{Chieffi}, A., {Dom{\'{\i}}nguez}, I., {Limongi}, M., \& {Straniero}, O. 2001,
  \apj, 554, 1159

\bibitem[{{Chon} {et~al.}(2021){Chon}, {Omukai}, \& {Schneider}}]{chon2021}
{Chon}, S., {Omukai}, K., \& {Schneider}, R. 2021, \mnras, 508, 4175

\bibitem[{{Choplin} {et~al.}(2021){Choplin}, {Siess}, \&
  {Goriely}}]{choplin2021}
{Choplin}, A., {Siess}, L., \& {Goriely}, S. 2021, \aap, 648, A119

\bibitem[{{Christlieb} {et~al.}(2004){Christlieb}, {Gustafsson}, {Korn},
  {Barklem}, {Beers}, {Bessell}, {Karlsson}, \&
  {Mizuno-Wiedner}}]{christlieb2004}
{Christlieb}, N., {Gustafsson}, B., {Korn}, A.~J., {et~al.} 2004, \apj, 603,
  708

\bibitem[{{Christlieb} {et~al.}(2002){Christlieb}, {Wisotzki}, \&
  {Gra{\ss}hoff}}]{christlieb2002}
{Christlieb}, N., {Wisotzki}, L., \& {Gra{\ss}hoff}, G. 2002, \aap, 391, 397

\bibitem[{{Cirillo} {et~al.}(2022){Cirillo}, {Piersanti}, \&
  {Straniero}}]{cirillo2022}
{Cirillo}, M., {Piersanti}, L., \& {Straniero}, O. 2022, Universe, 8, 44

\bibitem[{{Constantino} {et~al.}(2014){Constantino}, {Campbell}, {Gil-Pons}, \&
  {Lattanzio}}]{constantino2014}
{Constantino}, T., {Campbell}, S., {Gil-Pons}, P., \& {Lattanzio}, J. 2014,
  \apj, 784, 56

\bibitem[{{Cox} \& {Giuli}(1968)}]{coxgiuli1968}
{Cox}, J.~P. \& {Giuli}, R.~T. 1968, {Principles of stellar structure}

\bibitem[{{Cristallo} {et~al.}(2016){Cristallo}, {Karinkuzhi}, {Goswami},
  {Piersanti}, \& {Gobrecht}}]{cristallo2016}
{Cristallo}, S., {Karinkuzhi}, D., {Goswami}, A., {Piersanti}, L., \&
  {Gobrecht}, D. 2016, \apj, 833, 181

\bibitem[{{Cristallo} {et~al.}(2011){Cristallo}, {Piersanti}, {Straniero},
  {Gallino}, {Dom{\'{\i}}nguez}, {Abia}, {Di Rico}, {Quintini}, \&
  {Bisterzo}}]{cristallo2011}
{Cristallo}, S., {Piersanti}, L., {Straniero}, O., {et~al.} 2011, \apjs, 197,
  17

\bibitem[{{Cristallo} {et~al.}(2009){Cristallo}, {Piersanti}, {Straniero},
  {Gallino}, {Dom{\'{\i}}nguez}, \& {K{\"a}ppeler}}]{cristallo2009}
{Cristallo}, S., {Piersanti}, L., {Straniero}, O., {et~al.} 2009, \pasa, 26,
  139

\bibitem[{{Cui} {et~al.}(2012){Cui}, {Zhao}, {Chu}, {Li}, {Li}, {Zhang}, {Su},
  {Yao}, {Wang}, {Xing}, {Li}, {Zhu}, {Wang}, {Gu}, {Luo}, {Xu}, {Zhang},
  {Liu}, {Zhang}, {Yang}, {Cao}, {Chen}, {Chen}, {Chen}, {Chen}, {Chu}, {Feng},
  {Gong}, {Hou}, {Hu}, {Hu}, {Hu}, {Jia}, {Jiang}, {Jiang}, {Jiang}, {Jin},
  {Li}, {Li}, {Li}, {Liu}, {Liu}, {Lu}, {Mao}, {Men}, {Qi}, {Qi}, {Shi},
  {Tang}, {Tao}, {Wang}, {Wang}, {Wang}, {Wang}, {Wang}, {Wang}, {Wang},
  {Wang}, {Wang}, {Wang}, {Wang}, {Wang}, {Xu}, {Xu}, {Yang}, {Yu}, {Yuan},
  {Yuan}, {Zhai}, {Zhang}, {Zhang}, {Zhang}, {Zhao}, {Zhou}, {Zhou}, {Zhu}, \&
  {Zou}}]{cui2012}
{Cui}, X.-Q., {Zhao}, Y.-H., {Chu}, Y.-Q., {et~al.} 2012, Research in Astronomy
  and Astrophysics, 12, 1197

\bibitem[{{Cyburt} {et~al.}(2010){Cyburt}, {Amthor}, {Ferguson}, {Meisel},
  {Smith}, {Warren}, {Heger}, {Hoffman}, {Rauscher}, {Sakharuk}, {Schatz},
  {Thielemann}, \& {Wiescher}}]{cyb10}
{Cyburt}, R.~H., {Amthor}, A.~M., {Ferguson}, R., {et~al.} 2010, \apjs, 189,
  240

\bibitem[{{Dalton} {et~al.}(2012){Dalton}, {Trager}, {Abrams}, {Carter},
  {Bonifacio}, {Aguerri}, {MacIntosh}, {Evans}, {Lewis}, {Navarro}, {Agocs},
  {Dee}, {Rousset}, {Tosh}, {Middleton}, {Pragt}, {Terrett}, {Brock}, {Benn},
  {Verheijen}, {Cano Infantes}, {Bevil}, {Steele}, {Mottram}, {Bates},
  {Gribbin}, {Rey}, {Rodriguez}, {Delgado}, {Guinouard}, {Walton}, {Irwin},
  {Jagourel}, {Stuik}, {Gerlofsma}, {Roelfsma}, {Skillen}, {Ridings},
  {Balcells}, {Daban}, {Gouvret}, {Venema}, \& {Girard}}]{dalton2012}
{Dalton}, G., {Trager}, S.~C., {Abrams}, D.~C., {et~al.} 2012, in \procspie,
  Vol. 8446, Ground-based and Airborne Instrumentation for Astronomy IV, 84460P

\bibitem[{{Dardelet} {et~al.}(2014){Dardelet}, {Ritter}, {Prado}, {Heringer},
  {Higgs}, {Sandalski}, {Jones}, {Denisenkov}, {Venn}, {Bertolli}, {Pignatari},
  {Woodward}, \& {Herwig}}]{dardelet2014}
{Dardelet}, L., {Ritter}, C., {Prado}, P., {et~al.} 2014, in XIII Nuclei in the
  Cosmos (NIC XIII), 145

\bibitem[{{Dell'Agli} {et~al.}(2019){Dell'Agli}, {Valiante}, {Kamath},
  {Ventura}, \& {Garc{\'\i}a-Hern{\'a}ndez}}]{dellagli2019}
{Dell'Agli}, F., {Valiante}, R., {Kamath}, D., {Ventura}, P., \&
  {Garc{\'\i}a-Hern{\'a}ndez}, D.~A. 2019, \mnras, 486, 4738

\bibitem[{{Doherty} {et~al.}(2014{\natexlab{a}}){Doherty}, {Gil-Pons}, {Lau},
  {Lattanzio}, \& {Siess}}]{doherty2014a}
{Doherty}, C.~L., {Gil-Pons}, P., {Lau}, H.~H.~B., {Lattanzio}, J.~C., \&
  {Siess}, L. 2014{\natexlab{a}}, \mnras, 437, 195

\bibitem[{{Doherty} {et~al.}(2014{\natexlab{b}}){Doherty}, {Gil-Pons}, {Lau},
  {Lattanzio}, {Siess}, \& {Campbell}}]{doherty2014b}
{Doherty}, C.~L., {Gil-Pons}, P., {Lau}, H.~H.~B., {et~al.} 2014{\natexlab{b}},
  \mnras, 441, 582

\bibitem[{{Doherty} {et~al.}(2010){Doherty}, {Siess}, {Lattanzio}, \&
  {Gil-Pons}}]{doherty2010}
{Doherty}, C.~L., {Siess}, L., {Lattanzio}, J.~C., \& {Gil-Pons}, P. 2010,
  \mnras, 401, 1453

\bibitem[{{Fenner} {et~al.}(2003){Fenner}, {Gibson}, {Lee}, {Karakas},
  {Lattanzio}, {Chieffi}, {Limongi}, \& {Yong}}]{fenner2003}
{Fenner}, Y., {Gibson}, B.~K., {Lee}, H.~c., {et~al.} 2003, \pasa, 20, 340

\bibitem[{{Ferguson} {et~al.}(2005){Ferguson}, {Alexander}, {Allard}, {Barman},
  {Bodnarik}, {Hauschildt}, {Heffner-Wong}, \& {Tamanai}}]{ferguson2005}
{Ferguson}, J.~W., {Alexander}, D.~R., {Allard}, F., {et~al.} 2005, \apj, 623,
  585

\bibitem[{{Fishlock} {et~al.}(2014){Fishlock}, {Karakas}, {Lugaro}, \&
  {Yong}}]{fishlock2014}
{Fishlock}, C.~K., {Karakas}, A.~I., {Lugaro}, M., \& {Yong}, D. 2014, \apj,
  797, 44

\bibitem[{{Forestini} {et~al.}(1992){Forestini}, {Goriely}, {Jorissen}, \&
  {Arnould}}]{forestini1992}
{Forestini}, M., {Goriely}, S., {Jorissen}, A., \& {Arnould}, M. 1992, \aap,
  261, 157

\bibitem[{{Frebel} {et~al.}(2005){Frebel}, {Aoki}, {Christlieb}, {Ando},
  {Asplund}, {Barklem}, {Beers}, {Eriksson}, {Fechner}, {Fujimoto}, {Honda},
  {Kajino}, {Minezaki}, {Nomoto}, {Norris}, {Ryan}, {Takada-Hidai},
  {Tsangarides}, \& {Yoshii}}]{frebel2005}
{Frebel}, A., {Aoki}, W., {Christlieb}, N., {et~al.} 2005, \nat, 434, 871

\bibitem[{{Frebel} {et~al.}(2008){Frebel}, {Collet}, {Eriksson}, {Christlieb},
  \& {Aoki}}]{frebel2008}
{Frebel}, A., {Collet}, R., {Eriksson}, K., {Christlieb}, N., \& {Aoki}, W.
  2008, \apj, 684, 588

\bibitem[{{Frebel} \& {Norris}(2015)}]{frebel2015}
{Frebel}, A. \& {Norris}, J.~E. 2015, \araa, 53, 631

\bibitem[{{Frost} \& {Lattanzio}(1996)}]{frost1996}
{Frost}, C.~A. \& {Lattanzio}, J.~C. 1996, \apj, 473, 383

\bibitem[{{Fujimoto} {et~al.}(2000){Fujimoto}, {Ikeda}, \&
  {Iben}}]{fujimoto2000}
{Fujimoto}, M.~Y., {Ikeda}, Y., \& {Iben}, Jr., I. 2000, \apjl, 529, L25

\bibitem[{{Gibson} {et~al.}(2003){Gibson}, {Fenner}, {Renda}, {Kawata}, \&
  {Lee}}]{gibson2003}
{Gibson}, B.~K., {Fenner}, Y., {Renda}, A., {Kawata}, D., \& {Lee}, H.-c. 2003,
  \pasa, 20, 401

\bibitem[{{Gil-Pons} {et~al.}(2021){Gil-Pons}, {Doherty}, {Guti{\'e}rrez},
  {Campbell}, {Siess}, \& {Lattanzio}}]{gilpons2021}
{Gil-Pons}, P., {Doherty}, C.~L., {Guti{\'e}rrez}, J., {et~al.} 2021, \aap,
  645, A10

\bibitem[{{Gil-Pons} {et~al.}(2018){Gil-Pons}, {Doherty}, {Guti{\'e}rrez},
  {Siess}, {Campbell}, {Lau}, \& {Lattanzio}}]{gilpons2018}
{Gil-Pons}, P., {Doherty}, C.~L., {Guti{\'e}rrez}, J.~L., {et~al.} 2018, \pasa,
  35

\bibitem[{{Gil-Pons} {et~al.}(2013){Gil-Pons}, {Doherty}, {Lau}, {Campbell},
  {Suda}, {Guilani}, {Guti{\'e}rrez}, \& {Lattanzio}}]{gilpons2013}
{Gil-Pons}, P., {Doherty}, C.~L., {Lau}, H., {et~al.} 2013, \aap, 557, A106

\bibitem[{{Grevesse} {et~al.}(1996){Grevesse}, {Noels}, \&
  {Sauval}}]{grevesse1996}
{Grevesse}, N., {Noels}, A., \& {Sauval}, A.~J. 1996, in Astronomical Society
  of the Pacific Conference Series, Vol.~99, Cosmic Abundances, ed. S.~S.
  {Holt} \& G.~{Sonneborn}, 117

\bibitem[{{Hale} {et~al.}(2002){Hale}, {Champagne}, {Iliadis}, {Hansper},
  {Powell}, \& {Blackmon}}]{hale2002}
{Hale}, S.~E., {Champagne}, A.~E., {Iliadis}, C., {et~al.} 2002, \prc, 65,
  015801

\bibitem[{{Hale} {et~al.}(2004){Hale}, {Champagne}, {Iliadis}, {Hansper},
  {Powell}, \& {Blackmon}}]{hale2004}
{Hale}, S.~E., {Champagne}, A.~E., {Iliadis}, C., {et~al.} 2004, \prc, 70,
  045802

\bibitem[{{Hampel} {et~al.}(2016){Hampel}, {Stancliffe}, {Lugaro}, \&
  {Meyer}}]{Hampel2016}
{Hampel}, M., {Stancliffe}, R.~J., {Lugaro}, M., \& {Meyer}, B.~S. 2016, \apj,
  831, 171

\bibitem[{{Harris} {et~al.}(1983){Harris}, {Fowler}, {Caughlan}, \&
  {Zimmerman}}]{harris1983}
{Harris}, M.~J., {Fowler}, W.~A., {Caughlan}, G.~R., \& {Zimmerman}, B.~A.
  1983, \araa, 21, 165

\bibitem[{{Hartwig} {et~al.}(2019){Hartwig}, {Ishigaki}, {Klessen}, \&
  {Yoshida}}]{hartwig2019}
{Hartwig}, T., {Ishigaki}, M.~N., {Klessen}, R.~S., \& {Yoshida}, N. 2019,
  \mnras, 482, 1204

\bibitem[{{Herwig}(2004)}]{herwig2004}
{Herwig}, F. 2004, \apj, 605, 425

\bibitem[{{Herwig} {et~al.}(2003){Herwig}, {Langer}, \& {Lugaro}}]{herwig2003}
{Herwig}, F., {Langer}, N., \& {Lugaro}, M. 2003, \apj, 593, 1056

\bibitem[{{Herwig} {et~al.}(2011){Herwig}, {Pignatari}, {Woodward}, {Porter},
  {Rockefeller}, {Fryer}, {Bennett}, \& {Hirschi}}]{herwig2011}
{Herwig}, F., {Pignatari}, M., {Woodward}, P.~R., {et~al.} 2011, \apj, 727, 89

\bibitem[{{Iglesias} \& {Rogers}(1996)}]{igl96}
{Iglesias}, C.~A. \& {Rogers}, F.~J. 1996, \apj, 464, 943

\bibitem[{{Iliadis} {et~al.}(2001){Iliadis}, {D'Auria}, {Starrfield},
  {Thompson}, \& {Wiescher}}]{iliadis2001}
{Iliadis}, C., {D'Auria}, J.~M., {Starrfield}, S., {Thompson}, W.~J., \&
  {Wiescher}, M. 2001, \apjs, 134, 151

\bibitem[{{Ishigaki} {et~al.}(2018){Ishigaki}, {Tominaga}, {Kobayashi}, \&
  {Nomoto}}]{ishigaki2018}
{Ishigaki}, M.~N., {Tominaga}, N., {Kobayashi}, C., \& {Nomoto}, K. 2018, \apj,
  857, 46

\bibitem[{{Iwamoto}(2009)}]{iwamoto2009}
{Iwamoto}, N. 2009, \pasa, 26, 145

\bibitem[{Iwamoto {et~al.}(2005)Iwamoto, Umeda, Tominaga, Nomoto, \&
  Maeda}]{iwamoto2005}
Iwamoto, N., Umeda, H., Tominaga, N., Nomoto, K., \& Maeda, K. 2005, Science,
  309, 451

\bibitem[{{Izzard} {et~al.}(2006){Izzard}, {Lugaro}, {Illadis}, \&
  {Karakas}}]{izzard2006}
{Izzard}, R., {Lugaro}, M., {Illadis}, C., \& {Karakas}, A. 2006, in
  International Symposium on Nuclear Astrophysics - Nuclei in the Cosmos, 38.1

\bibitem[{{Izzard} {et~al.}(2009){Izzard}, {Glebbeek}, {Stancliffe}, \&
  {Pols}}]{izzard2009}
{Izzard}, R.~G., {Glebbeek}, E., {Stancliffe}, R.~J., \& {Pols}, O.~R. 2009,
  \aap, 508, 1359

\bibitem[{{Johnson} {et~al.}(2022){Johnson}, {Weinberg}, {Vincenzo}, {Bird}, \&
  {Griffith}}]{johnson2022}
{Johnson}, J.~W., {Weinberg}, D.~H., {Vincenzo}, F., {Bird}, J.~C., \&
  {Griffith}, E.~J. 2022, arXiv e-prints, arXiv:2202.04666

\bibitem[{{Jones} {et~al.}(2016){Jones}, {Ritter}, {Herwig}, {Fryer},
  {Pignatari}, {Bertolli}, \& {Paxton}}]{jones2016a}
{Jones}, S., {Ritter}, C., {Herwig}, F., {et~al.} 2016, \mnras, 455, 3848

\bibitem[{{Jorissen} \& {Arnould}(1989)}]{jorissen1989}
{Jorissen}, A. \& {Arnould}, M. 1989, \aap, 221, 161

\bibitem[{{Karakas}(2010)}]{karakas2010}
{Karakas}, A.~I. 2010, \mnras, 403, 1413

\bibitem[{{Karakas} \& {Lattanzio}(2014)}]{karakas2014}
{Karakas}, A.~I. \& {Lattanzio}, J.~C. 2014, \pasa, 31, e030

\bibitem[{{Karakas} {et~al.}(2006){Karakas}, {Lugaro}, {Wiescher},
  {G{\"o}rres}, \& {Ugalde}}]{karakas2006}
{Karakas}, A.~I., {Lugaro}, M.~A., {Wiescher}, M., {G{\"o}rres}, J., \&
  {Ugalde}, C. 2006, \apj, 643, 471

\bibitem[{{Keller} {et~al.}(2014){Keller}, {Bessell}, {Frebel}, {Casey},
  {Asplund}, {Jacobson}, {Lind}, {Norris}, {Yong}, {Heger}, {Magic}, {da
  Costa}, {Schmidt}, \& {Tisserand}}]{keller2014}
{Keller}, S.~C., {Bessell}, M.~S., {Frebel}, A., {et~al.} 2014, \nat, 506, 463

\bibitem[{{Keller} {et~al.}(2007){Keller}, {Schmidt}, {Bessell}, {Conroy},
  {Francis}, {Granlund}, {Kowald}, {Oates}, {Martin-Jones}, {Preston},
  {Tisserand}, {Vaccarella}, \& {Waterson}}]{keller2007}
{Keller}, S.~C., {Schmidt}, B.~P., {Bessell}, M.~S., {et~al.} 2007, \pasa, 24,
  1

\bibitem[{{Kim} \& {L{\'e}pine}(2022)}]{bokyung2022}
{Kim}, B. \& {L{\'e}pine}, S. 2022, \mnras, 510, 4308

\bibitem[{{Kobayashi} {et~al.}(2020){Kobayashi}, {Karakas}, \&
  {Lugaro}}]{kobayashi2020}
{Kobayashi}, C., {Karakas}, A.~I., \& {Lugaro}, M. 2020, \apj, 900, 179

\bibitem[{{Kobayashi} {et~al.}(2011){Kobayashi}, {Tominaga}, \&
  {Nomoto}}]{kob11}
{Kobayashi}, C., {Tominaga}, N., \& {Nomoto}, K. 2011, \apjl, 730, L14

\bibitem[{{Komiya} {et~al.}(2007){Komiya}, {Suda}, {Minaguchi}, {Shigeyama},
  {Aoki}, \& {Fujimoto}}]{komiya2007}
{Komiya}, Y., {Suda}, T., {Minaguchi}, H., {et~al.} 2007, \apj, 658, 367

\bibitem[{{Lattanzio}(1986)}]{lat86}
{Lattanzio}, J.~C. 1986, \apj, 311, 708

\bibitem[{{Lattanzio} {et~al.}(2015){Lattanzio}, {Siess}, {Church}, {Angelou},
  {Stancliffe}, {Doherty}, {Stephen}, \& {Campbell}}]{lattanzio2015}
{Lattanzio}, J.~C., {Siess}, L., {Church}, R.~P., {et~al.} 2015, \mnras, 446,
  2673

\bibitem[{{Lau} {et~al.}(2012){Lau}, {Gil-Pons}, {Doherty}, \&
  {Lattanzio}}]{lau2012}
{Lau}, H.~H.~B., {Gil-Pons}, P., {Doherty}, C., \& {Lattanzio}, J. 2012, \aap,
  542, A1

\bibitem[{{Lau} {et~al.}(2007){Lau}, {Stancliffe}, \& {Tout}}]{lau2007}
{Lau}, H.~H.~B., {Stancliffe}, R.~J., \& {Tout}, C.~A. 2007, \mnras, 378, 563

\bibitem[{{Lederer} \& {Aringer}(2009)}]{lederer2009}
{Lederer}, M.~T. \& {Aringer}, B. 2009, \aap, 494, 403

\bibitem[{{Lucatello} {et~al.}(2005){Lucatello}, {Tsangarides}, {Beers},
  {Carretta}, {Gratton}, \& {Ryan}}]{lucatello2005}
{Lucatello}, S., {Tsangarides}, S., {Beers}, T.~C., {et~al.} 2005, \apj, 625,
  825

\bibitem[{{Lugaro} {et~al.}(2003){Lugaro}, {Herwig}, {Lattanzio}, {Gallino}, \&
  {Straniero}}]{lugaro2003}
{Lugaro}, M., {Herwig}, F., {Lattanzio}, J.~C., {Gallino}, R., \& {Straniero},
  O. 2003, \apj, 586, 1305

\bibitem[{{Lugaro} {et~al.}(2004){Lugaro}, {Ugalde}, {Karakas}, {G{\"o}rres},
  {Wiescher}, {Lattanzio}, \& {Cannon}}]{lug04}
{Lugaro}, M., {Ugalde}, C., {Karakas}, A.~I., {et~al.} 2004, \apj, 615, 934

\bibitem[{{Marassi} {et~al.}(2014){Marassi}, {Chiaki}, {Schneider}, {Limongi},
  {Omukai}, {Nozawa}, {Chieffi}, \& {Yoshida}}]{marassi2014}
{Marassi}, S., {Chiaki}, G., {Schneider}, R., {et~al.} 2014, \apj, 794, 100

\bibitem[{{Marigo} \& {Aringer}(2009)}]{marigo2009}
{Marigo}, P. \& {Aringer}, B. 2009, \aap, 508, 1539

\bibitem[{{Martin} {et~al.}(2022{\natexlab{a}}){Martin}, {Ibata},
  {Starkenburg}, {Yuan}, {Malhan}, {Bellazzini}, {Viswanathan}, {Aguado},
  {Arentsen}, {Bonifacio}, {Carlberg}, {Gonz{\'a}lez Hern{\'a}ndez}, {Hill},
  {Jablonka}, {Kordopatis}, {Lardo}, {McConnachie}, {Navarro},
  {S{\'a}nchez-Janssen}, {Sestito}, {Thomas}, {Venn}, {Vitali}, \&
  {Voggel}}]{martin2022b}
{Martin}, N.~F., {Ibata}, R.~A., {Starkenburg}, E., {et~al.}
  2022{\natexlab{a}}, arXiv e-prints, arXiv:2201.01310

\bibitem[{{Martin} {et~al.}(2022{\natexlab{b}}){Martin}, {Venn}, {Aguado},
  {Starkenburg}, {Gonz{\'a}lez Hern{\'a}ndez}, {Ibata}, {Bonifacio}, {Caffau},
  {Sestito}, {Arentsen}, {Allende Prieto}, {Carlberg}, {Fabbro}, {Fouesneau},
  {Hill}, {Jablonka}, {Kordopatis}, {Lardo}, {Malhan}, {Mashonkina},
  {McConnachie}, {Navarro}, {S{\'a}nchez-Janssen}, {Thomas}, {Yuan}, \&
  {Mucciarelli}}]{martin2022a}
{Martin}, N.~F., {Venn}, K.~A., {Aguado}, D.~S., {et~al.} 2022{\natexlab{b}},
  \nat, 601, 45

\bibitem[{{Mel{\'e}ndez} \& {Cohen}(2007)}]{melendez2007}
{Mel{\'e}ndez}, J. \& {Cohen}, J.~G. 2007, \apjl, 659, L25

\bibitem[{{Mill{\'a}n-Irigoyen} {et~al.}(2020){Mill{\'a}n-Irigoyen},
  {Moll{\'a}}, \& {Ascasibar}}]{millan2019}
{Mill{\'a}n-Irigoyen}, I., {Moll{\'a}}, M., \& {Ascasibar}, Y. 2020, \mnras,
  494, 146

\bibitem[{{Moc{\'a}k} {et~al.}(2010){Moc{\'a}k}, {Campbell}, {M{\"u}ller}, \&
  {Kifonidis}}]{mocak2010}
{Moc{\'a}k}, M., {Campbell}, S.~W., {M{\"u}ller}, E., \& {Kifonidis}, K. 2010,
  \aap, 520, A114

\bibitem[{Mura-Guzmán {et~al.}(2020)Mura-Guzmán, Yong, Abate, Karakas,
  Kobayashi, Oh, Chun, \& Mace}]{muraguzman2020}
Mura-Guzmán, A., Yong, D., Abate, C., {et~al.} 2020, Monthly Notices of the
  Royal Astronomical Society, 498, 3549

\bibitem[{{Nomoto} {et~al.}(2013){Nomoto}, {Kobayashi}, \&
  {Tominaga}}]{nomoto2013}
{Nomoto}, K., {Kobayashi}, C., \& {Tominaga}, N. 2013, \araa, 51, 457

\bibitem[{{Nordlander} {et~al.}(2019){Nordlander}, {Bessell}, {Da Costa},
  {Mackey}, {Asplund}, {Casey}, {Chiti}, {Ezzeddine}, {Frebel}, {Lind},
  {Marino}, {Murphy}, {Norris}, {Schmidt}, \& {Yong}}]{nordlander2019}
{Nordlander}, T., {Bessell}, M.~S., {Da Costa}, G.~S., {et~al.} 2019, \mnras,
  488, L109

\bibitem[{{Osorio} {et~al.}(2022){Osorio}, {Aguado}, {Prieto}, {Hubeny}, \&
  {Gonz{\'a}lez Hern{\'a}ndez}}]{osorio2022}
{Osorio}, Y., {Aguado}, D.~S., {Prieto}, C.~A., {Hubeny}, I., \& {Gonz{\'a}lez
  Hern{\'a}ndez}, J.~I. 2022, \apj, 928, 173

\bibitem[{{Pols} {et~al.}(2012){Pols}, {Izzard}, {Stancliffe}, \&
  {Glebbeek}}]{pols2012}
{Pols}, O.~R., {Izzard}, R.~G., {Stancliffe}, R.~J., \& {Glebbeek}, E. 2012,
  \aap, 547, A76

\bibitem[{{Potekhin} {et~al.}(2015){Potekhin}, {Pons}, \&
  {Page}}]{potekhin2015}
{Potekhin}, A.~Y., {Pons}, J.~A., \& {Page}, D. 2015, \ssr, 191, 239

\bibitem[{{Prantzos} {et~al.}(2018){Prantzos}, {Abia}, {Limongi}, {Chieffi}, \&
  {Cristallo}}]{prantzos2018}
{Prantzos}, N., {Abia}, C., {Limongi}, M., {Chieffi}, A., \& {Cristallo}, S.
  2018, \mnras, 476, 3432

\bibitem[{{Ritter} {et~al.}(2018){Ritter}, {Herwig}, {Jones}, {Pignatari},
  {Fryer}, \& {Hirschi}}]{ritter2018}
{Ritter}, C., {Herwig}, F., {Jones}, S., {et~al.} 2018, \mnras, 480, 538

\bibitem[{{Roederer}(2017)}]{roederer2017}
{Roederer}, I.~U. 2017, \apj, 835, 23

\bibitem[{{Romano} {et~al.}(2017){Romano}, {Matteucci}, {Zhang},
  {Papadopoulos}, \& {Ivison}}]{romano2017}
{Romano}, D., {Matteucci}, F., {Zhang}, Z.~Y., {Papadopoulos}, P.~P., \&
  {Ivison}, R.~J. 2017, \mnras, 470, 401

\bibitem[{{Schneider} {et~al.}(2012){Schneider}, {Omukai}, {Bianchi}, \&
  {Valiante}}]{schneider12}
{Schneider}, R., {Omukai}, K., {Bianchi}, S., \& {Valiante}, R. 2012, \mnras,
  419, 1566

\bibitem[{{Sestito} {et~al.}(2021){Sestito}, {Buck}, {Starkenburg}, {Martin},
  {Navarro}, {Venn}, {Obreja}, {Jablonka}, \& {Macci{\`o}}}]{sestito2021}
{Sestito}, F., {Buck}, T., {Starkenburg}, E., {et~al.} 2021, \mnras, 500, 3750

\bibitem[{{Sharda} \& {Krumholz}(2022)}]{sharda2022}
{Sharda}, P. \& {Krumholz}, M.~R. 2022, \mnras, 509, 1959

\bibitem[{{Siess}(2007)}]{siess2007}
{Siess}, L. 2007, \aap, 476, 893

\bibitem[{{Siess} {et~al.}(2002){Siess}, {Livio}, \& {Lattanzio}}]{siess2002}
{Siess}, L., {Livio}, M., \& {Lattanzio}, J. 2002, \apj, 570, 329

\bibitem[{{Simon}(2019)}]{simon2019}
{Simon}, J.~D. 2019, \araa, 57, 375

\bibitem[{{Sk{\'u}lad{\'o}ttir} {et~al.}(2020){Sk{\'u}lad{\'o}ttir}, {Hansen},
  {Choplin}, {Salvadori}, {Hampel}, \& {Campbell}}]{skuladottir2020}
{Sk{\'u}lad{\'o}ttir}, {\'A}., {Hansen}, C.~J., {Choplin}, A., {et~al.} 2020,
  \aap, 634, A84

\bibitem[{{Spite} {et~al.}(2021){Spite}, {Spite}, \& {Barbuy}}]{spite2021}
{Spite}, M., {Spite}, F., \& {Barbuy}, B. 2021, \aap, 652, A97

\bibitem[{{Spitoni} {et~al.}(2017){Spitoni}, {Vincenzo}, \&
  {Matteucci}}]{spitoni2017}
{Spitoni}, E., {Vincenzo}, F., \& {Matteucci}, F. 2017, \aap, 599, A6

\bibitem[{{Stancliffe}(2006)}]{stancliffe2006}
{Stancliffe}, R.~J. 2006, \mnras, 370, 1817

\bibitem[{{Stancliffe} {et~al.}(2011){Stancliffe}, {Dearborn}, {Lattanzio},
  {Heap}, \& {Campbell}}]{stancliffe2011}
{Stancliffe}, R.~J., {Dearborn}, D.~S.~P., {Lattanzio}, J.~C., {Heap}, S.~A.,
  \& {Campbell}, S.~W. 2011, \apj, 742, 121

\bibitem[{{Starkenburg} {et~al.}(2014){Starkenburg}, {Shetrone}, {McConnachie},
  \& {Venn}}]{starkenburg2014}
{Starkenburg}, E., {Shetrone}, M.~D., {McConnachie}, A.~W., \& {Venn}, K.~A.
  2014, \mnras, 441, 1217

\bibitem[{{Suda} {et~al.}(2008){Suda}, {Katsuta}, {Yamada}, {Suwa}, {Ishizuka},
  {Komiya}, {Sorai}, {Aikawa}, \& {Fujimoto}}]{suda2008}
{Suda}, T., {Katsuta}, Y., {Yamada}, S., {et~al.} 2008, \pasj, 60, 1159

\bibitem[{{Suda} {et~al.}(2013){Suda}, {Komiya}, {Yamada}, {Katsuta}, {Aoki},
  {Gil-Pons}, {Doherty}, {Campbell}, {Wood}, \& {Fujimoto}}]{suda2013}
{Suda}, T., {Komiya}, Y., {Yamada}, S., {et~al.} 2013, \mnras, 432, 46

\bibitem[{{Tamura} {et~al.}(2016){Tamura}, {Takato}, {Shimono}, {Moritani},
  {Yabe}, {Ishizuka}, {Ueda}, {Kamata}, {Aghazarian}, {Arnouts}, {Barban},
  {Barkhouser}, {Borges}, {Braun}, {Carr}, {Chabaud}, {Chang}, {Chen}, {Chiba},
  {Chou}, {Chu}, {Cohen}, {de Almeida}, {de Oliveira}, {de Oliveira}, {Dekany},
  {Dohlen}, {dos Santos}, {dos Santos}, {Ellis}, {Fabricius}, {Ferrand},
  {Ferreira}, {Golebiowski}, {Greene}, {Gross}, {Gunn}, {Hammond}, {Harding},
  {Hart}, {Heckman}, {Hirata}, {Ho}, {Hope}, {Hovland}, {Hsu}, {Hu}, {Huang},
  {Jaquet}, {Jing}, {Karr}, {Kimura}, {King}, {Komatsu}, {Le Brun}, {Le
  F{\`e}vre}, {Le Fur}, {Le Mignant}, {Ling}, {Loomis}, {Lupton}, {Madec},
  {Mao}, {Marrara}, {Mendes de Oliveira}, {Minowa}, {Morantz}, {Murayama},
  {Murray}, {Ohyama}, {Orndorff}, {Pascal}, {Pereira}, {Reiley}, {Reinecke},
  {Ritter}, {Roberts}, {Schwochert}, {Seiffert}, {Smee}, {Sodre}, {Spergel},
  {Steinkraus}, {Strauss}, {Surace}, {Suto}, {Suzuki}, {Swinbank}, {Tait},
  {Takada}, {Tamura}, {Tanaka}, {Tresse}, {Verducci}, {Vibert}, {Vidal},
  {Wang}, {Wen}, {Yan}, \& {Yasuda}}]{tamura2016}
{Tamura}, N., {Takato}, N., {Shimono}, A., {et~al.} 2016, in Society of
  Photo-Optical Instrumentation Engineers (SPIE) Conference Series, Vol. 9908,
  Ground-based and Airborne Instrumentation for Astronomy VI, ed. C.~J.
  {Evans}, L.~{Simard}, \& H.~{Takami}, 99081M

\bibitem[{{Tashibu} {et~al.}(2017){Tashibu}, {Yasuda}, \&
  {Kozasa}}]{tashibu2017}
{Tashibu}, S., {Yasuda}, Y., \& {Kozasa}, T. 2017, \mnras, 466, 1709

\bibitem[{{Tsujimoto} \& {Bekki}(2012)}]{tsujimoto2012}
{Tsujimoto}, T. \& {Bekki}, K. 2012, \apj, 747, 125

\bibitem[{{Tumlinson}(2010)}]{tumlinson2010}
{Tumlinson}, J. 2010, \apj, 708, 1398

\bibitem[{{Umeda} \& {Nomoto}(2003)}]{umeda2003}
{Umeda}, H. \& {Nomoto}, K. 2003, \nat, 422, 871

\bibitem[{{Vangioni} \& {Olive}(2019)}]{vangioni2019}
{Vangioni}, E. \& {Olive}, K.~A. 2019, \mnras, 484, 3561

\bibitem[{{Vassiliadis} \& {Wood}(1993)}]{vassiliadis1993}
{Vassiliadis}, E. \& {Wood}, P.~R. 1993, \apj, 413, 641

\bibitem[{{Ventura} {et~al.}(2018){Ventura}, {D'Antona}, {Imbriani}, {Di
  Criscienzo}, {Dell'Agli}, \& {Tailo}}]{ventura2018}
{Ventura}, P., {D'Antona}, F., {Imbriani}, G., {et~al.} 2018, \mnras, 477, 438

\bibitem[{{Ventura} {et~al.}(2001){Ventura}, {D'Antona}, {Mazzitelli}, \&
  {Gratton}}]{ventura2001}
{Ventura}, P., {D'Antona}, F., {Mazzitelli}, I., \& {Gratton}, R. 2001, \apjl,
  550, L65

\bibitem[{{Ventura} {et~al.}(2021){Ventura}, {Dell'Agli}, {Romano}, {Tosi},
  {Limongi}, {Chieffi}, {Castellani}, {Tailo}, {Lugaro}, {Marini}, \&
  {Yag{\"u}e Lopez}}]{ventura2021}
{Ventura}, P., {Dell'Agli}, F., {Romano}, D., {et~al.} 2021, \aap, 655, A6

\bibitem[{{Ventura} {et~al.}(2013){Ventura}, {Di Criscienzo}, {Carini}, \&
  {D'Antona}}]{ventura2013}
{Ventura}, P., {Di Criscienzo}, M., {Carini}, R., \& {D'Antona}, F. 2013,
  \mnras, 431, 3642

\bibitem[{{Ventura} {et~al.}(2015){Ventura}, {Karakas}, {Dell'Agli}, {Boyer},
  {Garc{\'\i}a-Hern{\'a}ndez}, {Di Criscienzo}, \& {Schneider}}]{ventura2015a}
{Ventura}, P., {Karakas}, A.~I., {Dell'Agli}, F., {et~al.} 2015, \mnras, 450,
  3181

\bibitem[{{Ventura} {et~al.}(2016){Ventura}, {Karakas}, {Dell'Agli},
  {Garc{\'{\i}}a-Hern{\'a}ndez}, {Boyer}, \& {Di Criscienzo}}]{ventura2016a}
{Ventura}, P., {Karakas}, A.~I., {Dell'Agli}, F., {et~al.} 2016, \mnras, 457,
  1456

\bibitem[{{Vescovi} {et~al.}(2022){Vescovi}, {Cristallo}, {Palmerini}, {Abia},
  \& {Busso}}]{vescovi2022}
{Vescovi}, D., {Cristallo}, S., {Palmerini}, S., {Abia}, C., \& {Busso}, M.
  2022, in European Physical Journal Web of Conferences, Vol. 260, European
  Physical Journal Web of Conferences, 11009

\bibitem[{{White} \& {Springel}(2000)}]{white2000}
{White}, S. D.~M. \& {Springel}, V. 2000, in The First Stars, ed. A.~{Weiss},
  T.~G. {Abel}, \& V.~{Hill}, 327

\bibitem[{{Woodward} {et~al.}(2015){Woodward}, {Herwig}, \&
  {Lin}}]{woodward2015}
{Woodward}, P.~R., {Herwig}, F., \& {Lin}, P.-H. 2015, \apj, 798, 49

\bibitem[{{Yanny} {et~al.}(2009){Yanny}, {Rockosi}, {Newberg}, {Knapp},
  {Adelman-McCarthy}, {Alcorn}, {Allam}, {Allende Prieto}, {An}, {Anderson},
  {Anderson}, {Bailer-Jones}, {Bastian}, {Beers}, {Bell}, {Belokurov},
  {Bizyaev}, {Blythe}, {Bochanski}, {Boroski}, {Brinchmann}, {Brinkmann},
  {Brewington}, {Carey}, {Cudworth}, {Evans}, {Evans}, {Gates}, {G{\"a}nsicke},
  {Gillespie}, {Gilmore}, {Nebot Gomez-Moran}, {Grebel}, {Greenwell}, {Gunn},
  {Jordan}, {Jordan}, {Harding}, {Harris}, {Hendry}, {Holder}, {Ivans},
  {Ivezi{\v c}}, {Jester}, {Johnson}, {Kent}, {Kleinman}, {Kniazev},
  {Krzesinski}, {Kron}, {Kuropatkin}, {Lebedeva}, {Lee}, {French Leger},
  {L{\'e}pine}, {Levine}, {Lin}, {Long}, {Loomis}, {Lupton}, {Malanushenko},
  {Malanushenko}, {Margon}, {Martinez-Delgado}, {McGehee}, {Monet}, {Morrison},
  {Munn}, {Neilsen}, {Nitta}, {Norris}, {Oravetz}, {Owen}, {Padmanabhan},
  {Pan}, {Peterson}, {Pier}, {Platson}, {Re Fiorentin}, {Richards}, {Rix},
  {Schlegel}, {Schneider}, {Schreiber}, {Schwope}, {Sibley}, {Simmons},
  {Snedden}, {Allyn Smith}, {Stark}, {Stauffer}, {Steinmetz}, {Stoughton},
  {SubbaRao}, {Szalay}, {Szkody}, {Thakar}, {Sivarani}, {Tucker}, {Uomoto},
  {Vanden Berk}, {Vidrih}, {Wadadekar}, {Watters}, {Wilhelm}, {Wyse}, {Yarger},
  \& {Zucker}}]{yanny2009}
{Yanny}, B., {Rockosi}, C., {Newberg}, H.~J., {et~al.} 2009, \aj, 137, 4377

\bibitem[{{Zackrisson} {et~al.}(2011){Zackrisson}, {Rydberg}, {Schaerer},
  {{\"O}stlin}, \& {Tuli}}]{zackrisson2011}
{Zackrisson}, E., {Rydberg}, C.-E., {Schaerer}, D., {{\"O}stlin}, G., \&
  {Tuli}, M. 2011, \apj, 740, 13

\end{thebibliography}
%
%

\begin{appendix}

\section{List of species and nuclear reactions used in the stellar evolution code \texttt{MONSTAR}}\label{sec:reactions}

The species included in the structure evolution code \texttt{MONSTAR} are: \iso{1}H, \iso{3}He, \iso{4}He, \iso{12}C, \iso{14}N, \iso{16}O and Z$_{other}$ (all the species of mass higher than \iso{16}O).
The nucleosynthesis code \texttt{MONSOON} considers the following 77 species:
g (the {\it gallino} particle, used as a proxy for light s-process isotopes), n, 
 \iso{1-2}H, \iso{3-4}He, \iso{7}Li, \iso{7}Be. \iso{8}B,
\iso{12-14}C, \iso{13-15}N, \iso{14-19}O, \iso{17-20}F, \iso{19-23}Ne, \iso{21-24}Na, \iso{23-27}Mg, \iso{25-28}Al, $^{26-}$Al and $^{26*}$Al (in the ground and metastable state, respectively), \iso{27-33}Si, \iso{29-34}P, \iso{32-35}S, \iso{54-61}Fe, \iso{59-61}Co, \iso{58-62}Ni.

The detailed list of nuclear reactions used in \texttt{MONSTAR} are shown in Table \ref{tab:rrates}.

\begin{table*}[h]
    \begin{center}
    \caption{Nuclear reactions considered in our stellar structure evolution code, \texttt{MONSTAR}.} \end{center}
\centering
\vspace{-0.5cm}
    \begin{tabular}{llll}
    \hline
            & {\bf Source} & {\bf Reactions} & {\bf Assumed reactions}\\[2pt]
             \hline
             \hline
    pp-chains & \cite{harris1983} & \iso{1}H(p,e$^+\:\nu_e$)\iso{2}H(p,$\gamma$)\iso{3}He & 3\iso{1}H $\rightarrow$ \iso{3}He\\[5pt]
   &  &    \iso{3}He(\iso{3}He,2p)\iso{4}He & 2\iso{3}He $\rightarrow$ \iso{4}He + 2\iso{1}H  \\
   &  &  $\rm ^3He(\alpha,\gamma)^7Be\:
   \Bigg\{ \begin{array}{c}
        \rm(e^-\:\nu)^7Li(p,\gamma)^8Be(\alpha)^4He\\
        \rm(p,\gamma)^8B(e^+\:\nu)^8Be(\alpha)^4He
   \end{array} \Bigg .$
   & \iso{3}He, \iso{4}He, \iso{1}H $\rightarrow$ 2\:\iso{4}He\\
         \\[0pt]
         \hline 
    CNO-cycle & CF88     & \iso{12}C(p,$\gamma$)\iso{13}N(e$^+\:\nu$)\iso{13}C(p,$\gamma$)\iso{14}N & \iso{12}C, 2\iso{1}H $\rightarrow$ \iso{14}N\\
    & \cite{champagne2005}    & \iso{14}N(p,$\gamma$)\iso{15}O(e$^+\:\nu$)\iso{15}N(p,$\alpha$)\iso{12}C & \iso{14}N, 2\iso{1}H $\rightarrow$ \iso{12}C, \iso{4}He \\
    & CF88    & \iso{16}O(p,$\gamma$)\iso{17}F(e$^+\:\nu$)\iso{17}O(p,$\alpha$)\iso{14}N &
         \iso{16}O, 2\iso{1}H $\rightarrow$ \iso{14}N, \iso{4}He \\
         \\[-10pt]
          & & & \\
         \hline
    He-burning & CF88, Ang99     & \iso{4}He(2$\alpha$,$\gamma$)\iso{12}C & 3\iso{4}He $\rightarrow$ \iso{12}C \\
    &     & \iso{12}C($\alpha$,$\gamma$)\iso{16}O &  \iso{12}C, \iso{4}He $\rightarrow$ \iso{16}O\\
    &     & \iso{14}N($\alpha$,$\gamma$)\iso{18}F(e$^+\:\nu$)\iso{18}O($\alpha$, $\gamma$)\iso{22}Ne & \iso{14}N, 2\iso{4}He  $\rightarrow$ Z$_{other}$\\
         \\[-10pt]
          & & & \\
         \hline
    C-burning & CF88, Ang99    & \iso{12}C(\iso{12}C,$\alpha$)\iso{20}Ne & 2\iso{12}C$\rightarrow$ Z$_{other}$\\
    \end{tabular}
    \tablefoot{The second column shows the source from which the reaction rates were taken, the third column shows the full reaction chains, and the fourth column shows the abbreviated reaction set used. The abbreviated reaction chains account for all the energy generation whilst minimising computational cost. C-burning in the version of the code we are using considers the `no-sodium approximation', according to which only the reaction \iso{12}C(\iso{12}C,$\alpha$)\iso{20}Ne is taken into account. For a more realistic treatment of C-burning, see \cite{doherty2010}. \iso{20}Ne resulting from this process is included in Z$_{other}$. CF88 and Ang99 refer to \cite{caughlan1988} and \cite{angulo1999}, respectively.  \\}
    \label{tab:rrates}
\end{table*}

\onecolumn
\section{Abundances patterns in terms of [X/Fe]} 
\begin{table*}[h]
    \centering
    \caption{Abundances pattern of selected elements in terms of [X/Fe].}
\begin{tabular}{l c c c c c c c c c c c c c}
$M_\mathrm{ini}/$\msun & $<log_{10}(^{7}Li)>$ & C & N & O & F & Ne & Na & Mg & Al & Si & P & S \\[1pt]
\hline\\
3.0 &   2.50 &   4.30 &   4.97 &   2.55 &   4.30 &   3.50 &   3.29 &   3.22 &   2.40 &   1.36 &   1.69 &    4.0$\times$10$^{-2}$ \\
3.0-dil &   1.47 &   3.27 &   3.93 &   1.52 &   3.27 &   2.46 &   2.26 &   2.18 &   1.38 &   0.48 &   0.73 &    3.8$\times$10$^{-3}$ \\
4.0 &   1.43 &   3.55 &   5.03 &   2.31 &   3.34 &   3.22 &   3.19 &   3.00 &   2.48 &   1.22 &   1.72 &    4.2$\times$10$^{-2}$ \\
4.0-dil &   0.65 &   2.68 &   4.16 &   1.44 &   2.47 &   2.34 &   2.31 &   2.13 &   1.62 &   0.49 &   0.89 &    5.9$\times$10$^{-3}$ \\
5.0 &   0.89 &   3.44 &   4.98 &   2.29 &   2.90 &   2.94 &   2.92 &   2.93 &   2.57 &   1.13 &   1.71 &    5.7$\times$10$^{-2}$ \\
5.0-dil &   0.33 &   2.66 &   4.20 &   1.53 &   2.13 &   2.17 &   2.15 &   2.16 &   1.81 &   0.49 &   0.98 &    1.0$\times$10$^{-2}$ \\
6.0 &   0.29 &   3.37 &   4.63 &   2.03 &   2.34 &   2.37 &   2.37 &   2.62 &   2.33 &   0.94 &   1.60 &    4.2$\times$10$^{-2}$ \\
6-0-dil &   0.08 &   2.67 &   3.93 &   1.35 &   1.65 &   1.68 &   1.68 &   1.92 &   1.64 &   0.41 &   0.94 &    8.8$\times$10$^{-3}$ \\
7.0 &   0.12 &   2.79 &   4.14 &   1.50 &   1.80 &   1.46 &   1.23 &   1.60 &   1.52 &   0.58 &   0.51 &    6.9$\times$10$^{-3}$ \\
7.0-dil &   0.03 &   2.15 &   3.50 &   0.90 &   1.18 &   0.86 &   0.67 &   0.99 &   0.92 &   0.21 &   0.18 &    1.6$\times$10$^{-3}$ \\
7.5 &   1.67 &   2.74 &   4.16 &   1.63 &   1.47 &   1.34 &   0.95 &   1.52 &   1.34 &   0.40 &   0.53 &    4.2$\times$10$^{-3}$ \\
7.5-dil &   1.08 &   2.13 &   3.54 &   1.04 &   0.89 &   0.78 &   0.46 &   0.94 &   0.78 &   0.13 &   0.19 &    1.0$\times$10$^{-3}$ \\
\multicolumn{1}{c}{}
\end{tabular}
\tablefoot{ Data are either given by the ejecta, or under the assumption that 1$\%$ of this matter was homogeneously diluted in the surface 0.2 \msun{} of an unevolved $Z=10^{-6}$ star. Lithium abundance is shown as $<log_{10}(^{7}Li)>=Log_{10}(N(Li)/N(H))+12$. Stellar winds for these results are from \cite{bloecker1995}, with parameter $\eta=0.02$.}
\label{tab:abuns6}
\end{table*}

 \begin{table*}[ht]
    \centering
    \caption{Abundances pattern of selected elements in terms of [X/Fe] as in Figure \ref{fig:feh} $Z=10^{-7}$ star.  }
\begin{tabular}{l c c c c c c c c c c c c c}
$M_\mathrm{ini}/$\msun & $<log_{10}(^{7}Li)>$ & C & N & O & F & Ne & Na & Mg & Al & Si & P & S \\[1pt]
\hline\\
3.0 &   2.11 &   5.29 &   6.23 &   3.86 &   5.80 &   4.58 &   4.37 &   4.23 &   3.44 &   2.16 &   2.32 &    2.1E-01 \\
3.0-dil &   1.12 &   4.28 &   5.21 &   2.84 &   4.78 &   3.56 &   3.35 &   3.21 &   2.42 &   1.17 &   1.32 &    2.5E-02 \\
4.0 &   1.11 &   4.52 &   5.93 &   3.29 &   4.18 &   4.08 &   4.07 &   3.98 &   3.46 &   2.15 &   2.70 &    3.1E-01 \\
4.0-dil &   0.48 &   3.75 &   5.16 &   2.52 &   3.41 &   3.31 &   3.30 &   3.21 &   2.69 &   1.39 &   1.93 &    7.1E-02 \\
5.0 &   0.73 &   4.17 &   5.80 &   3.13 &   3.58 &   3.64 &   3.63 &   3.66 &   3.47 &   1.86 &   2.52 &    2.7E-01 \\
5.0-dil &   0.24 &   3.40 &   5.03 &   2.35 &   2.80 &   2.87 &   2.85 &   2.89 &   2.69 &   1.12 &   1.76 &    5.9E-02 \\
6.0 &   0.03 &   4.31 &   5.83 &   3.18 &   3.36 &   3.55 &   3.48 &   3.66 &   3.56 &   1.87 &   2.44 &    2.5E-01 \\
6.0-dil &   0.01 &   3.61 &   5.13 &   2.48 &   2.66 &   2.85 &   2.78 &   2.96 &   2.86 &   1.20 &   1.75 &    6.1E-02 \\
7.0 &  -0.20 &   4.09 &   5.47 &   2.81 &   3.58 &   2.95 &   2.77 &   3.02 &   2.96 &   1.47 &   1.57 &    7.6E-02 \\
7-0-dil &  -0.04 &   3.45 &   4.83 &   2.17 &   2.94 &   2.31 &   2.13 &   2.38 &   2.33 &   0.87 &   0.96 &    1.9E-02 \\
7.5 &  -0.13 &   3.71 &   5.07 &   2.39 &   2.34 &   2.32 &   2.13 &   2.54 &   2.46 &   1.20 &   1.37 &    7.5E-02 \\
7.5-dil &  -0.03 &   3.09 &   4.46 &   1.78 &   1.73 &   1.71 &   1.53 &   1.92 &   1.85 &   0.67 &   0.81 &    1.9E-02 \\
8.0 &   1.62 &   3.64 &   5.18 &   2.59 &   1.96 &   2.16 &   1.93 &   2.36 &   2.24 &   0.89 &   1.18 &    7.1E-02 \\
8.0-dil &   1.05 &   3.04 &   4.59 &   1.99 &   1.37 &   1.58 &   1.35 &   1.77 &   1.66 &   0.43 &   0.66 &    1.9E-02 \\
\end{tabular}
\tablefoot{Information is provided as in Table \ref{tab:abuns6}.\\}
\label{tab:abuns7}
\end{table*}

 \begin{table*}[ht]
    \centering
    \caption{Abundances pattern of selected elements in terms of [X/Fe] as in Figure \ref{fig:feh} $Z=210^{-8}$ star.}
\begin{tabular}{l c c c c c c c c c c c c c}
$M_\mathrm{ini}/$\msun & $<log_{10}(^{7}Li)>$ & C & N & O & F & Ne & Na & Mg & Al & Si & P & S \\[1pt]
\hline\\
3.0 &   2.09 &   5.94 &   7.13 &   4.81 &   6.08 &   5.56 &   5.36 &   5.10 &   4.36 &   3.26 &   3.68 &    1.4E+00 \\
3.0-dil &   1.11 &   4.93 &   6.12 &   3.80 &   5.06 &   4.55 &   4.35 &   4.09 &   3.35 &   2.25 &   2.67 &    5.1E-01 \\
4.0 &   1.23 &   5.58 &   7.05 &   4.47 &   5.44 &   5.10 &   5.02 &   4.90 &   4.37 &   3.00 &   3.47 &    9.8E-01 \\
4.0-dil &   0.49 &   4.71 &   6.18 &   3.60 &   4.57 &   4.23 &   4.14 &   4.03 &   3.50 &   2.13 &   2.60 &    3.3E-01 \\

6.0  &  -0.00 &   5.21 &   6.70 &   4.01 &   4.22 &   4.43 &   4.38 &   4.53 &   4.41 &   2.77 &   3.44 &    9.6E-01 \\
6.0-dil &  -0.00 &   4.51 &   6.01 &   3.31 &   3.52 &   3.73 &   3.68 &   3.83 &   3.71 &   2.07 &   2.74 &    4.2E-01 \\
7.0 &  -0.37 &   5.04 &   6.50 &   3.81 &   3.78 &   4.03 &   3.93 &   4.20 &   4.19 &   2.62 &   3.12 &    7.5E-01 \\
7.0-dil &  -0.06 &   4.40 &   5.86 &   3.17 &   3.14 &   3.39 &   3.28 &   3.56 &   3.55 &   1.98 &   2.48 &    3.1E-01 \\
7.5 &  -0.50 &   4.70 &   6.36 &   3.77 &   3.26 &   3.85 &   3.58 &   3.96 &   4.08 &   2.74 &   2.87 &    6.2E-01 \\
7.5-dil &  -0.08 &   4.08 &   5.74 &   3.15 &   2.63 &   3.22 &   2.96 &   3.34 &   3.46 &   2.11 &   2.25 &    2.4E-01 \\
8.0 &  -1.10 &   4.72 &   6.07 &   3.43 &   3.10 &   3.31 &   3.10 &   3.53 &   3.43 &   2.30 &   2.42 &    4.5E-01 \\
8.0-dil &  -0.12 &   4.13 &   5.47 &   2.84 &   2.51 &   2.72 &   2.51 &   2.93 &   2.84 &   1.71 &   1.83 &    1.6E-01 \\
8.5 &   0.41 &   4.64 &   5.98 &   3.31 &   3.23 &   3.11 &   2.83 &   3.27 &   3.16 &   1.90 &   2.05 &    3.7E-01 \\
8.5-dil &   0.15 &   4.07 &   5.40 &   2.74 &   2.66 &   2.54 &   2.25 &   2.70 &   2.59 &   1.35 &   1.49 &    1.3E-01 \\
\end{tabular}
\tablefoot{Information is provided as in Table \ref{tab:abuns6}.\\}
\label{tab:abuns8}
\end{table*}

 \begin{table*}[ht]
    \centering
    \caption{Abundances pattern of selected elements in terms of [X/Fe] as in Figure \ref{fig:feh} $Z=10^{-10}$ star.}
\begin{tabular}{l c c c c c c c c c c c c c}
$M_\mathrm{ini}/$\msun & $<log_{10}(^{7}Li)>$ & C & N & O & F & Ne & Na & Mg & Al & Si & P & S \\[1pt]
\hline\\
3.0 &   2.21 &   8.28 &   9.26 &   6.76 &   8.54 &   7.90 &   7.68 &   7.23 &   6.53 &   5.25 &   5.32 &    2.6E+00 \\
3.0-dil &   1.23 &   7.27 &   8.25 &   5.75 &   7.53 &   6.89 &   6.67 &   6.22 &   5.52 &   4.24 &   4.31 &    1.6E+00 \\
3.5 &   3.06 &   8.01 &   8.29 &   6.14 &   7.43 &   6.45 &   6.05 &   5.71 &   4.71 &   4.09 &   5.47 &    3.9E+00 \\
3.5-dil &   2.14 &   7.08 &   7.36 &   5.21 &   6.50 &   5.53 &   5.12 &   4.78 &   3.78 &   3.16 &   4.54 &    2.9E+00 \\
3.75 &   1.81 &   7.34 &   8.74 &   6.94 &   7.06 &   7.19 &   6.91 &   6.56 &   6.63 &   6.31 &   8.21 &    6.8E+00 \\
3.75-dil &   0.95 &   6.44 &   7.84 &   6.04 &   6.16 &   6.29 &   6.01 &   5.66 &   5.73 &   5.41 &   7.31 &    5.9E+00 \\
4.0 &   1.28 &   7.54 &   9.07 &   6.45 &   7.47 &   7.27 &   7.21 &   7.00 &   6.48 &   5.18 &   5.65 &    3.0E+00 \\
4.0-dil &   0.54 &   6.67 &   8.19 &   5.57 &   6.60 &   6.40 &   6.34 &   6.13 &   5.61 &   4.31 &   4.78 &    2.1E+00 \\
6.0 &  -0.28 &   7.16 &   8.73 &   5.98 &   6.27 &   6.49 &   6.44 &   6.55 &   6.43 &   4.75 &   5.38 &    2.8E+00 \\
6.0-dil &  -0.04 &   6.46 &   8.03 &   5.28 &   5.56 &   5.78 &   5.74 &   5.85 &   5.72 &   4.05 &   4.68 &    2.1E+00 \\
7.0 &  -0.54 &   6.84 &   8.58 &   5.78 &   5.69 &   6.13 &   5.99 &   6.12 &   6.17 &   4.48 &   4.80 &    2.2E+00 \\
7.0-dil &  -0.08 &   6.20 &   7.94 &   5.13 &   5.04 &   5.49 &   5.34 &   5.47 &   5.52 &   3.84 &   4.15 &    1.5E+00 \\
7.5 &  -0.69 &   6.97 &   8.47 &   5.83 &   6.18 &   5.91 &   5.70 &   6.06 &   6.02 &   4.50 &   4.68 &    2.1E+00 \\
7.5-dil &  -0.09 &   6.35 &   7.85 &   5.21 &   5.56 &   5.29 &   5.08 &   5.44 &   5.40 &   3.88 &   4.06 &    1.5E+00 \\
8.0 &  -2.30 &   5.00 &   6.46 &   3.15 &  -0.17 &   0.65 &  -0.05 &  -0.32 &   0.07 &   0.52 &   0.72 &    4.0E-01 \\
8.0-dil &  -0.13 &   4.41 &   5.86 &   2.56 &  -0.04 &   0.28 &  -0.01 &  -0.06 &   0.02 &   0.20 &   0.32 &    1.4E-01 \\
8.5 &   0.05 &   5.85 &   7.43 &   4.24 &   1.59 &   3.67 &   3.23 &   2.79 &   3.22 &   2.14 &   1.20 &    4.1E-01 \\
8.5-dil &   0.01 &   5.27 &   6.85 &   3.66 &   1.04 &   3.10 &   2.65 &   2.22 &   2.64 &   1.57 &   0.69 &    1.5E-01 \\
\end{tabular}
\tablefoot{Information is provided as in Table \ref{tab:abuns6}.\\}
\label{tab:abuns10}
\end{table*}
\end{appendix}
\end{document}